\documentclass{jfm}
\pdfoutput=1
\usepackage{graphicx}
\usepackage{newtxtext}
\usepackage{newtxmath}
\usepackage{natbib}
\usepackage{hyperref}
\usepackage[font=small,justification=justified,format=plain]{caption}
\usepackage{subcaption}
\usepackage{dblfloatfix}
\usepackage{float}
\usepackage{varioref}
\usepackage{appendix}

\hypersetup{
    colorlinks = true,
    urlcolor   = blue,
    citecolor  = black,
}

\newcommand{\RomanNumeralCaps}[1]
\linenumbers

\graphicspath{{Figs/}{Figs/axisymmetric/}{Figs/planar/}{Figs/3D/}{Figs/others/}}

\title{Linear stability of a rotating liquid column revisited}

\author{Pulkit Dubey\aff{1} \aff{3}
   Anubhab Roy\aff{2}
   \and Ganesh Subramanian\aff{3}
   \corresp{\email{sganesh@jncasr.ac.in}}
   }

\affiliation{
   \aff{1} Department of Mathematics and Statistics, University of New Hampshire, Durham, NH 03824, USA
   \aff{2} Department of Applied Mechanics, Indian Institute of Technology Madras, Chennai, Tamil Nadu 600036, India
   \aff{3} Engineering Mechanics Unit, JNCASR, Bangalore, Karnataka 560064, India
   }

\begin{document}

\maketitle

\begin{abstract}
	We revisit the somewhat classical problem of the linear stability of a
rigidly rotating liquid column in this communication. Although literature
pertaining to this problem dates back to 1959, the relation between inviscid
and viscous stability criteria has not yet been clarified. While the viscous
criterion for stability, given by $We < n^2 + k^2 -1$, is both necessary and
sufficient, this relation has only been shown to be sufficient in the
inviscid case. Here, $We = \rho \Omega^2 a^3 / \gamma$ is the Weber number
and measures the relative magnitudes of the centrifugal and surface tension
forces, with $\Omega$ being the angular velocity of the rigidly rotating
column, $a$ the column radius, $\rho$ the density of the fluid, and $\gamma$
the surface tension coefficient; $k$ and $n$ denote the axial and azimuthal
wavenumbers of the imposed perturbation. We show that the subtle difference
between the inviscid and viscous criteria arises from the surprisingly
complicated picture of inviscid stability in the $We-k$ plane. For all $n >
1$, the viscously unstable region, corresponding to $We > n^2 + k^2-1$,
contains an infinite hierarchy of inviscidly stable islands ending in cusps,
with a dominant leading island. Only the dominant island, now infinite in
extent along the $We$ axis, persists for $n=1$. This picture may be
understood, based on the underlying eigenspectrum, as arising from the
cascade of coalescences between a retrograde mode, that is the continuation
of the cograde surface-tension-driven mode across the zero Doppler frequency
point, and successive retrograde Coriolis modes constituting an infinite
hierarchy.

\end{abstract}

\begin{keywords}
	Linear stability, Rotating liquid columns, Cusp catastrophe 
\end{keywords}

\section{Introduction}
\label{sec:introduction}
	
This article discusses the linear stability of a rigidly rotating liquid column.
The limit of zero rotation corresponds to the classical Rayleigh-Plateau problem
analyzed first, in the inviscid limit, by \cite{plateau_1873} and later by
\cite{rayleigh_1878}. The subsequent literature on the rigidly rotating liquid
column [\cite{hocking_michael_1959, hocking_1960, gillis_1962, pedley_1967}]
primarily focused on the necessary and/or sufficient condition for instability,
although later \cite{weidman_1997} examined the dominant unstable modes for the
inviscid rotating column based on growth-rate calculations. More recently,
\cite{kubitschek_weidman_2007} obtained the dominant modes for the viscous rotating column, and organized their
results based on the wavenumber of the dominant perturbation, in a parameter
plane consisting of the Weber number (a dimensionless measure of rotation
defined below) and the column Reynolds number. The boundaries demarcating the crossover of the dominant mode in this plane converged smoothly to the inviscid predictions of \cite{weidman_1997} for $Re\to \infty$. The later experiments of
\cite{kubitschek_2007} were consistent with the modal crossover boundaries obtained in \cite{kubitschek_weidman_2007}. While all of the above results show the expected destabilizing effect of rotation, in terms of a larger range of
wavenumbers turning unstable with an increase in the column angular velocity, on
account of centrifugal forces, there remains a difference between the viscous
and inviscid criteria for instability obtained in the early literature\,[\cite{hocking_michael_1959, hocking_1960, gillis_1962, pedley_1967}]. Although the expression for the stability threshold (see eq.
\ref{eq:3d_stability_criterion} below) remains the same in both cases, it has
been shown to be necessary and sufficient in the presence of viscosity
[\cite{gillis_1962}], but only serves as a sufficient condition in the inviscid
limit [\cite{pedley_1967, weidman_1994, henderson_2002}]. In this article, we
re-examine the instability of a rigidly rotating liquid column, with an emphasis
on the entire inviscid spectrum, including the neutral modes. The analysis sheds
new light on this problem, showing that inviscid unstable modes arise from an
infinite hierarchy of coalescences between pairs of dispersion curves just above
the viscous threshold. The resulting intricate picture helps explain the
aforementioned difference between the nature of the inviscid and viscous
threshold criteria.

Perturbations to a liquid column may be characterized in terms of their
azimuthal $(n)$ and axial $(k)$ wavenumbers, and accordingly, may be classified
as axisymmetric ($n = 0, k\neq 0$), planar ($n \neq 0,k = 0$), and helical\,(spiral) or
three-dimensional perturbations ($n \neq 0,  k \neq 0$). Starting in section 2,
we study the dispersion curves, and the associated stability thresholds, for a
rigidly rotating column of liquid subject to each of the aforementioned classes
of perturbations. The nature of the dispersion curves is a function of the Weber
number, $We = \rho \Omega^2 a^3 / \gamma$, a dimensionless parameter that
compares the relative importance of centrifugal and surface tension forces;
here, $\rho$ is the density, $\Omega$ the column angular velocity, $a$ the
column radius, and $\gamma$ the coefficient of surface tension. For helical
perturbations, we analyze the dispersion curves in the $We-k$ plane for
different fixed $n$'s. Note that, following early work by Hocking
[\cite{hocking_michael_1959, hocking_1960}], $We^{-1}$ has often been referred
to as the Hocking parameter ($L$); for instance, the aforementioned efforts of \cite{weidman_1997} and \cite{kubitschek_weidman_2007} presented their results for the dominant inviscid modes as a function of $L$, and the dominant viscous modes in the $L-Re$ plane, respectively. In what follows, we stick to $We$. In the
next paragraph, we begin by recapitulating the well known results for the
classical case of a stationary liquid column ($We = 0$).

The classical Rayleigh-Plateau instability is one of a stationary liquid column
to sufficiently long wavelength axisymmetric perturbations, and explains the
spontaneous breakup of a (slow) jet into nearly uniformly sized droplets (see
\cite{chandrasekhar_1981}; for sufficiently slow speeds, the shear at the
air-water interface is unimportant, and the jet may be made equivalent to a
stationary column via a Galilean transformation). \cite{plateau_1873} concluded,
via a quasi-static analysis, that perturbations with an axial wavelength greater
than the circumference of the column ($ka<1$ or, if the wavenumber is scaled
with the column radius, $k < 1$) act to destabilize the column by decreasing the
total interfacial area. \cite{rayleigh_1878} then accounted for both inertia and
surface tension, obtaining the following dispersion relation for small amplitude
Fourier mode perturbations, proportional to $e^{i(kx + n\theta - \sigma t)}$, in
the inviscid limit
\begin{equation}
   \sigma^2 = k\frac{I_n'(k)}{I_n(k)}(k^2+n^2-1),
   \label{eq:RP_3d_dispersion}
\end{equation}
where $\sigma = \sigma_r + i\sigma_i$ has been scaled with $\sqrt{\gamma/\rho a^3}$
($\sigma_r$ being the frequency and $\sigma_i$ being the growth rate), and the
axial wavenumber $(k)$ is now scaled with $1/a$; $I_n(k)$ is the modified Bessel
function of the first kind with the prime denoting differentiation. From (1.1),
only axisymmetric perturbations are found to be unstable ($\sigma_i > 0$) for $k
< 1$, with the maximum growth rate corresponding to $k = 0.697$. The growing and
decaying modes transform to a pair of neutral modes across $k = 1$, the latter
corresponding to capillary waves propagating in opposite directions along the
axis of the column. For $k \rightarrow \infty$, one obtains $\sigma =
\sqrt{\gamma k^{3} / \rho}$ regardless of $n$, which is the dispersion relation
for capillary waves propagating on an infinite plane interface.

\cite{hocking_michael_1959} and \cite{hocking_1960} first investigated the
effects of rotation on a liquid column subject to planar and axisymmetric
perturbations, obtaining the necessary and sufficient criteria for stability.
For the axisymmetric case, the authors obtained the criterion $We < k^2 -1$,
regardless of viscosity. For the planar case, the authors found the inviscid
criterion to be $We < n(n+1)$, while that for any finite viscosity to be $We <
n^2 -1$ (see also \cite{gillis_1961}). \cite{gillis_1962} later studied
three-dimensional perturbations of a viscous rotating column and concluded that
the latter criterion generalizes to $We < k^2 + n^2 - 1$. \cite{pedley_1967}
showed that although the aforementioned viscous criterion remains relevant in
the inviscid limit, it only serves as a sufficient condition for stability.
Thus, for inviscid columns, a necessary and sufficient condition is not yet
known, and clarifying the above difference between the viscous and inviscid
criteria is the subject of this effort. As stated above, our focus on the entire
eigenspectrum allows us to understand in detail the regions in parameter space
corresponding to the inviscid unstable modes while also pointing to the
necessary and sufficient criterion for inviscid stability.

While the main findings of the present effort pertain to helical perturbations,
we nevertheless consider all three classes of perturbations mentioned above, in
sequence, and a complete picture of linear stability emerges as a consequence.
Thus, section \ref{sec:rotating_column} below starts off with a brief
description of the linear stability formulation, which is then followed by
subsections pertaining to axisymmetric (section \ref{sec:axisym_ptb}) and planar
(section \ref{sec:planar_ptb}) perturbations. We detail our new findings for
three-dimensional perturbations in section \ref{sec:3d_ptb}. In the conclusions
section (section \ref{sec:conclusion}), 
we show that our findings with regard to the non-trivial nature of the inviscid
spectrum carry over to the case where the interfacial cohesion underlying
surface tension is replaced by a volumetric cohesion mechanism, that of
self-gravitation, instead. This makes our findings relevant to the astrophysical
scenario, and we end with a few pertinent comments in this regard.

\section{The Rotating Liquid Column}
\label{sec:rotating_column}

The rigidly rotating columnar base state corresponds to $u_r = 0$, $u_\theta =
\Omega r$, $u_z = 0,$ and $p = p_0 + \frac{\rho \Omega^2 r^2}{2}$ for $r < a$,
where $p_o$ is an arbitrary baseline pressure on account of incompressibility.
The governing linearized equations for small-amplitude perturbations may be
derived in the usual way from the Euler equations, with kinematic (radial
velocity) and dynamic (pressure) boundary conditions at the column free surface.
The equations governing inviscid evolution have already been written down and
solved in earlier efforts [\cite{hocking_michael_1959, hocking_1960,
weidman_1997}], and in what follows, we directly examine the resulting
dispersion relations. Note that the density of the exterior fluid is assumed to
be small relative to that of the liquid column, and its influence on column
oscillations is neglected.

\subsection{Axisymmetric Perturbations}
\label{sec:axisym_ptb}
	
The dispersion relation for axisymmetric perturbations was obtained by Hocking
(1960), and is given by
\begin{equation}
   We \; \sigma^{2} \sqrt{\frac{4}{\sigma^{2}}-1} 
   \frac{J_{0}(\alpha)}{J_{1}(\alpha)} - k\left(k^{2}-1\right) + We \; k=0,
   \label{eq:axisymmetric_dispersion}
\end{equation}
where $\alpha = k\sqrt{\frac{4}{\sigma^2} - 1}$. Here, as before, $k$ is scaled 
with $1/a$ but $\sigma$ is scaled with $\Omega$ as opposed to 
$\sqrt{\gamma k^3 / \rho}$ used in eq. \ref{eq:RP_3d_dispersion}. Using eq. 
\ref{eq:axisymmetric_dispersion}, Hocking obtained the necessary and sufficient 
criterion for stability to be
\begin{equation}
   We < k^2 - 1
   \label{eq:axisymmetric_criterion}
\end{equation}
indicating that centrifugal forces destabilize the system, increasing the 
interval of unstable wavenumbers from $(0,1)$ in the non-rotating case to 
$(0,\sqrt{1+We})$ in the rotating case.

% Axisymmetric dispersion curves for RP, Rankine and liquid column%

\begin{figure}
   \centering
\begin{subfigure}{0.48\textwidth}
  \centering
  \includegraphics[width=\linewidth]{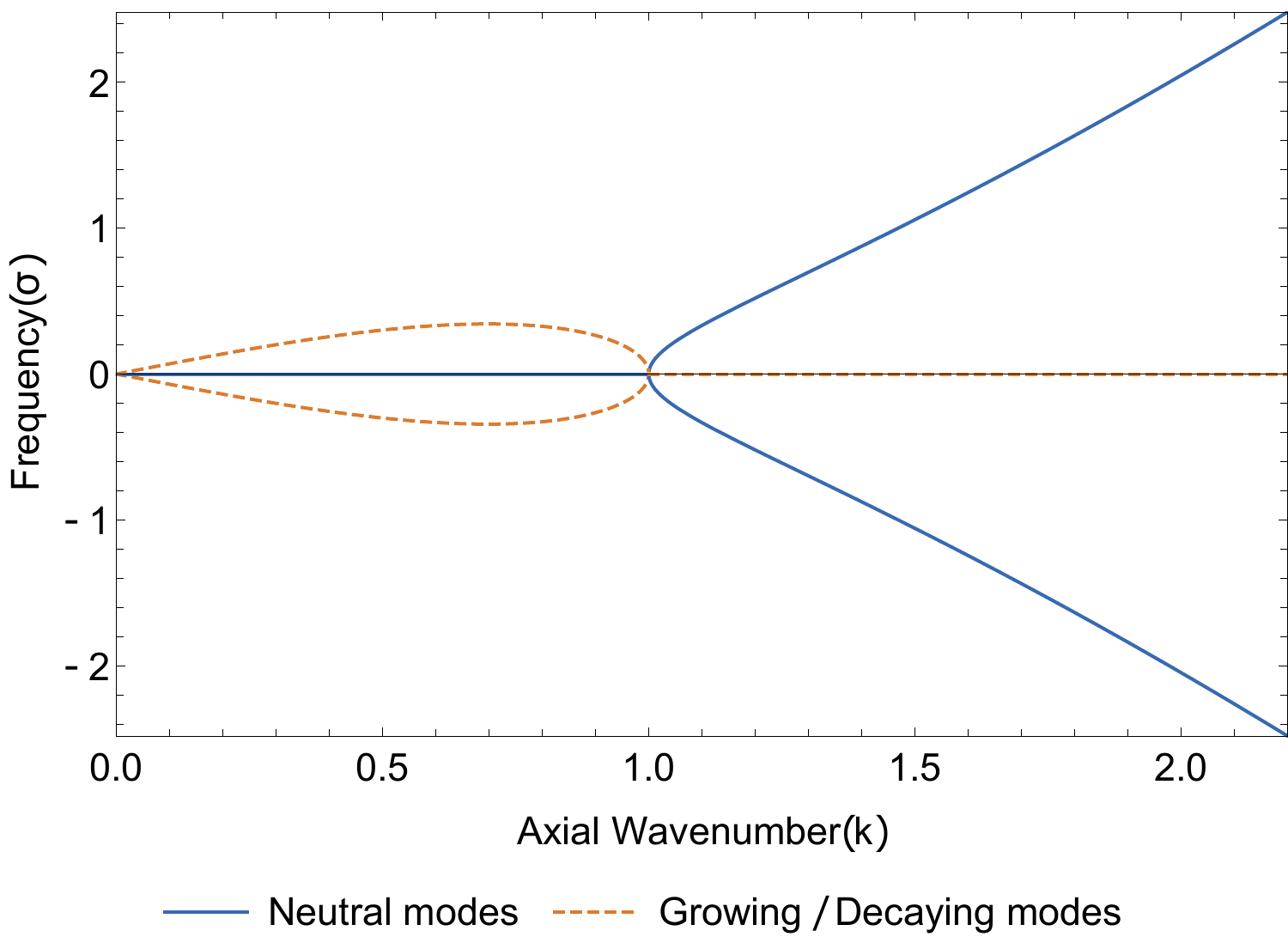}
	\caption {Rayleigh-Plateau problem}
   \label{fig:axisymmetric_dispersion_curves-a}
\end{subfigure}
\begin{subfigure}{0.48\textwidth}
  \centering
  \includegraphics[width=\linewidth]{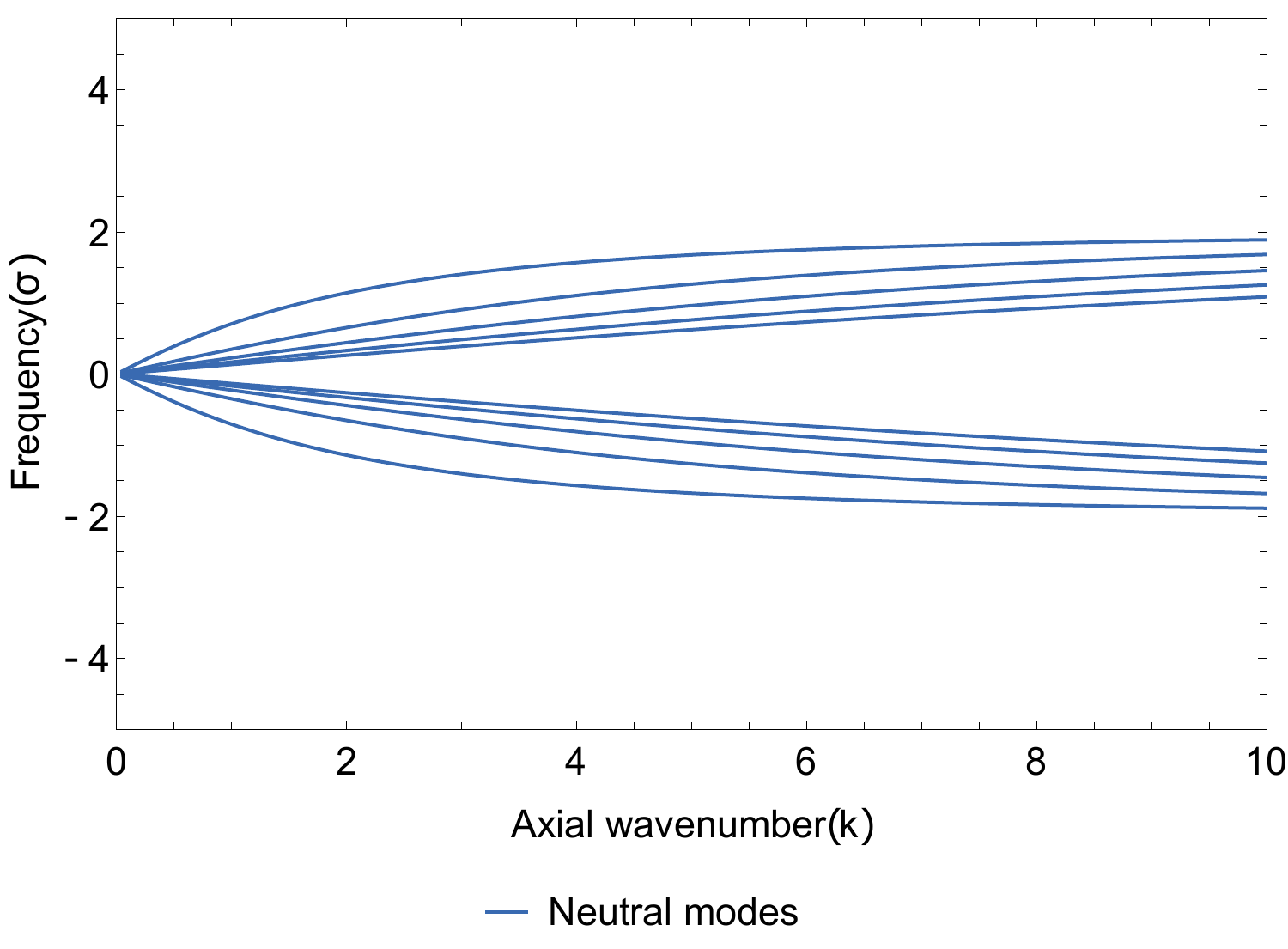}
    \caption{Rankine vortex}
    \label{fig:axisymmetric_dispersion_curves-b}
\end{subfigure}
\begin{subfigure}{0.98\textwidth}
   \centering
   \includegraphics[width=\linewidth]{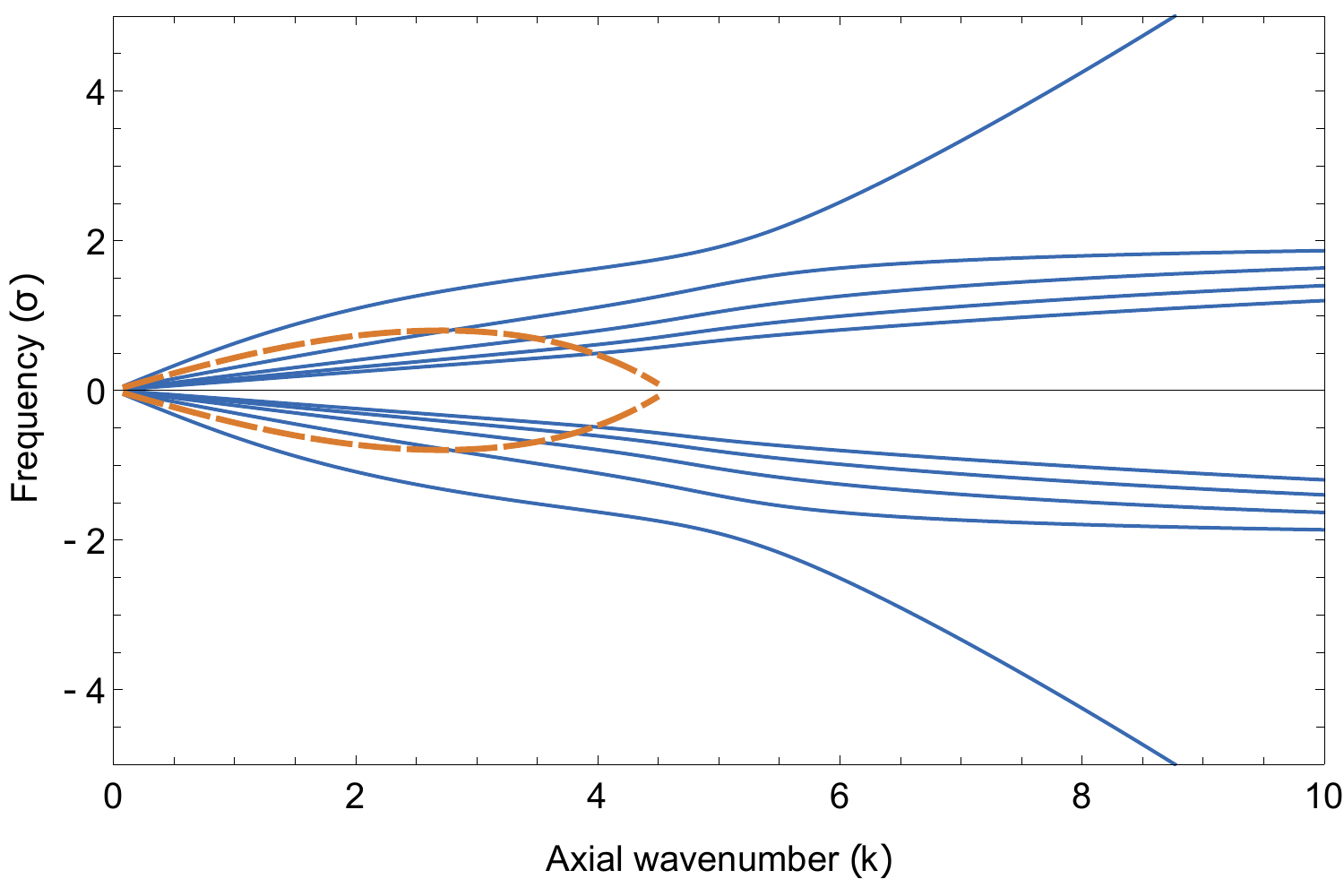}
    \caption{Rotating liquid column}
    \label{fig:axisymmetric_dispersion_curves-c}
 \end{subfigure}
\caption{The dispersion curves for the Rayleigh-Plateau problem (top-left), the
Rankine vortex (top-right) and the rotating liquid column with $We = 20$
(bottom). Blue curves denote neutral modes while the dotted orange curves denote
growing/decaying modes. Only the first ten of the infinite number of neutral modes are
shown for the Rankine vortex and the rotating column.}
\label{fig:axisymmetric_dispersion_curves}
\end{figure}

\begin{figure}
   \centering
\begin{subfigure}{0.48\textwidth}
  \centering
  \includegraphics[width=\linewidth]{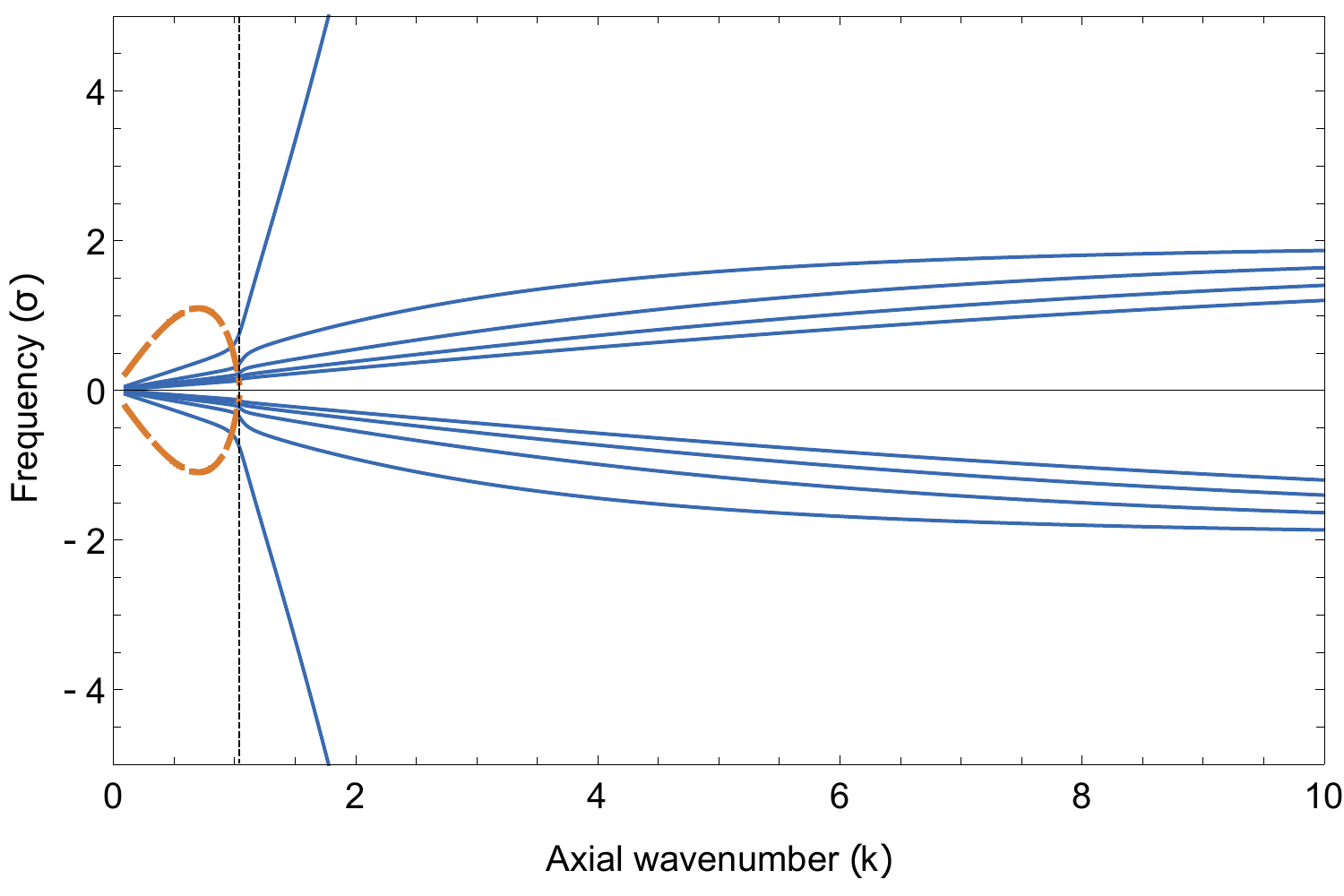}
  \caption{$We = 0.1$}
%  \label{fig:}
\end{subfigure}
\begin{subfigure}{0.48\textwidth}
  \centering
  \includegraphics[width=\linewidth]{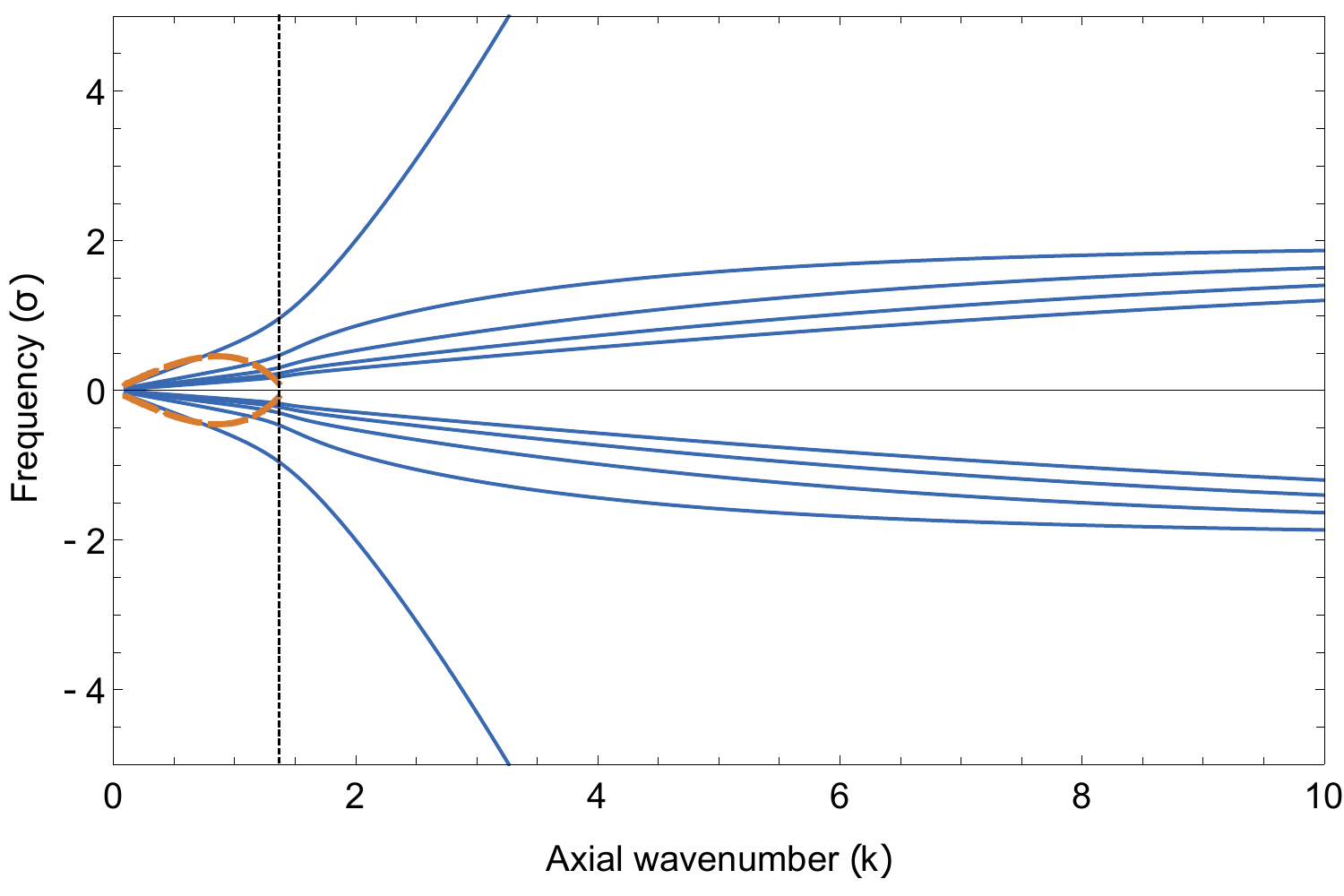}
	\caption{$We = 1$}
%  \label{fig:}
\end{subfigure}
\begin{subfigure}{0.48\textwidth}
   \centering
   \includegraphics[width=\linewidth]{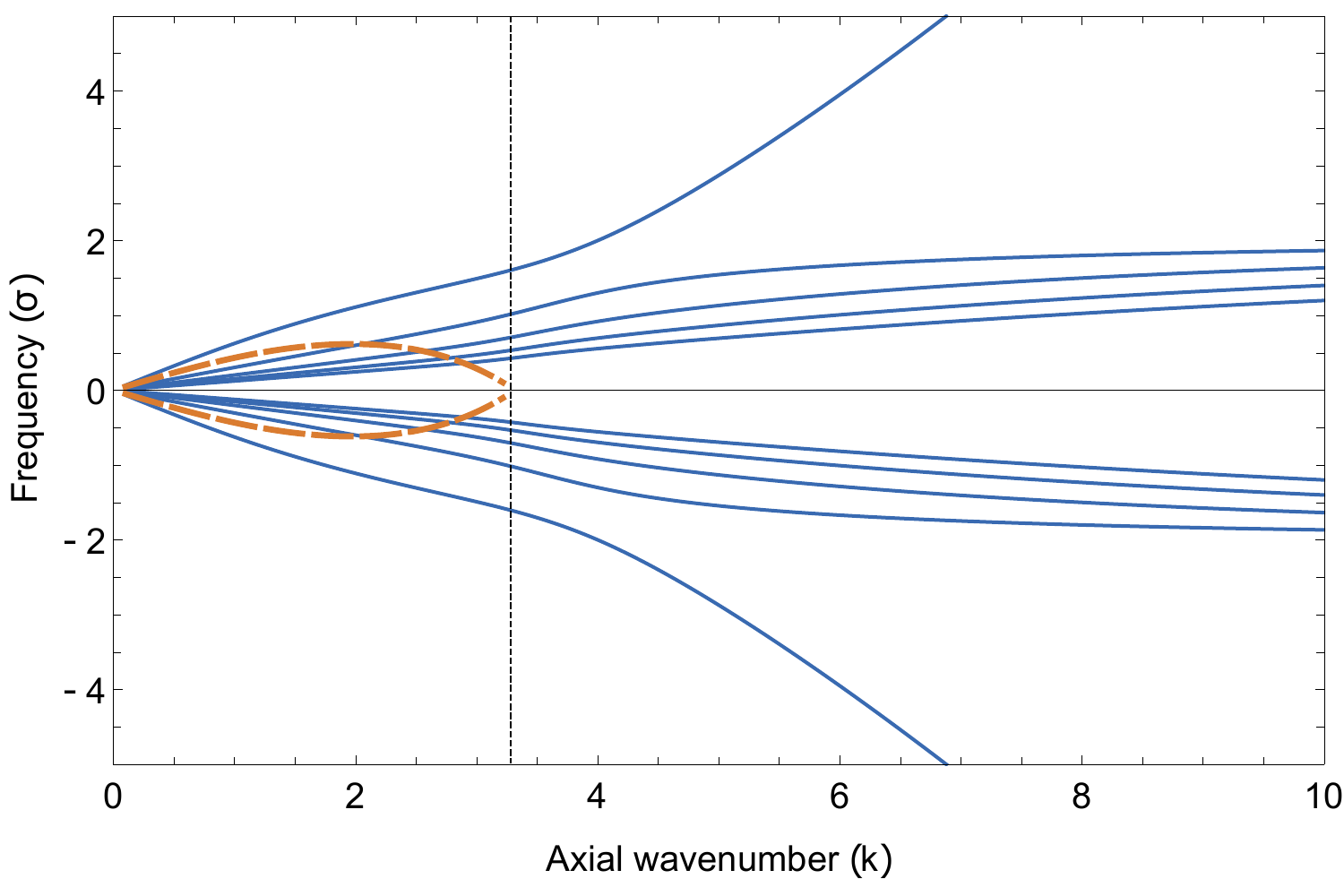}
    \caption{$We = 10$}
 %  \label{fig:}
 \end{subfigure}
\begin{subfigure}{0.48\textwidth}
  \centering
  \includegraphics[width=\linewidth]{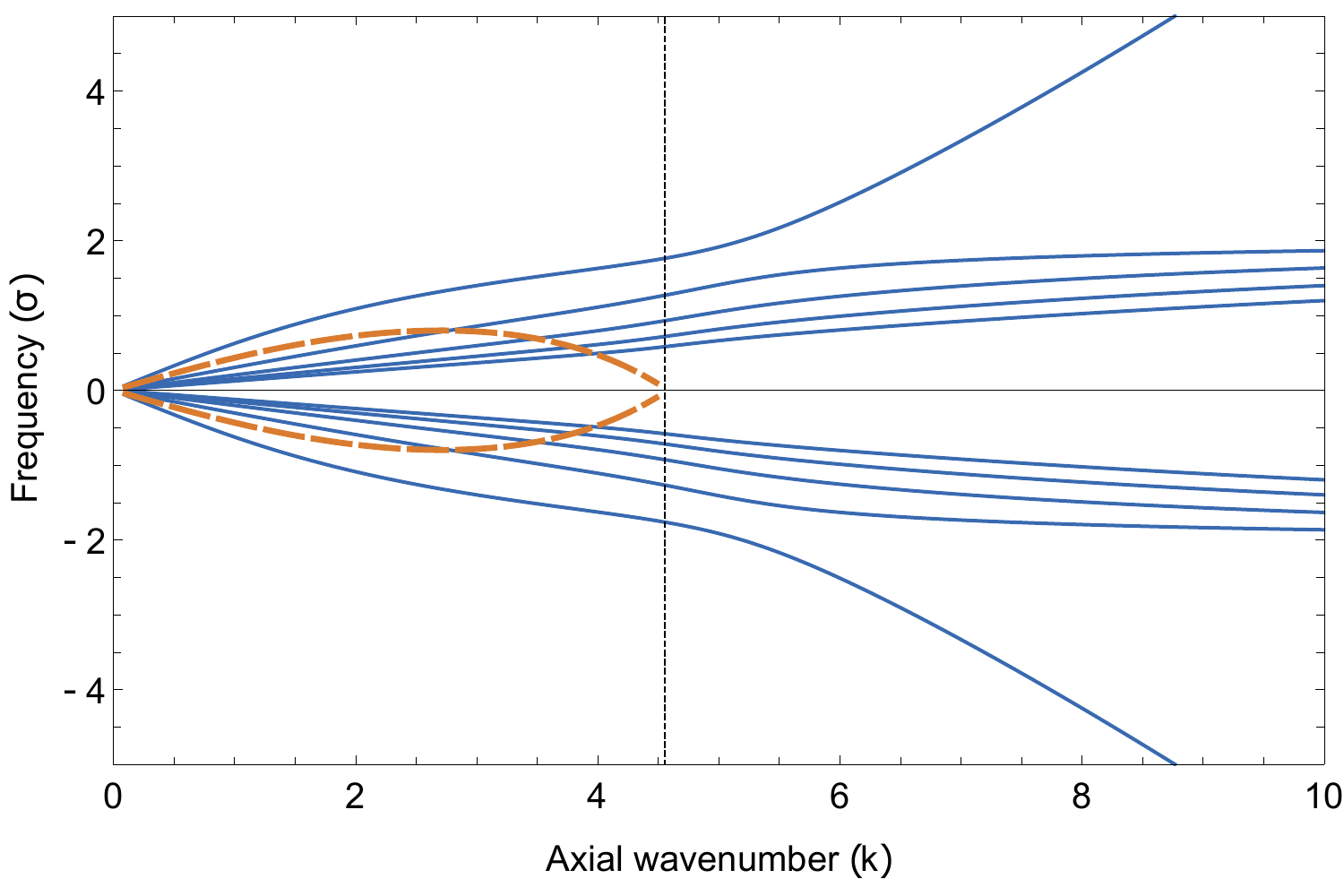}
	\caption{$We = 20$}
%  \label{fig:}
\end{subfigure}
\caption{The dispersion curves of the rotating liquid column with varying $We$.
Blue curves denote neutral modes while dashed orange curves denote growing /
decaying modes. A vertical dashed line drawn at $k = \sqrt{1+We}$ indicates the
axial wavenumber beyond which the effects of surface tension on the neutral modes
become significant.}
\label{fig:axisymmetric_dispersion_We}
\end{figure}

Fig. \ref{fig:axisymmetric_dispersion_curves-c} shows the dispersion curves for
the axisymmetrically perturbed column obtained from eq.
\ref{eq:axisymmetric_dispersion}. The spectrum is seen to borrow its traits from
two constituent cases - the Rayleigh-Plateau configuration involving only
surface tension (Fig. \ref{fig:axisymmetric_dispersion_curves-a}) and the
Rankine vortex involving only rotation (Fig.
\ref{fig:axisymmetric_dispersion_curves-b}). Much like the Rankine vortex, the
rotating liquid column supports an infinite sequence of primarily
Coriolis-force-driven modes (henceforth referred to as the Coriolis modes), with
the inner dispersion curves corresponding to perturbations with an increasingly
fine-scaled radial structure; note that the Rankine vortex has recently been
shown to also possess a continuous spectrum on account of the irrotational shear
in the column exterior [\cite{roy_subramanian_2014}, \cite{roy_2021}]. The
Coriolis modes in both these problems have frequencies $\sigma_r \in (-2,2)$.
Additionally, the rotating liquid column has two modes that owe their origin to
surface tension akin to the Rayleigh-Plateau configuration (henceforth referred
to as the capillary modes). Fig. \ref{fig:axisymmetric_dispersion_curves-c}
shows the family of Coriolis modes along with the two capillary modes, and their
variation with $k$, for $We = 20$. Note that the pair of capillary modes follow
the characteristic $k^{3/2}$ scaling for large $k$, while the Coriolis mode
frequencies approach $\sigma_r = \pm 2$ in this limit. The effect of $We$ on the
dispersion curves is illustrated in Fig. \ref{fig:axisymmetric_dispersion_We}.
For $k \ll \sqrt{We + 1}$, Coriolis forces dominate the spectrum, and it is only
for $k > \sqrt{We + 1}$ that the effects of surface tension become apparent,
with the pair of capillary mode dispersion curves transitioning to the
aforementioned $k^{3/2}$ scaling. For $We \gg k^2-1$, eq.
\ref{eq:axisymmetric_dispersion} reduces to $\sigma^{2}
\sqrt{\frac{4}{\sigma^{2}}-1} \frac{J_{0}(\alpha)}{J_{1}(\alpha)} + k = 0$, a
$We$-independent dispersion relation; the transition $k$ has receded to infinity
and the entire spectrum is governed by Coriolis forces.

% Axisymmetric dispersion curves of the liquid column - variation with We %

The description above clearly shows that there is a qualitative change in the
nature of the eigenspectrum with the onset of rotation, as evident from
comparing Figs. \ref{fig:axisymmetric_dispersion_curves-a} and
\ref{fig:axisymmetric_dispersion_curves-c}, and points to the stationary column
$(We \rightarrow 0)$ being a singular limiting case. The interfacial dynamics
for the stationary column is entirely determined by the pair of capillary modes
(whether neutral or unstable), which are now irrotational. For these modes, the
non-dimensional frequency ($\sigma$) in eq. \ref{eq:axisymmetric_dispersion}
diverges in the limit of zero rotation. Thus, $\alpha$ reduces to $ik$ and one
recovers the Rayleigh-Plateau dispersion relation given in eq.
\ref{eq:RP_3d_dispersion}. In contrast, the Coriolis modes, which remain vortical
in the zero-rotation limit, may be regarded as a complete set of stationary
vortical perturbations that do not perturb the column free surface. For these
modes, the non-dimensional frequency $\sigma$ remains finite (implying that the
dimensional one vanishes as $O(\Omega)$), and therefore, so does $\alpha$. The
dispersion relation reduces to $\mathit{We}\, J_0(\alpha)/J_1(\alpha) = k(k^2 -
1)$, implying that $J_1(\alpha) = 0$ governs the Coriolis mode frequencies for
$We=0$; this also ensures that the radial velocity vanishes at $r = 1$ (see eq. 3.2a in \cite{kubitschek_weidman_2007}),
consistent with the aforementioned requirement of an unperturbed free surface.
An analogous limiting scenario prevails for any non-zero $n$, with the limiting
dispersion relation being given by $\sigma-n = \frac{2n\,J_n(\alpha)}{\alpha
J_n'(\alpha)}$; thus, the Coriolis modes, in all cases, reduce to stationary
vortical perturbations as $We \rightarrow 0$, and are irrelevant to the dynamics
of free surface disturbances.

Finally, it needs mentioning that while the growing and decaying modes for the
rotating column exist over a wider range of wavenumbers given by $0 < k <
\sqrt{1+We}$, they do not emerge as a result of coalescence of neutral modes, as
was the case for the Rayleigh-Plateau problem (where the so-called `principle of
exchange of stability' holds); instead, as evident from the dashed dispersion
curves in Figs. \ref{fig:axisymmetric_dispersion_curves} and
\ref{fig:axisymmetric_dispersion_We}, they arise independently as the stability
threshold is crossed. Importantly, since the eigenvalues associated with these
modes are purely imaginary, the state of neutral stability is one of rigid
rotation, and therefore, as pointed out by \cite{rosenthal_1962}, viscosity
cannot alter the threshold. Thus, $We = k^2 - 1$ remains the threshold
regardless of the column Reynolds number (which may be defined as $Re = \Omega
a^2/\nu$, $\nu$ being the kinematic viscosity).

\subsection{Planar Perturbations}
\label{sec:planar_ptb}
	Next, we examine planar perturbations, the inviscid dispersion relation for
which was first obtained by \cite{hocking_michael_1959} and is given by
\begin{equation}
   \sigma = n-1 \pm \sqrt{\frac{(n-1)(n(n+1)-We)}{We}},
   \label{eq:planar_dispersion}
\end{equation}
which is algebraic (quadratic) instead of transcendental, as was the case for
axisymmetric perturbations (see eq. \ref{eq:axisymmetric_dispersion}), and as is
also the case for helical perturbations (see eq.
\ref{eq:3D_rotating_column_dispersion}). The existence of only a pair of planar
modes is because the infinite number of Coriolis modes degenerate to $\sigma =
n$ for $k \rightarrow 0$ in the inviscid limit; see Figs
\ref{fig:axisymmetric_dispersion_curves-c} and \ref{fig:3D_regime1}. For any
$We$, the two planar-wave frequencies are symmetrically distributed about $n-1$.
For small $We$, they are asymptotically large (of $O(We^{-1/2})$, which
corresponds to a dimensional frequency of $\sqrt{\gamma/\rho a^3}$, as expected,
and approach each other with increasing $We$. For $n \geq 2$, the modes coalesce
at $\sigma = n-1$ for $We = n(n+1)$, the inviscid threshold mentioned earlier,
and become complex valued for $We > n(n+1)$, implying instability.

Unlike the axisymmetric case, viscosity has a profound effect on the stability
of the rotating column to planar perturbations. To see this, let $\Lambda =
\frac{n(n^2-1)}{We} - n$ so that the pair of inviscid frequencies above may be
written as $\sigma = n-1 \pm \sqrt{1+\Lambda}$ and the inviscid threshold
corresponds to $\Lambda = -1$. For large but finite $Re$, \cite{hocking_1960}
obtained the following expressions for the planar-wave frequencies, corrected
for viscous effects, again as the solutions of a quadratic equation 
\begin{equation}
   \sigma = n - 1 \pm \sqrt{1+\Lambda} -
   \frac{i}{Re} \left(2n(n-1)\frac{-1\pm\sqrt{1+\Lambda}}{\pm \sqrt{1+\Lambda}}\right).
   \label{eq:planar_viscous}
\end{equation}
Note that the $O(1/Re)$ scaling for the viscous correction, implied by eq.
\ref{eq:planar_viscous}, is not valid near the inviscid threshold as the
expression within brackets diverges for $\Lambda \to -1$. One may nevertheless
derive the requirement for viscous instability. This corresponds to
$\sigma$ in eq. \ref{eq:planar_viscous} having a negative imaginary part which
in turn translates to $\Lambda = 0$ (bounded away from the aforementioned
breakdown value), or $We = n^2-1$. This threshold was shown to be applicable for
columns of all finite $Re$ [\cite{gillis_1961}]. Rather counterintuitively, on
one hand, the viscous threshold does not depend on $Re$, leading to the
aforementioned discontinuous jump in the threshold $We$ from $n(n+1)$ to
$n^2-1$. On the other hand, the viscous threshold for stability is less than the
inviscid threshold, implying the destabilizing influence of viscosity. The
latter behavior is attributed to the phase difference between pressure and
displacement waves. The two waves are exactly out of phase in the inviscid
limit. This is no longer true in presence of viscous effects which allow for a
net work done during a single oscillation. The role of viscosity in modifying
the phase difference, and thereby inducing exponential growth, is reminiscent of
the  Miles mechanism that accounts for the growth of wind-driven gravity waves
[\cite{miles_1957, benjamin_1959}]; although, on account of the rigidly rotating base state,
there isn't the complicating effect of a critical layer [\cite{miles_1957}] in the present problem. 

\begin{figure}
   \begin{subfigure}{0.49\textwidth}
      \centering
 	   \includegraphics[width=\textwidth]{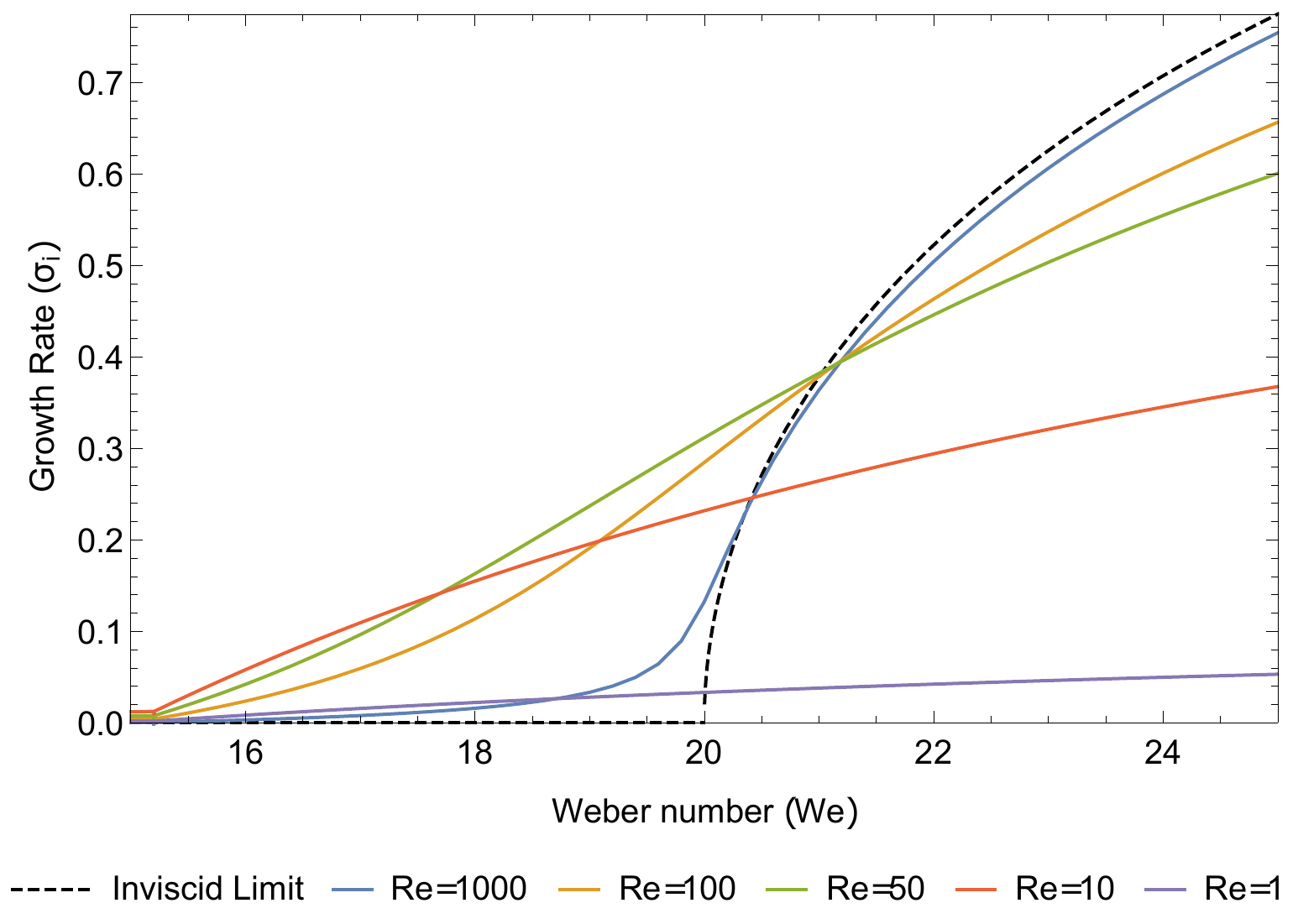}
 	   \caption{Growth rates vs $We$ for various $Re$.}
 	   \label{fig:planar_growth_rate_vs_We}
   \end{subfigure}
   \begin{subfigure}{0.49\textwidth}
      \centering
      \includegraphics[width=\textwidth]{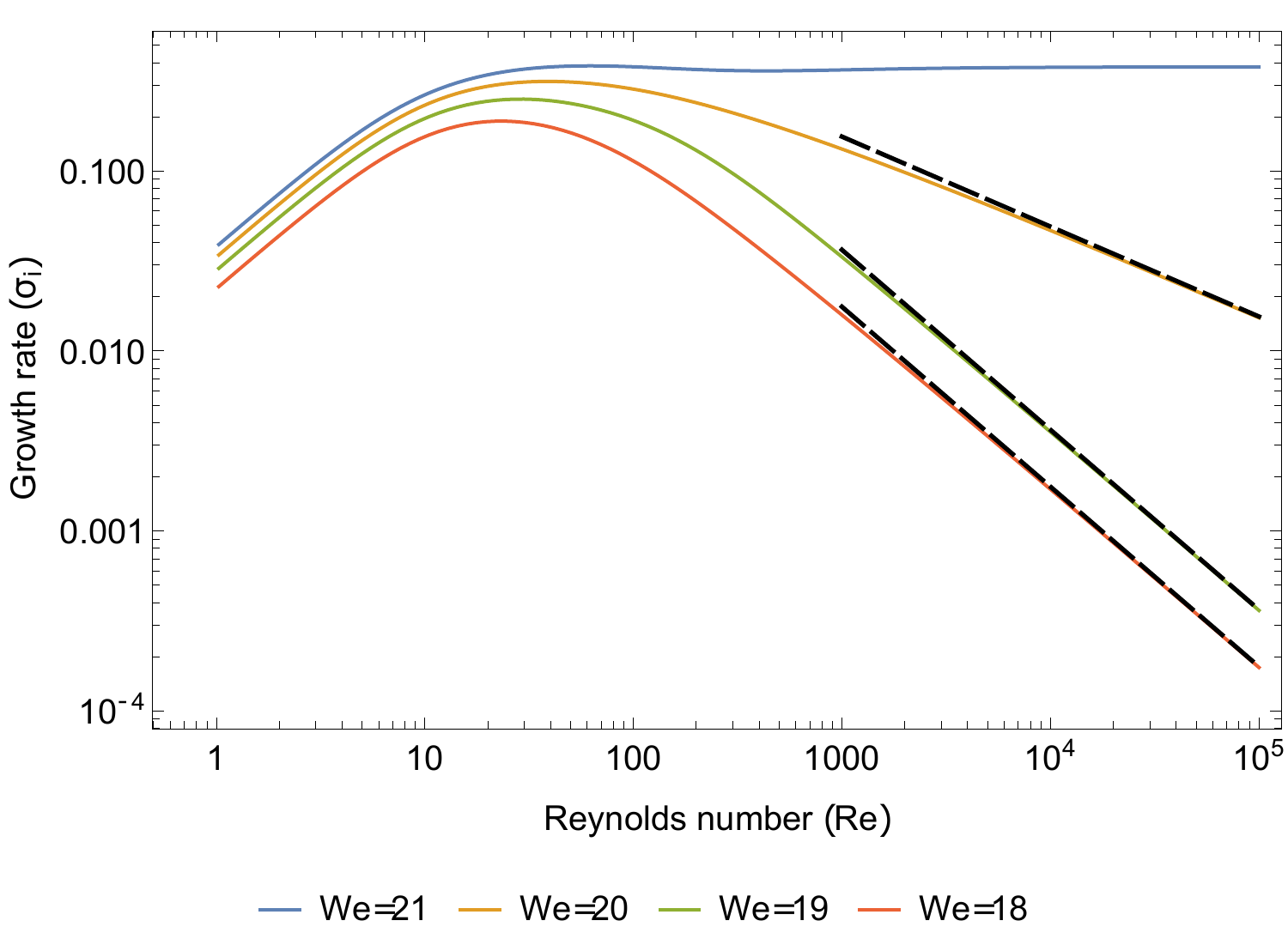}
      \caption{Growth rates vs $Re$ for various $We$.}
      \label{fig:planar_growth_rate_vs_Re}
   \end{subfigure}
   \caption{The two figures plot the growth rates of planar perturbations as a
   function of $We$ and $Re$, respectively, for $n = 4$. Note the discontinuous
   change in the stability criterion from $We = 20$, for the inviscid case (the
   dotted black curve in (a)) to $We = 15$ for any finite $Re$. Dashed black
   lines in (b) denote the large $Re$ asymptotes obtained from the planar viscous
   dispersion relation due to \cite{hocking_1960}.}
   \label{fig:planar_growth_rate_with_We_Re}
\end{figure}

Figs. \ref{fig:planar_growth_rate_vs_We} and \ref{fig:planar_growth_rate_vs_Re}
show that although the stability threshold changes discontinuously for any
finite $Re$, the growth rates in the interval between the inviscid and viscous
thresholds ($n^2 - 1 < We < n(n+1)$) scale viscously, and therefore, decrease to
zero for $Re \rightarrow \infty$ (see also the figure in \cite{gillis_1961} and
Fig. 6c in \cite{kubitschek_weidman_2007}). For $We \in (n^2-1, n(n+1))$, the
growth rates scale as $Re^{-1}$, while remaining $O(1)$ for $We > n(n+1)$.
Within a small $O(Re^{-2})$ interval around $We = n(n+1)$, the growth rates
exhibit a slower decay of $O(Re^{-1/2})$ for $Re \rightarrow \infty$, consistent
with the singular role of viscosity in the neighborhood of the inviscid
threshold, implied by eq. \ref{eq:planar_viscous}; see Fig.
\ref{fig:planar_growth_rate_vs_Re}. Finally, it is worth mentioning that the
viscous dispersion relation is a transcendental one even for planar modes
[\cite{hocking_1960}]. Thus, for any finite $Re$, there exist an infinite number
of planar modes, and importantly, they remain non-degenerate. It may be shown
that all but two of these modes (the two being governed by the corrected
quadratic derived by Hocking, given by eq. \ref{eq:planar_viscous}) have
$\sigma_r \rightarrow n\Omega$, with $\sigma_i \sim O(Re^{-1})$ in the limit $Re
\rightarrow \infty$. The zero Doppler frequency limit $(\sigma_r - n\Omega = 0)$
suggests that these remaining modes correspond to the limiting forms of the
planar Coriolis modes for large but finite $Re$. In fact, for any finite $Re$,
the $\sigma_i$'s for the Coriolis modes form an infinite sequence asymptoting to
$-\infty$ with increasing modal index, this being consistent with the fact that
the finer-scaled modes must exhibit progressively greater (viscous) decay rates.

% Planar growth rates - variation with Re and We %

\subsection{Three-dimensional Perturbations}
\label{sec:3d_ptb}
	The effect of rotation on three-dimensional perturbations is well understood
only in the presence of viscosity. As shown first by \cite{gillis_1962}, the
necessary and sufficient criterion for the stability of the rotating column in
the presence of viscosity is 
\begin{equation}
We < k^2 + n^2 -1.
\label{eq:3d_stability_criterion}
\end{equation}
It is easily seen that eq. \ref{eq:3d_stability_criterion} reduces to the
corresponding criteria for viscous planar and axisymmetric perturbations for $n
= 0$ and $k = 0$, respectively. For inviscid stability, however, this criterion
has only been shown to be a sufficient one [\cite{pedley_1967}]. The relevance
of the same threshold $We (=k^2 + n^2-1)$ in the presence and absence of
viscosity is because, as will be seen below, similar to the axisymmetric case,
the system is in a state of rigid-body rotation at this $We$. However, the
change from a necessary and sufficient condition in the viscous case to only a
sufficient one in the inviscid limit points to the possibly subtle relation
between the inviscid and viscous stability scenarios.

\begin{figure}
   \centering
\begin{subfigure}{0.48\textwidth}
	\centering
	\includegraphics[width=\linewidth]{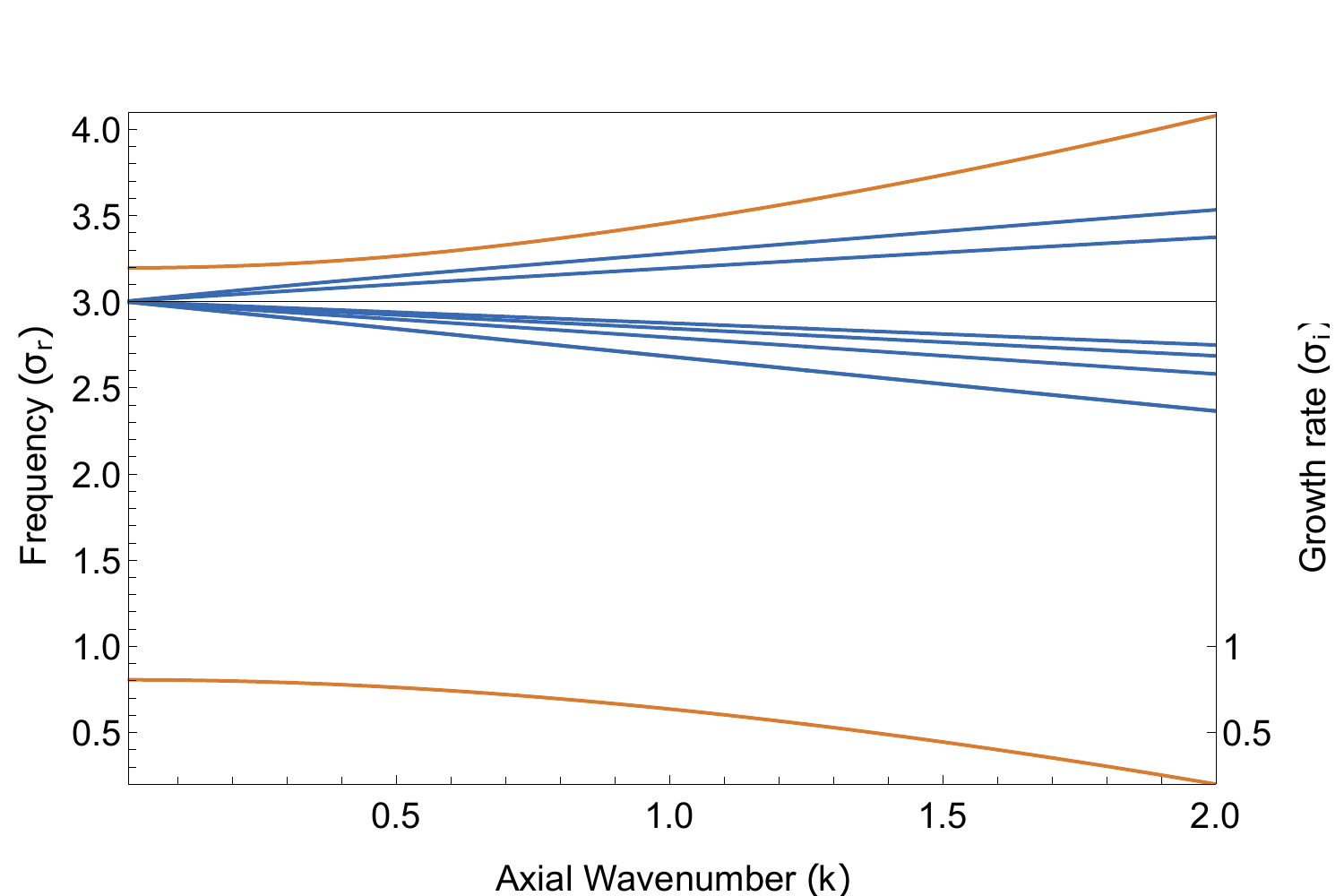}
	\caption{$We = 7$}
	\label{fig:3D_regime1}
\end{subfigure}
\begin{subfigure}{0.48\textwidth}
	\centering
	\includegraphics[width=\linewidth]{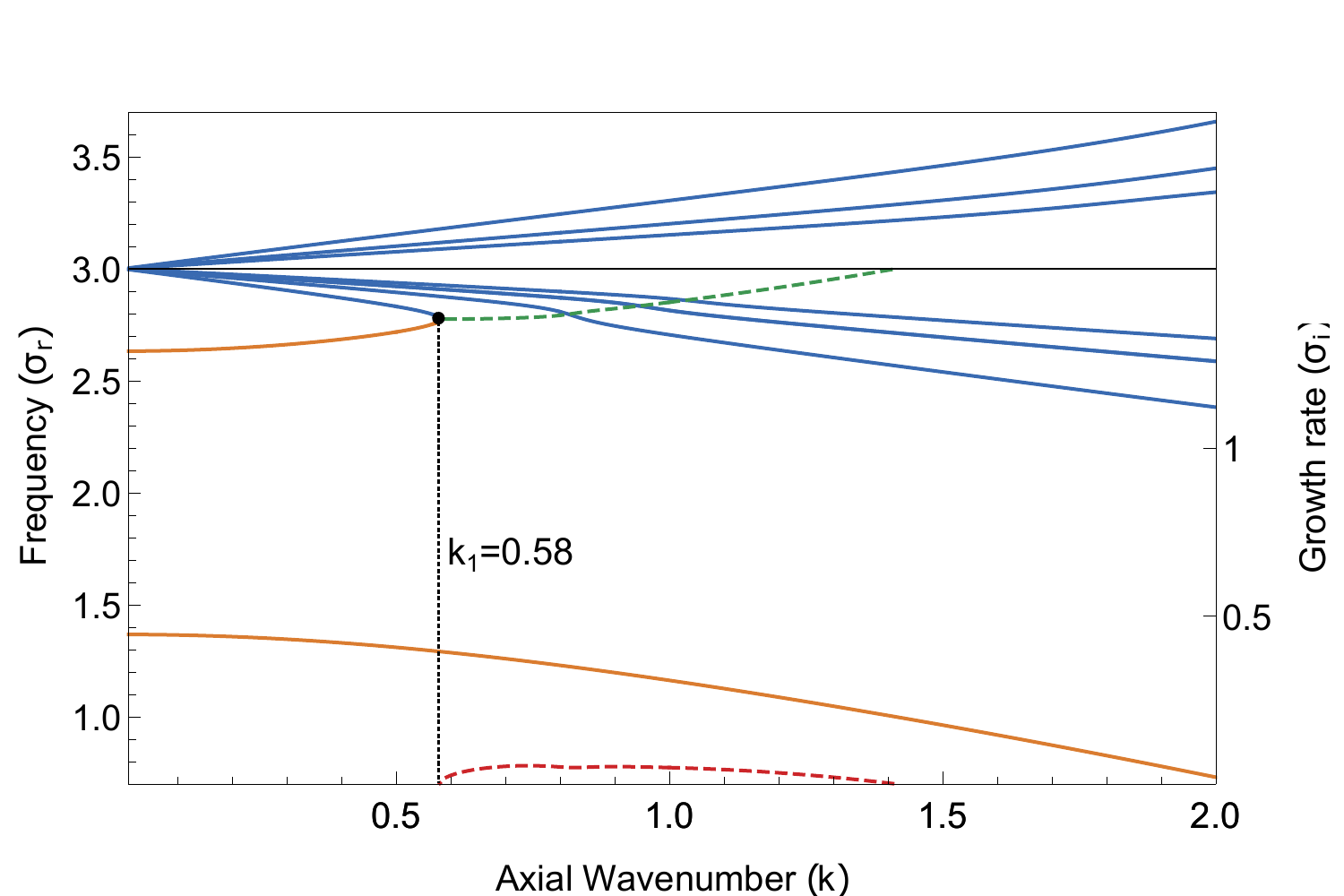}
	\caption{$We = 10$}
	\label{fig:3D_regime2}
\end{subfigure}
\begin{subfigure}{0.48\textwidth}
	\centering
	\includegraphics[width=\linewidth]{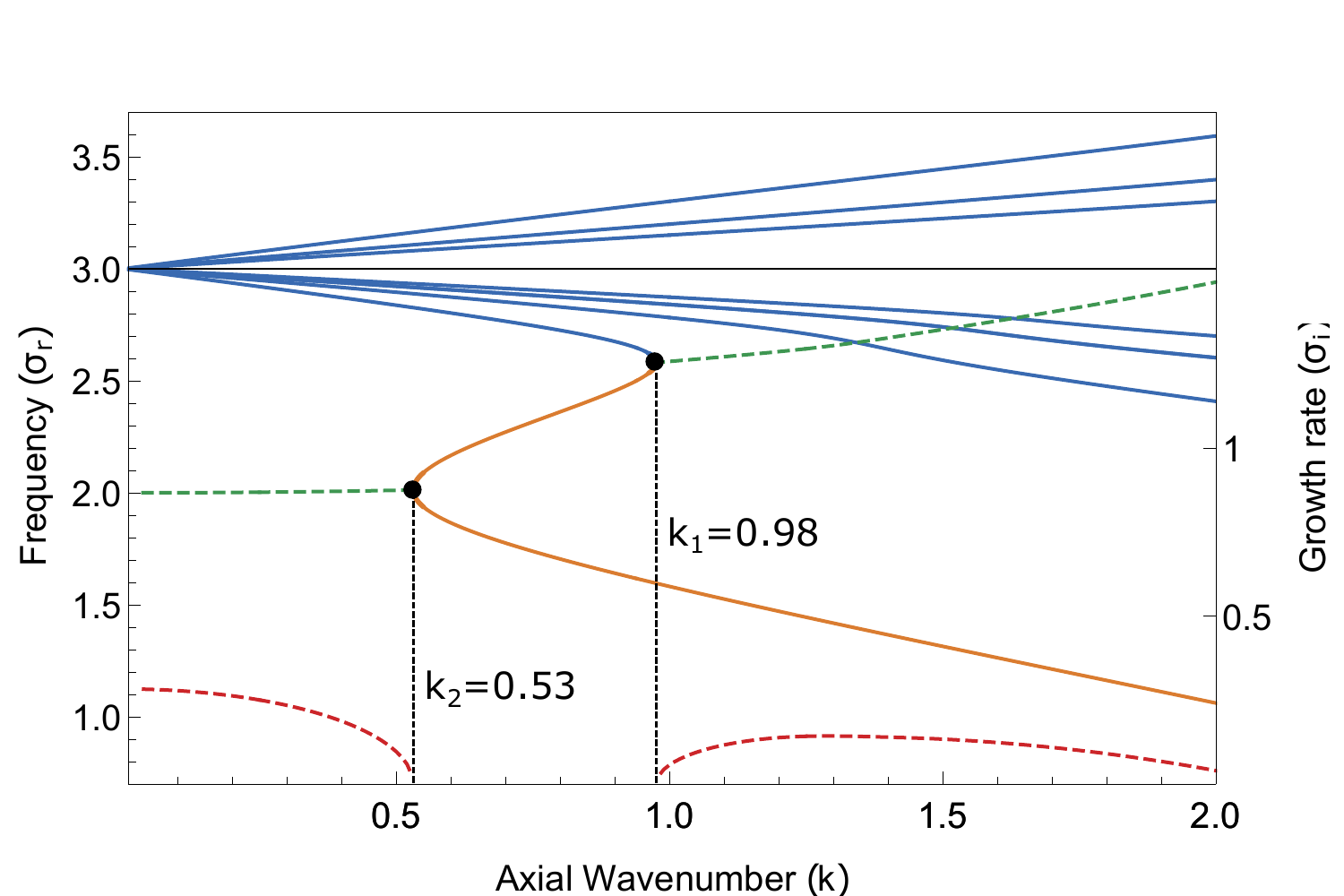}
	\caption{$We = 12.5$}
	\label{fig:3D_regime3}
\end{subfigure}
\begin{subfigure}{0.48\textwidth}
	\centering
	\includegraphics[width=\linewidth]{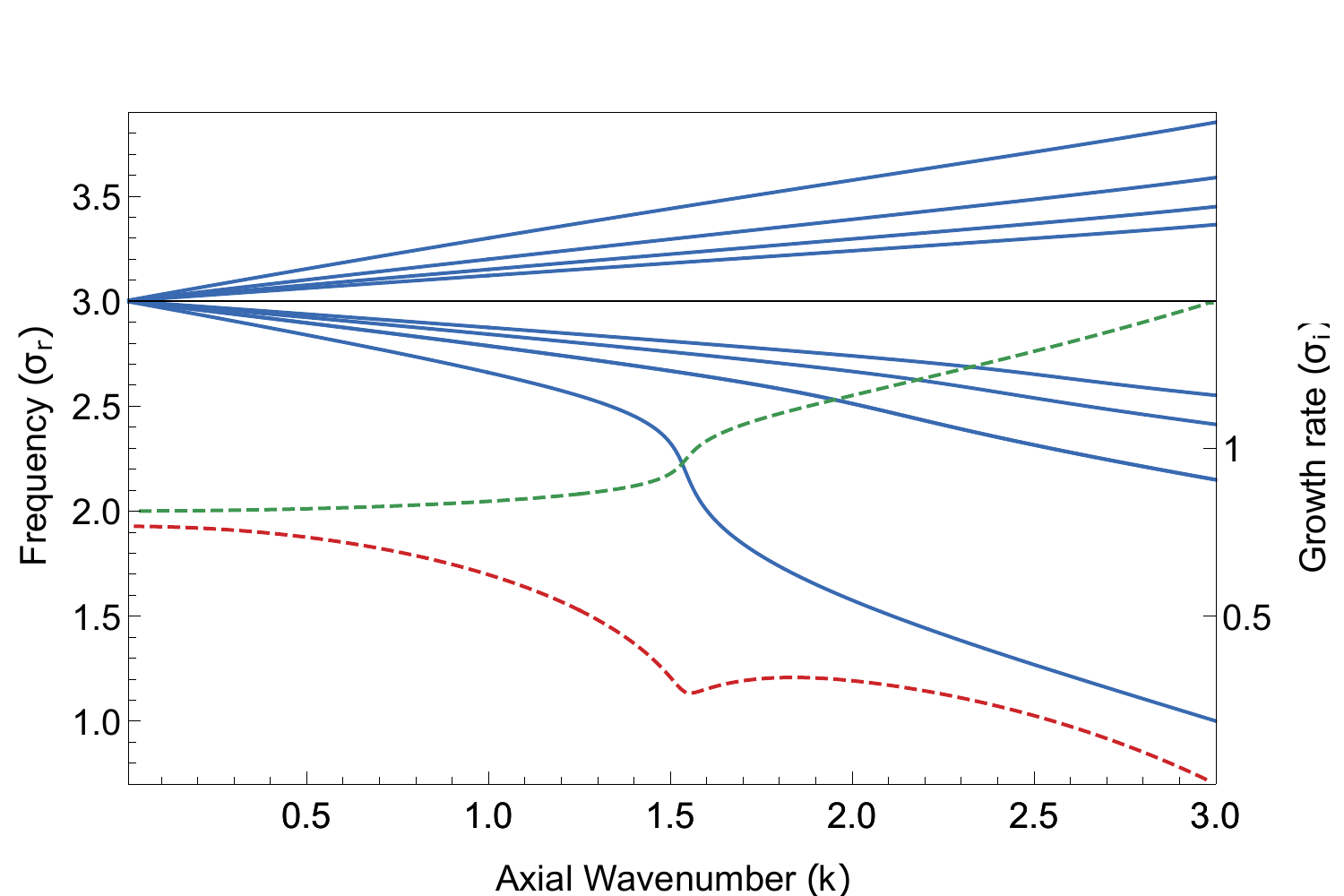}
	\caption{$We = 17$}
	\label{fig:3D_regime4}
\end{subfigure}
\caption{The inviscid dispersion curves corresponding to the different
$We$-regimes for $n = 3$: (a) $We < n^2-1$ (b) $n^2 - 1 < We < n(n+1)$, (c)
$n(n+1) < We < We_{\text{cusp}}$ and (d) $We > We_{\text{cusp}}$;
$We_{\text{cusp}} \approx 16 $ for $n=3$. The neutral Coriolis modes are shown
in blue and the neutral capillary modes in orange. The real part of the unstable
eigenvalues are shown as dashed green curves while the imaginary parts are shown
as dashed red curves.}
\label{fig:3D_dispersion_curves}
\end{figure}

The dispersion relation for the inviscid rotating column has, in fact, already
been given by \cite{weidman_1997} as 
\begin{equation}
\alpha \frac{J_{n-1}(\alpha)}{J_{n}(\alpha)} - 
\frac{We \left(4-(\sigma-n)^{2}\right)}{We + 1-n^{2}-k^{2}} - 
n\left(1+\frac{2}{(\sigma-n)}\right)=0,
\label{eq:3D_rotating_column_dispersion}
\end{equation}
where $\alpha = k\sqrt{\frac{4}{(\sigma-n)^2}-1}$. It will be shown below
that, for a given $n > 1$, the spectrum as governed by eq.
\ref{eq:3D_rotating_column_dispersion} changes qualitatively with increasing
$We$. This is illustrated in Fig. \ref{fig:3D_dispersion_curves} which shows
four sets of dispersion curves, for $n = 3$, each corresponding to a different
regime. 

% We regimes for n = 3 

% We regimes for n = 4

\begin{figure}
   \centering
   \begin{subfigure}{0.49\textwidth}
      \centering
      \includegraphics[width=\linewidth]{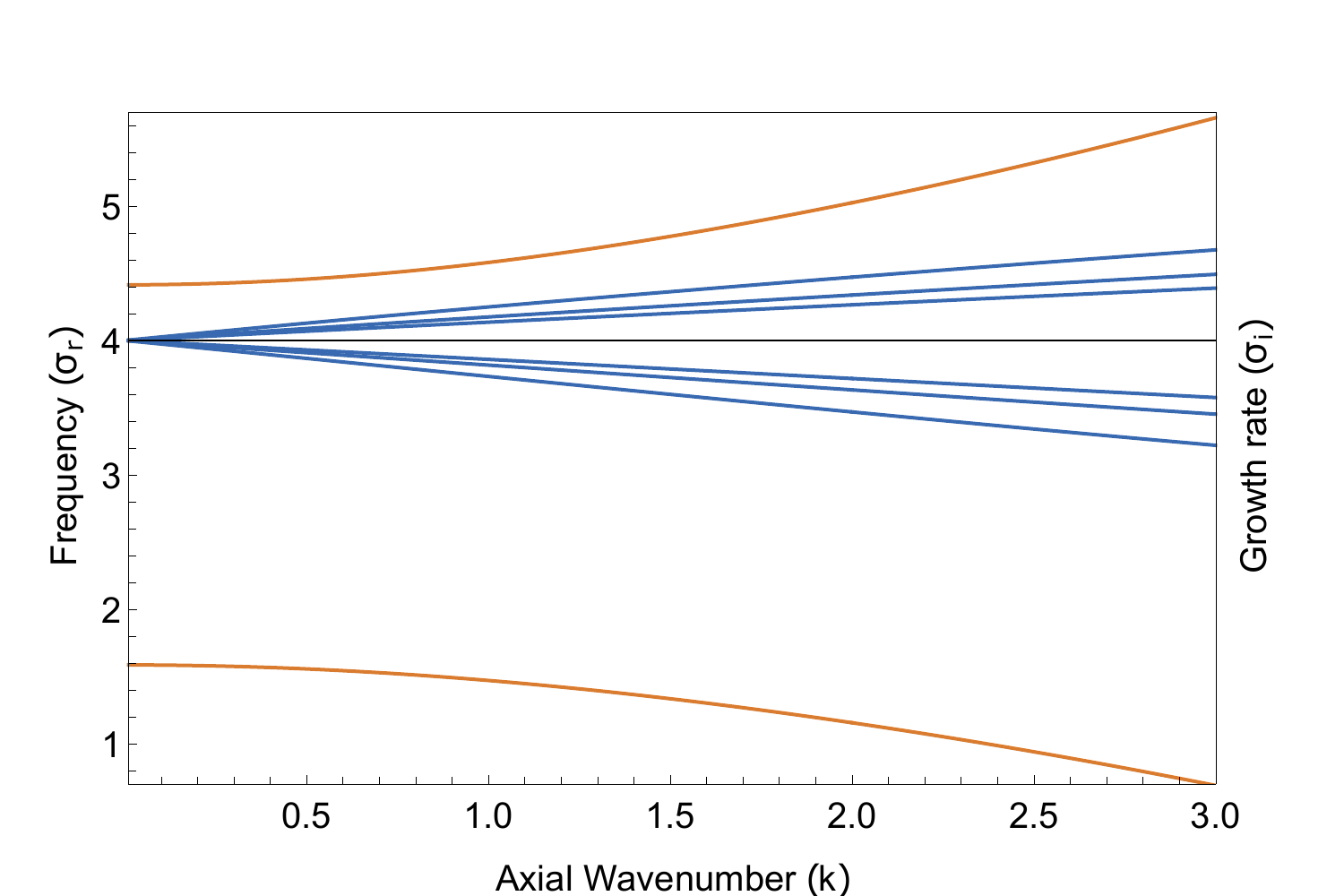}
      \caption{$We=12$}
      \label{fig:3D_regimes_n4-a}
   \end{subfigure}
   \begin{subfigure}{0.49\textwidth}
      \centering
      \includegraphics[width=\linewidth]{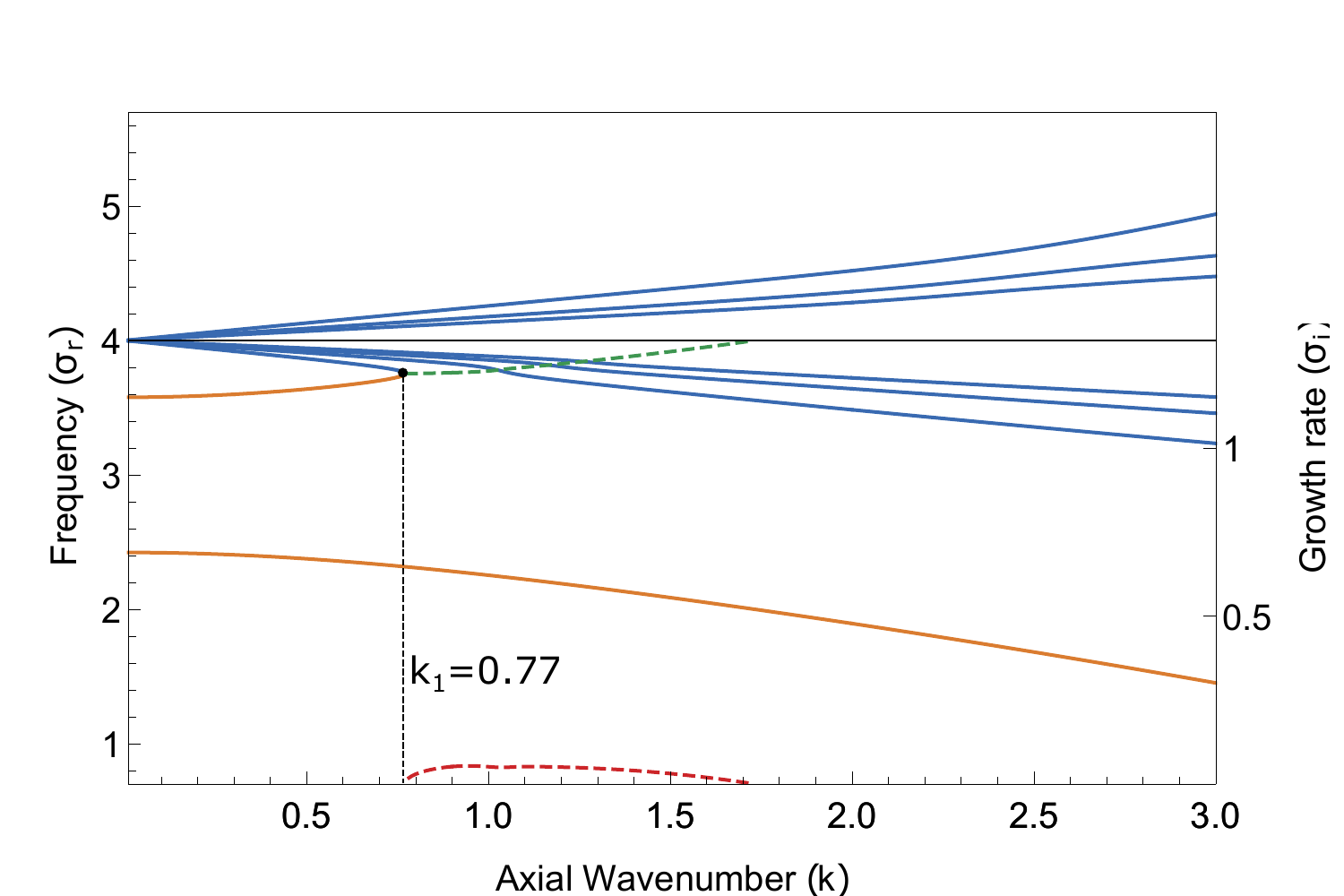}
      \caption{$We=18$}
      \label{fig:3D_regimes_n4-b}
   \end{subfigure}
   \begin{subfigure}{0.49\textwidth}
      \centering
      \includegraphics[width=\linewidth]{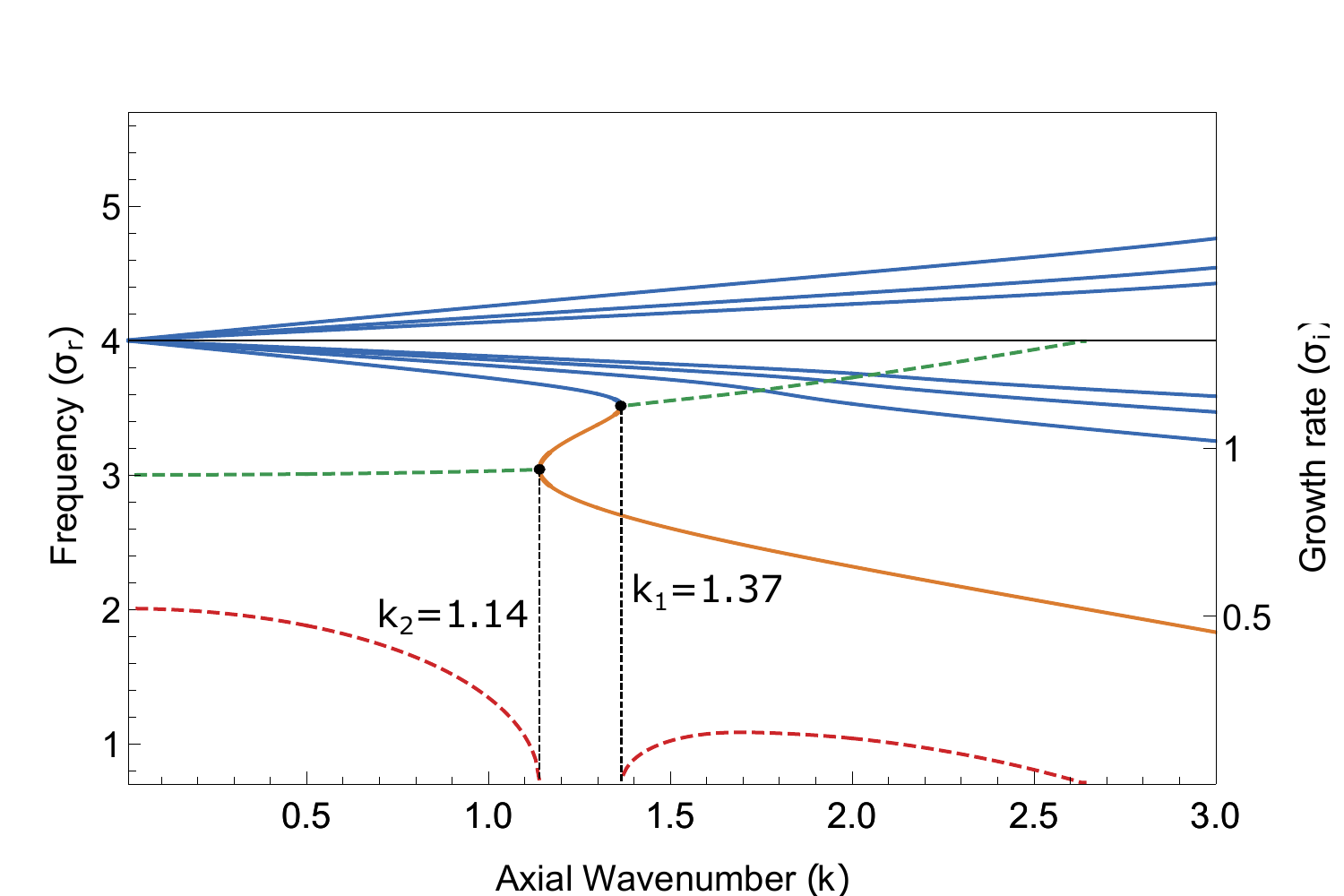}
      \caption{$We=22$}
      \label{fig:3D_regimes_n4-c}
   \end{subfigure}
   \begin{subfigure}{0.49\textwidth}
      \centering
      \includegraphics[width=\linewidth]{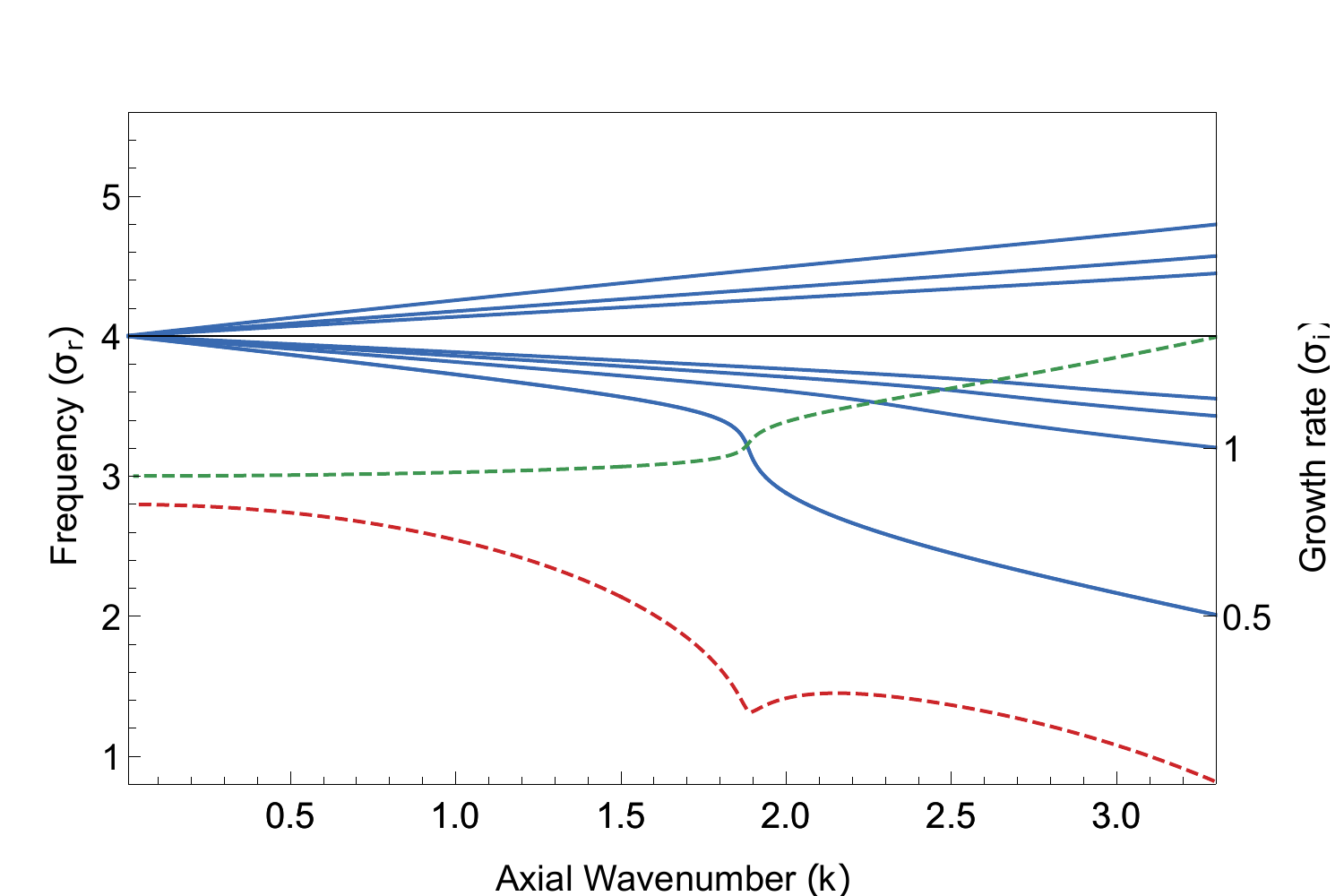}
      \caption{$We=26$}
      \label{fig:3D_regimes_n4-d}
   \end{subfigure}
   \caption{The inviscid dispersion curves corresponding to the different
   $We$-regimes for $n=4$: (a) $We < n^2-1$ (b) $n^2 - 1 < We < n(n+1)$, (c)
   $n(n+1) < We < We_{cusp}$ and (d) $We > We_{cusp}$; $We_{\text{cusp}} \approx
   25.5$ for $n=4$. The neutral Coriolis modes are shown in blue and the neutral
   capillary modes in orange. The real part of the unstable eigenvalues are
   shown as dashed green curves while the imaginary parts are shown as dashed
   red curves.}
   \label{fig:3D_regimes_n4}
\end{figure}

The first regime, $We < n^2-1$ ($=8$ for $n=3$), corresponds to the case where
the rotating column is stable since $We$ is below the aforestated viscous
threshold; the dispersion curves for this $We$ are shown in Fig.
\ref{fig:3D_regime1}. Similar to the axisymmetric case (see Fig.
\ref{fig:axisymmetric_dispersion_curves}), the spectrum consists of a pair of
capillary modes (orange), and an infinite hierarchy of Coriolis modes (blue),
the first few of which are shown in the figure. As $We$ is increased, the
capillary branches move towards each other with the upper branch moving down
towards smaller $\sigma$. This is in accordance with eq.
\ref{eq:planar_dispersion} above, which shows that the two capillary branch
frequencies in the planar limit $(k = 0)$ approach $n - 1$ with $We$ approaching
the inviscid threshold. Since the Coriolis modes degenerate to $\sigma = n$ in
the planar limit $(k = 0)$, this motion of the upper capillary branch would seem
to cause it to cross the Coriolis modes with increasing $We$. The coalescences
of the retrograde capillary mode with the retrograde Coriolis dispersion curves,
that result after the crossing, would then appear to lead to the emergence of
unstable modes at higher $We$. Note that such a scenario was not possible in the
axisymmetric case, where the frequency in the limit $k \rightarrow 0$ was
identically zero regardless of the particular dispersion curve (capillary or
Coriolis) or $We$; see Fig. \ref{fig:axisymmetric_dispersion_We}.

The second regime, $n^2-1 < We < n(n+1)$, is where, as already seen for $k = 0$,
the column is unstable (to planar perturbations) only in the presence of
viscosity. As shown in Fig. \ref{fig:3D_regime2}, the upper capillary branch
appears to have moved below the line corresponding to $\sigma = n$, and now
suffers a coalescence with the lowermost Coriolis branch at $k_1 \approx 0.58$.
This coalescence is accompanied by the pair of eigenvalues becoming
complex-valued for larger $k$, implying instability. The instability continues
until the column approaches a state of rigid rotation, corresponding to $We =
k^2+n^2-1$, and the corresponding wavenumber is therefore given by
$\sqrt{We-n^2+1}$. Thus, the lone interval of instability in this regime is
given by $k_1<k<\sqrt{We-n^2+1}$ with the upper limit being $\sqrt{2}$ for $n=3$
and the chosen $We$ in Fig. \ref{fig:3D_regime2}.

At $We = n(n+1)$, the pair of capillary branches coalesce at $k = 0$, the
corresponding value of $\sigma$ being $(n-1)$, as predicted by eq.
\ref{eq:planar_dispersion}. Thus, for $We$ larger than $n(n+1)$, the
eigenspectrum exhibits two coalescences, resulting in the unstable intervals $0
< k < k_2$ and $k_1 < k < \sqrt{We-n^2+1}$ (for the values chosen in Fig.
\ref{fig:3D_regime3}, $k_1 \approx 0.98$ and $k_2 \approx 0.53$). The first
interval corresponds to the unstable mode that results from the coalesced pair
of capillary modes, and its lower limit ($k_2 = 0$ at $We = n(n+1)$) denotes the
onset of inviscid instability to planar perturbations. The two coalescences
lead to an intervening stable interval given by $k_2 < k < k_1$ (Fig.
\ref{fig:3D_regime3}). As a result, the composite dispersion curve that
describes the neutral mode is now a combination of portions of the original
Coriolis and capillary branches, and has a hysteretic character, as evident from
the dispersion curve bending back in the aforementioned stable interval. Although, strictly speaking,
the participating Coriolis and capillary modes lose their identity, and only a
composite curve remains, an intuitive association of its parts with the original
curves is clear. We, therefore, continue to color portions of the composite
curve based on the underlying `parent' curves. With
increasing $We$, $k_1$ and $k_2$ approach each other and the hysteretic region
shrinks and eventually vanishes. The critical $We$ where hysteresis vanishes is
termed $We_{\text{cusp}}$; as explained below, this is because the disappearance
of the hysteretic region is marked by a cusp in the $We-k$ plane. The third
regime can then be defined as $n(n+1) < We < We_{\text{cusp}}$. For $n=3$,
$We_{\text{cusp}} \approx 16$ (see Fig. \ref{fig:n3_islands}).

The fourth regime corresponds to $We > We_{\text{cusp}}$ when the fold in the
dispersion curve, and thence, the intermediate stable wavenumber interval
vanishes. There is now only a single unstable interval corresponding to $0 < k <
\sqrt{We-n^2+1}$ (Fig. \ref{fig:3D_regime4}), consistent with the aforementioned
viscous criterion.

The results presented above agree quantitatively with the restricted observations of
\cite{weidman_1997} who obtained the (inviscid)\,growth rates for $n=1, 2$ and $3$, as a function of $k$, for
$We=10$. Consider Fig. 5a in \cite{weidman_1997}, where the authors present 
growth rates for $n=2$ and $3$. Since $10 > We_\text{cusp}$ for $n = 2$, the growth-rates correspond to the fourth regime according to the classification above. Nevertheless, the local minimum seen in the growth rate curve for $n=2$ (akin to the dotted red curves in Fig.
\ref{fig:3D_regime4} and \ref{fig:3D_regimes_n4-d}) is reminiscent of the hysteresis that occurs
at a smaller $We$. The growth rate curve for $n=3$ presented in the same figure
matches quantitatively with the dotted red curve in \ref{fig:3D_regime2}. Since
$n(n+1) < 10 < We_{\text{cusp}}$ for $n = 3$, this growth rate behavior 
behavior corresponding to the second regime described above. Their observations for $n=1$ can be
similarly understood based on the discussion presented in the appendix. Thus, while our results are consistent with the earlier findings in \cite{weidman_1997}, our focus on the entire inviscid eigenspectrum allows us to move beyond growth rates calculations for specific $We$'s, and thereby, infer the implications of the growth rate behavior on the relation between the inviscid and viscous stability criteria.

% We-k-sigma schematic for cusp formation

\begin{figure}
   \centering
   \includegraphics[width=\textwidth]{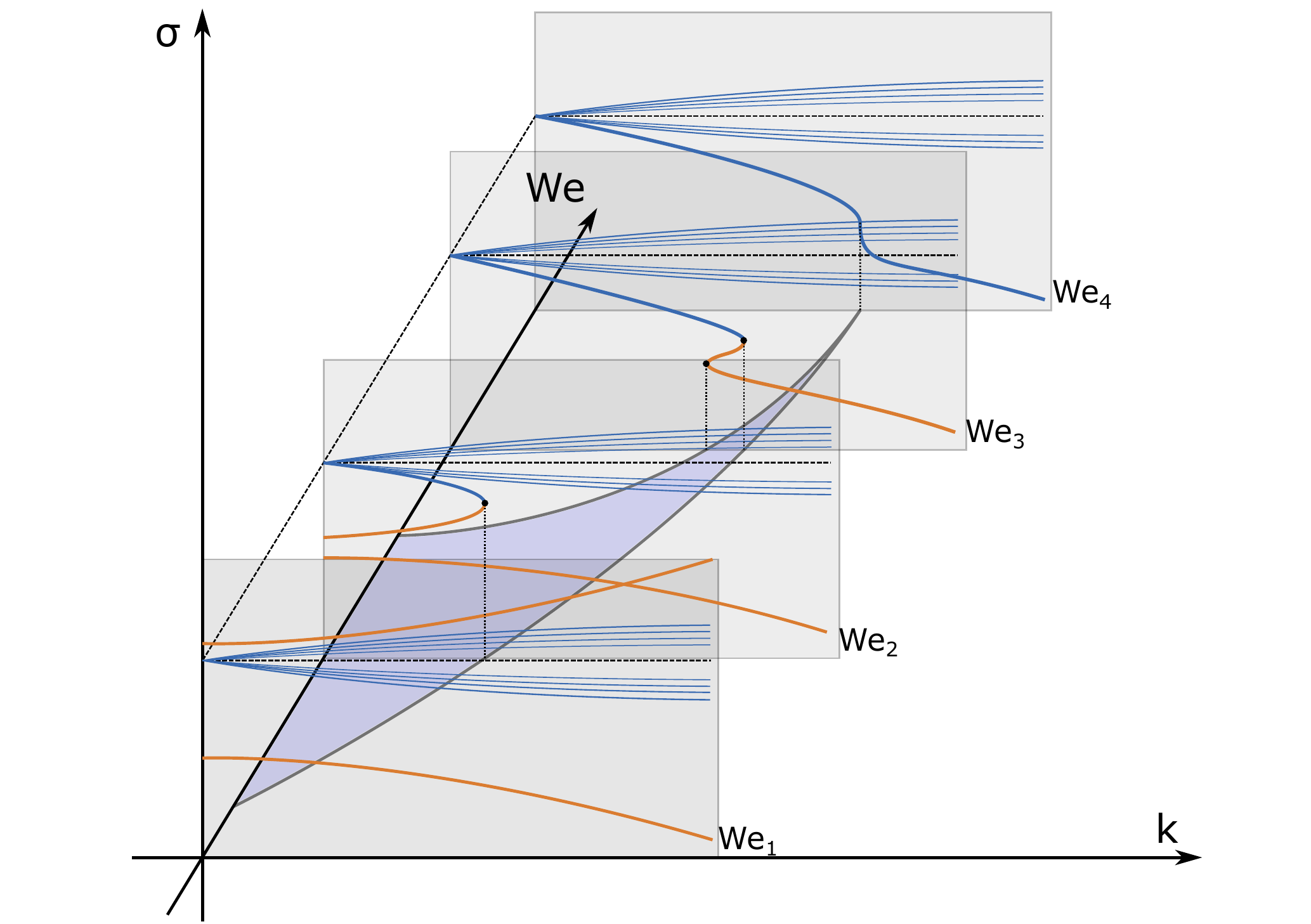}
   \caption{A schematic showing the loci of the pairs of folding points,
   associated with the hysteretic dispersion curves, which leads
   to the inviscidly stable island in the $We-k$ plane.} 
   \label{fig:cusp_formation}
\end{figure}

% Stable islands in the We-k plane

\begin{figure}
   \centering
   \begin{subfigure}{.48\textwidth}
      \centering
      \includegraphics[width=\textwidth]{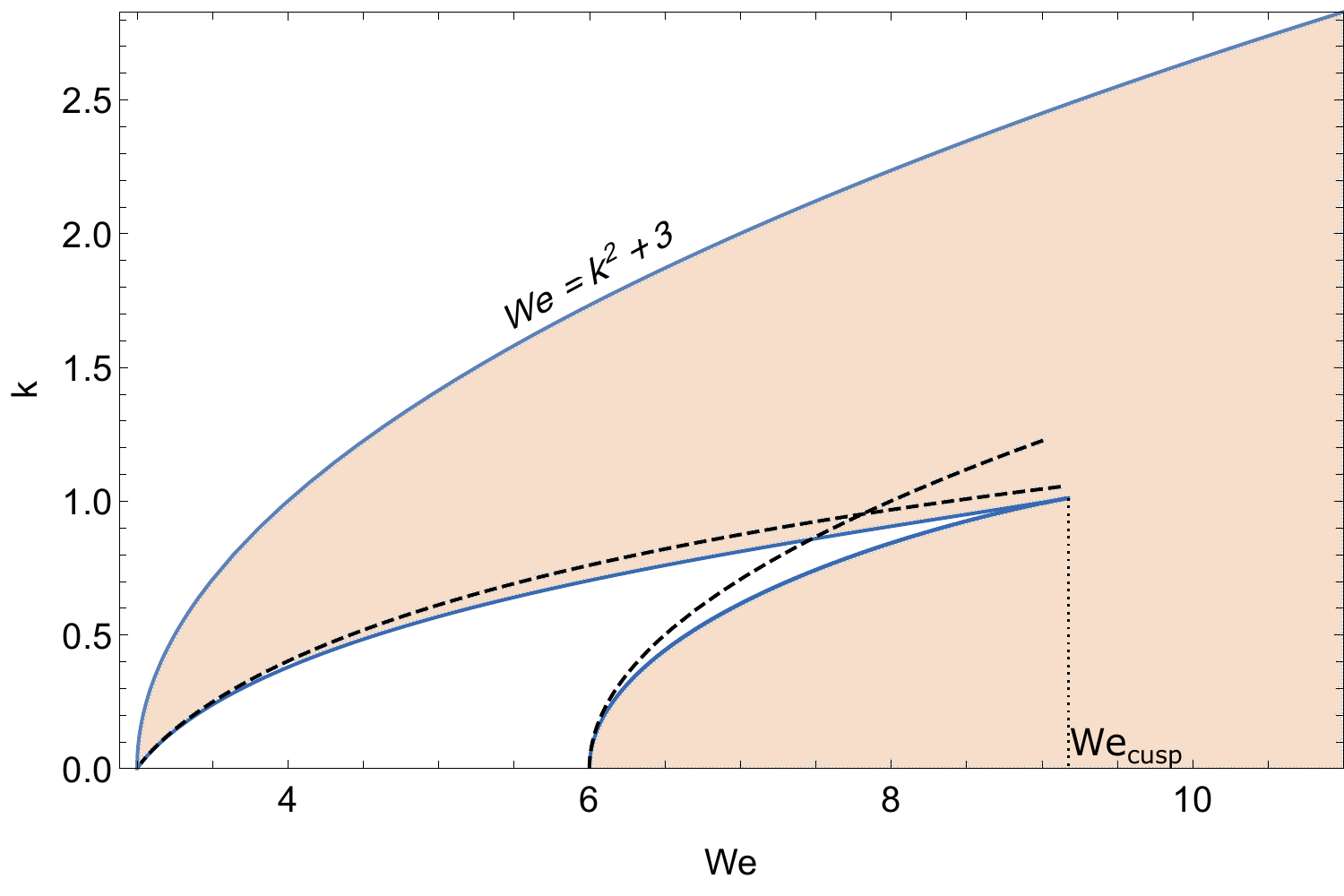}
      \caption{$n=2$}
      \label{fig:n2_islands}
   \end{subfigure}
   \hfill
   \begin{subfigure}{.48\textwidth}
      \centering
      \includegraphics[width=\textwidth]{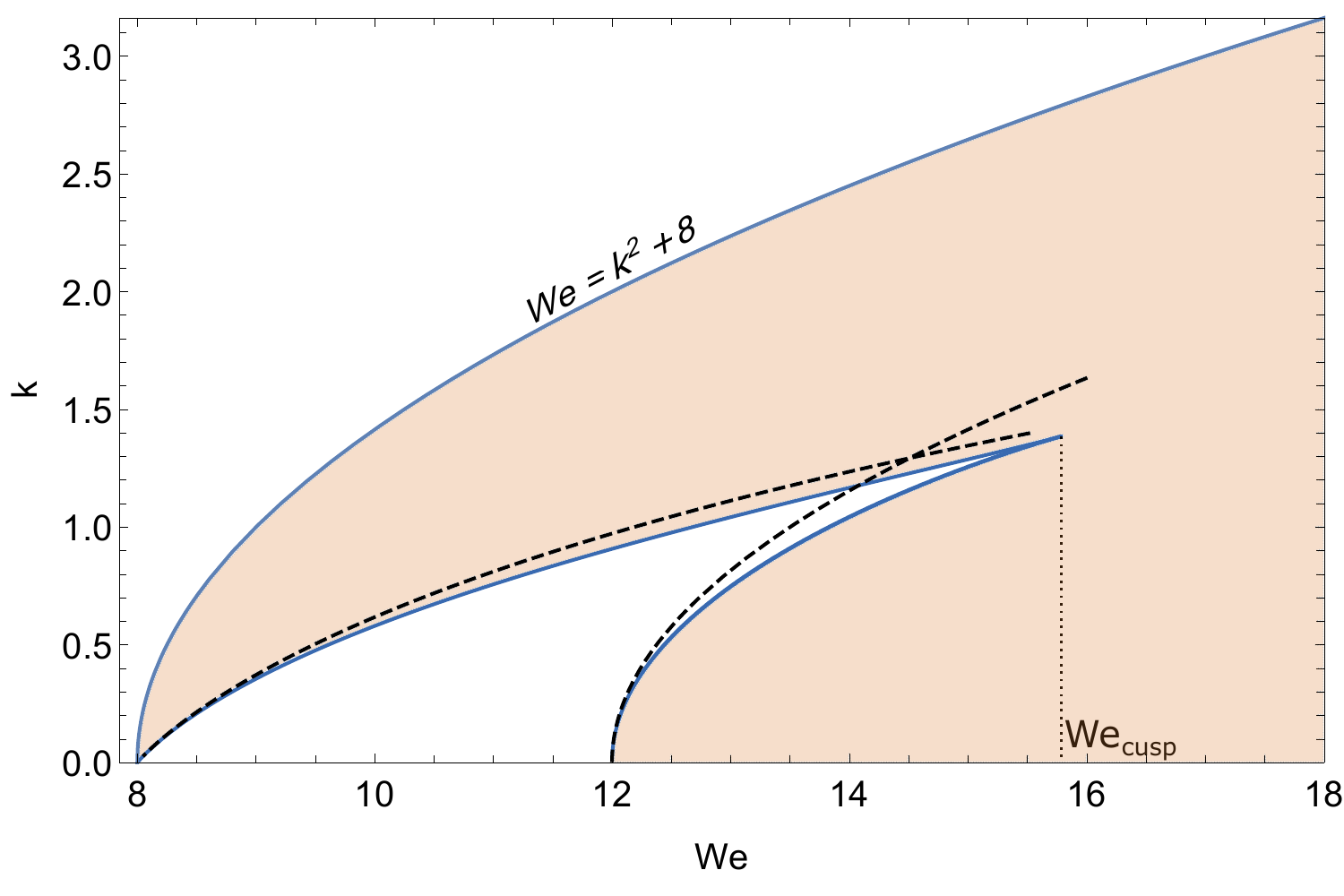}
      \caption{$n=3$}
      \label{fig:n3_islands}
   \end{subfigure}
   \hfill
   \begin{subfigure}{.48\textwidth}
      \centering
      \includegraphics[width=\textwidth]{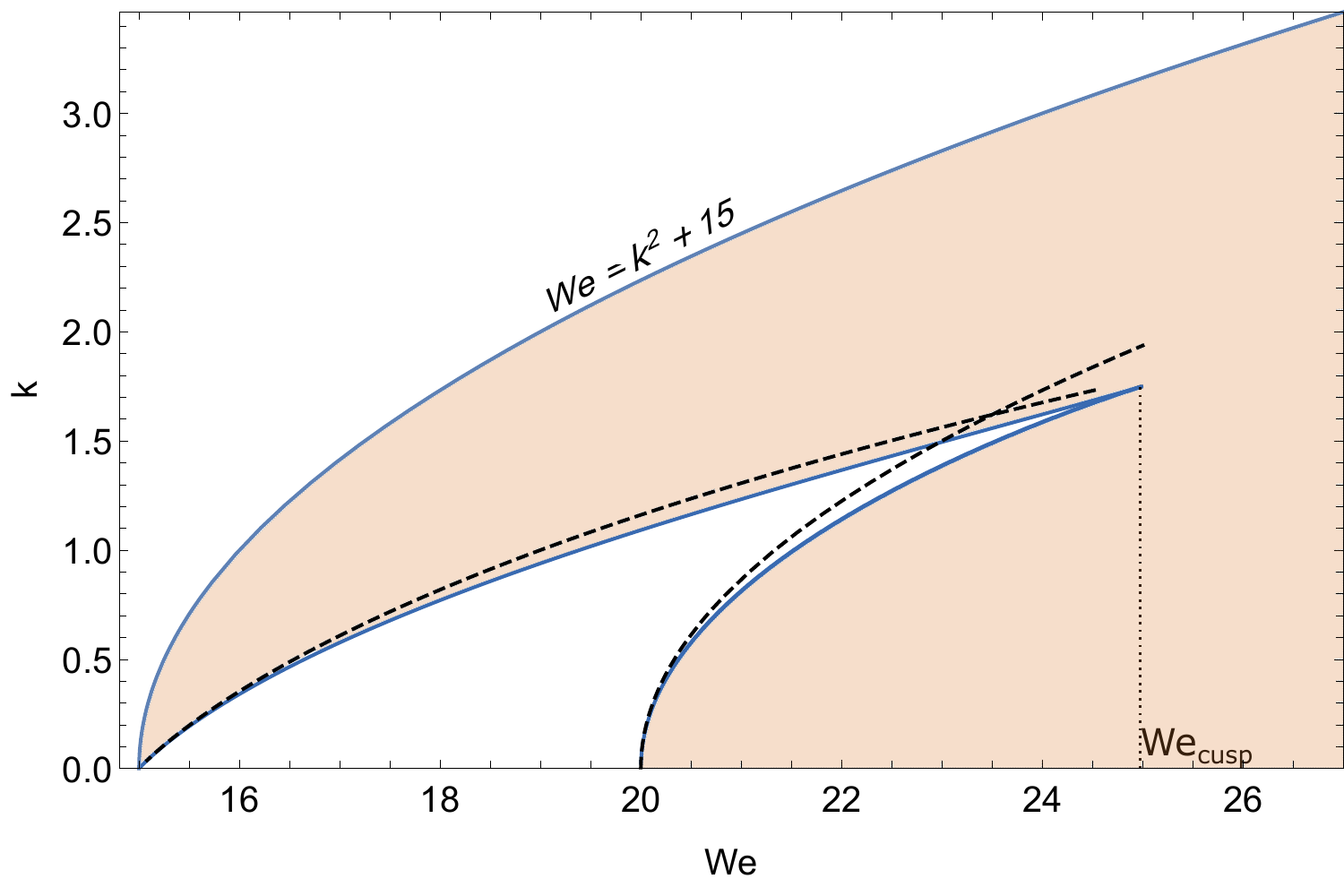}
      \caption{$n=4$}
      \label{fig:n4_islands}
   \end{subfigure}
   \hfill
   \begin{subfigure}{.48\textwidth}
      \centering
      \includegraphics[width=\textwidth]{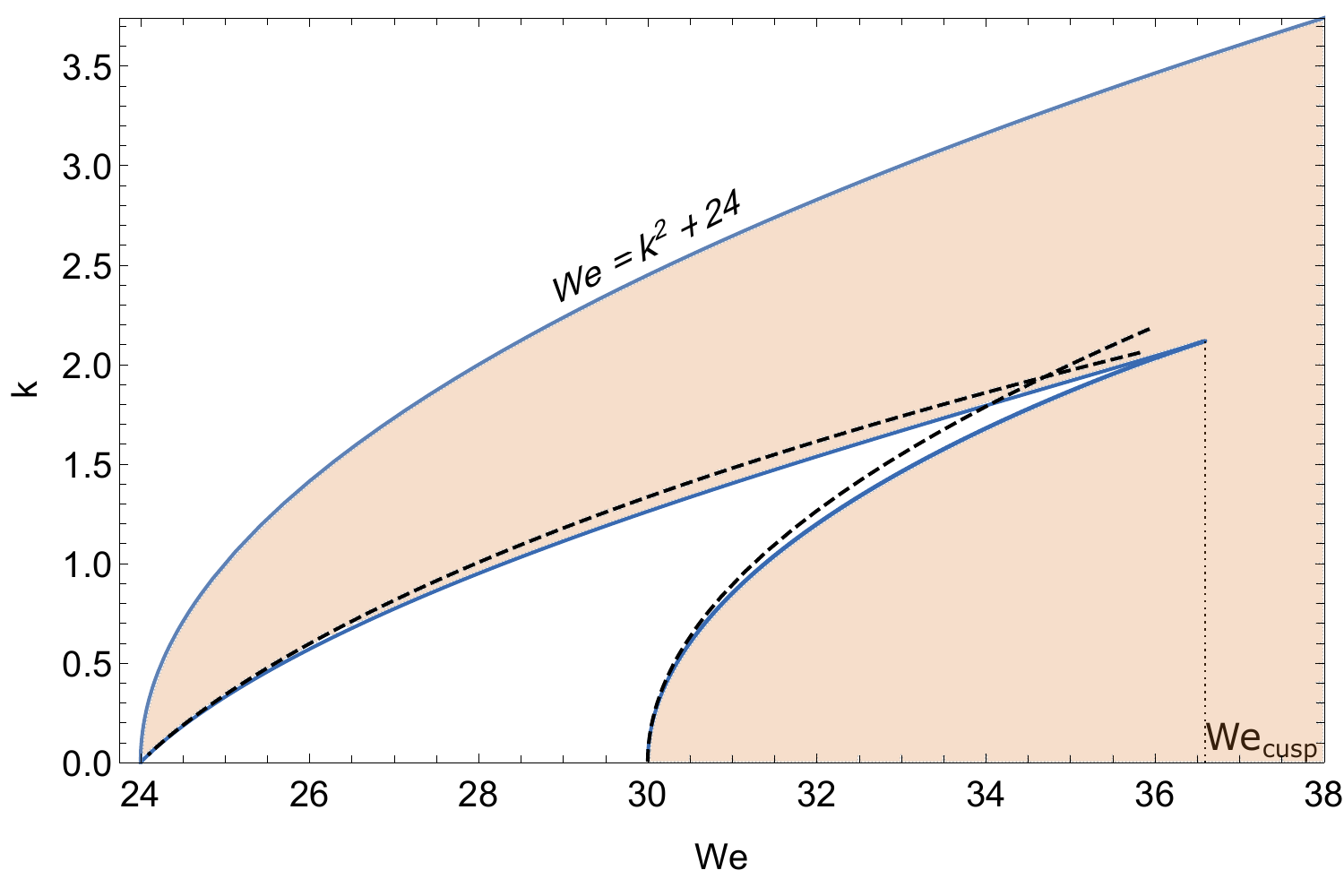}
      \caption{$n=5$}
      \label{fig:n5_islands}
   \end{subfigure}
\caption{Depiction of the stable islands in the $We-k$ plane for $n =2, 3, 4$
and $5$. Blue curves denote both, the outer boundary ($We = n^2+k^2-1$) and the
numerically obtained boundaries of the inviscidly stable island within; the
shaded region between these boundaries denotes the inviscidly unstable region.
The dashed black curves denote the small-$k$ asymptotes for the island
boundaries; the lower branch asymptote is $k \thicksim
\sqrt{\frac{n^2-1}{n(n+1)}(We-n(n+1))}$ and that for the upper branch is
obtained by simultaneously solving $d\sigma/dk = \infty$ with eq.
\ref{eq:3D_rotating_column_dispersion} for small $k$.}
\label{fig:stability_islands}
\end{figure}

The general behavior of the dispersion curves with increasing $We$, highlighted
above, holds for all $n$'s greater than unity. Fig. \ref{fig:3D_regimes_n4}
shows an analogous behavior of the eigenspectrum for $n = 4$ with
$We_{\text{cusp}} \approx 25$. The dispersion curves, such as those in Figs.
\ref{fig:3D_dispersion_curves} and \ref{fig:3D_regimes_n4}, may be stacked upon
one another, along the $We$-axis, so as to demarcate the regions of inviscid
stability in the $We-k$ plane for each $n$. Fig. \ref{fig:cusp_formation} shows
schematically how this may be achieved (the unstable wavenumber ranges have been
omitted for clarity). With varying $We$, projections of the pair of folding
points associated with each hysteretic dispersion curve (the black dots in Figs.
\ref{fig:3D_regime3}, \ref{fig:3D_regimes_n4-c} and \ref{fig:cusp_formation}),
that mark the intermediate stable interval in the $\sigma-k$ plane, yield the
two branches of a stable island in the $We-k$ plane. The picture is that of a
cusp catastrophe (\cite{zeeman_1976}), implying that the aforementioned pair of
branches terminates in a cusp. The points of coalescence between a capillary
mode and the lowest retrograde Coriolis mode (Fig. \ref{fig:3D_regime2} and
\ref{fig:3D_regimes_n4-b}) yield the upper branch of the stable island, while
those between the two capillary modes (Fig. \ref{fig:3D_regime3} and
\ref{fig:3D_regimes_n4-c}) yield the lower branch. The cusp-shaped islands of
inviscid stability in the $We-k$ plane, for $n = 2, 3, 4$ and $5$, are shown in
Fig. \ref{fig:stability_islands}. While closed form expressions for the
boundaries of these islands are not available, one may nevertheless obtain their
small-$k$ approximations. For the lower branch, one has $\sigma \rightarrow n-1$
for $k \to 0$; the resulting limiting form of the dispersion relation gives the
required approximation as $k \thicksim \sqrt{\frac{n^2-1}{n(n+1)}(We-n(n+1))}$.
For the upper branch, however, $(\sigma-n)/k$ remains $\mathcal O(1)$ as $k \to
0$. Since this remains true for all of the Coriolis mode branches, the upper
branch asymptote is obtained by exploiting the fact that the slope of the
hysteretic dispersion curve diverges at the turning points. Thus, simultaneously
solving eq. \ref{eq:3D_rotating_column_dispersion} with $d\sigma/dk \to \infty$
for $k \rightarrow 0$ yields the small-$k$ approximation for the upper branch. These approximations
have been shown as dashed black curves in Fig. \ref{fig:stability_islands},
where they are seen to compare well to the numerically determined island
boundaries well beyond the rigorous interval of validity $(k \ll 1)$. While one
expects the $We-k$ plane to remain similar in form for $n > 5$, the scenario for
$n = 1$ is essentially different, and is analyzed in the Appendix. In this case,
there exists only one stable island that extends to infinity along the $We$
axis. The exceptional behavior is not entirely unexpected, given that the limit
$k \to 0$ is a singular one - $n=1$ corresponds to a mere translation of the
rotating column in the planar limit.

%splitting of the capillary mode at We = n^2-1

\begin{figure}
    \centering
    \begin{subfigure}{.48\textwidth}
       \centering
       \includegraphics[width=\linewidth]{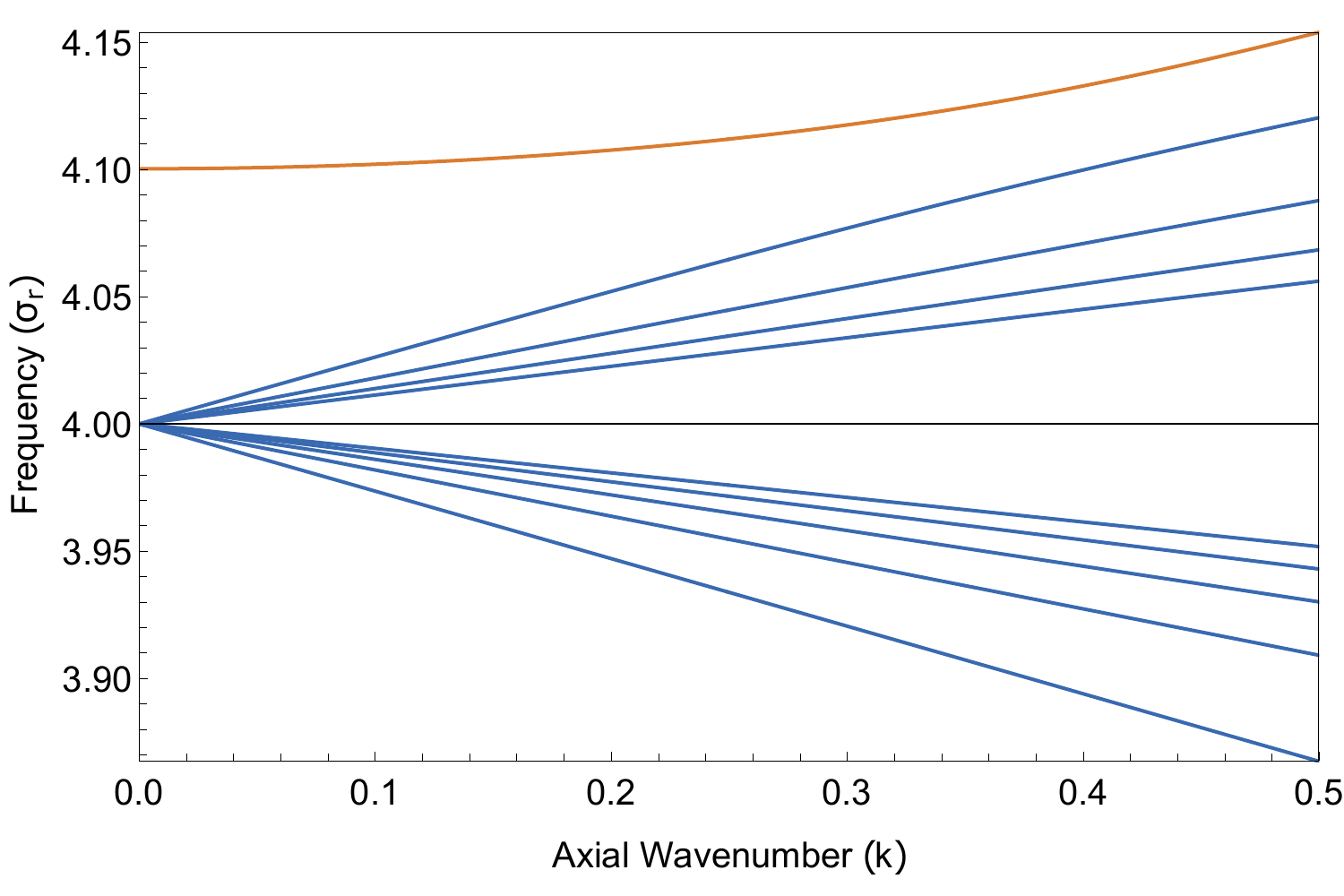}
       \caption{$We = 14.25$}
       \label{fig:mode_splitting-a}
    \end{subfigure}
    \hfill
    \begin{subfigure}{.48\textwidth}
       \centering
       \includegraphics[width=\linewidth]{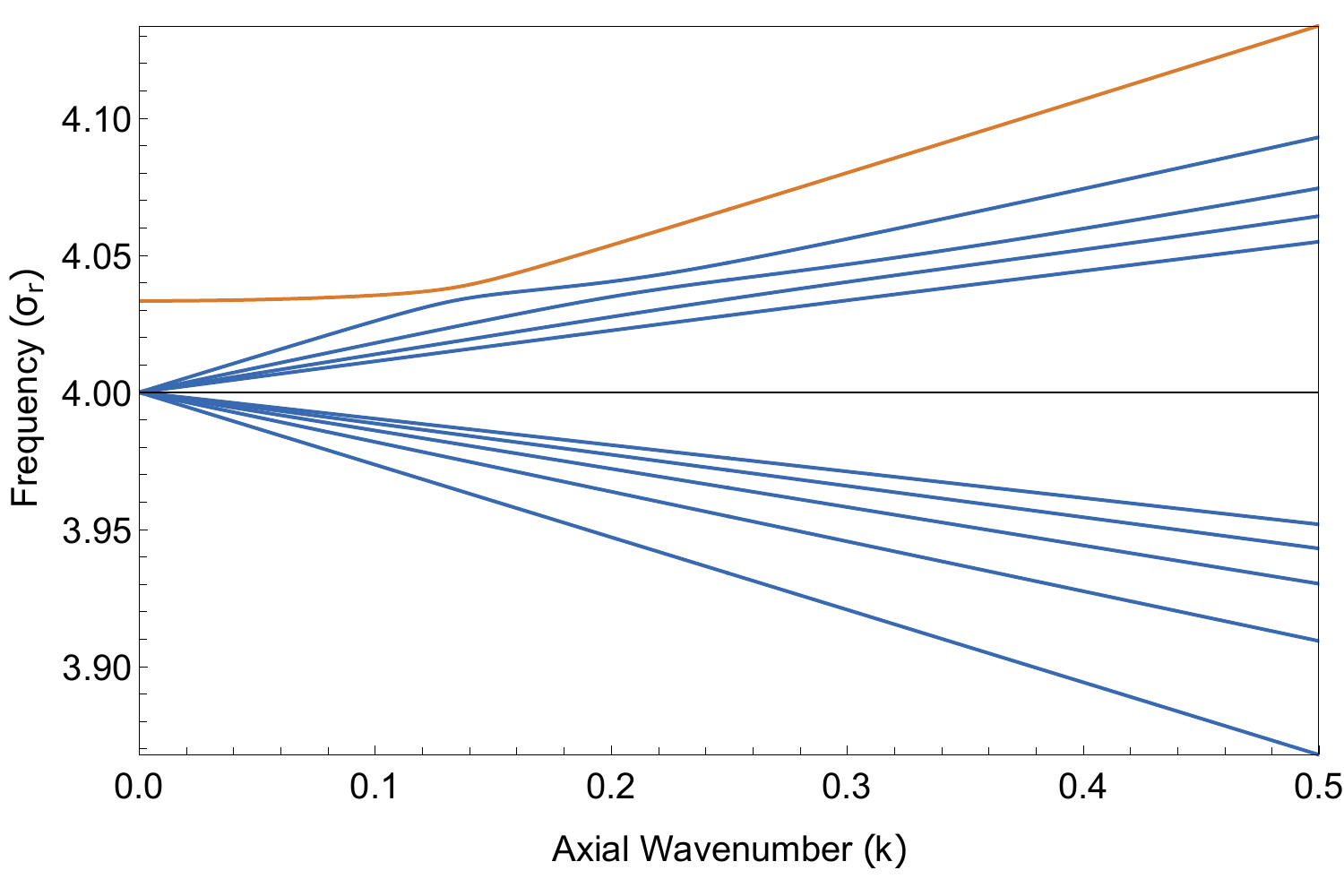}
       \caption{$We = 14.75$}
       \label{fig:mode_splitting-b}
    \end{subfigure}
    \hfill
    \begin{subfigure}{.48\textwidth}
       \centering
       \includegraphics[width=\linewidth]{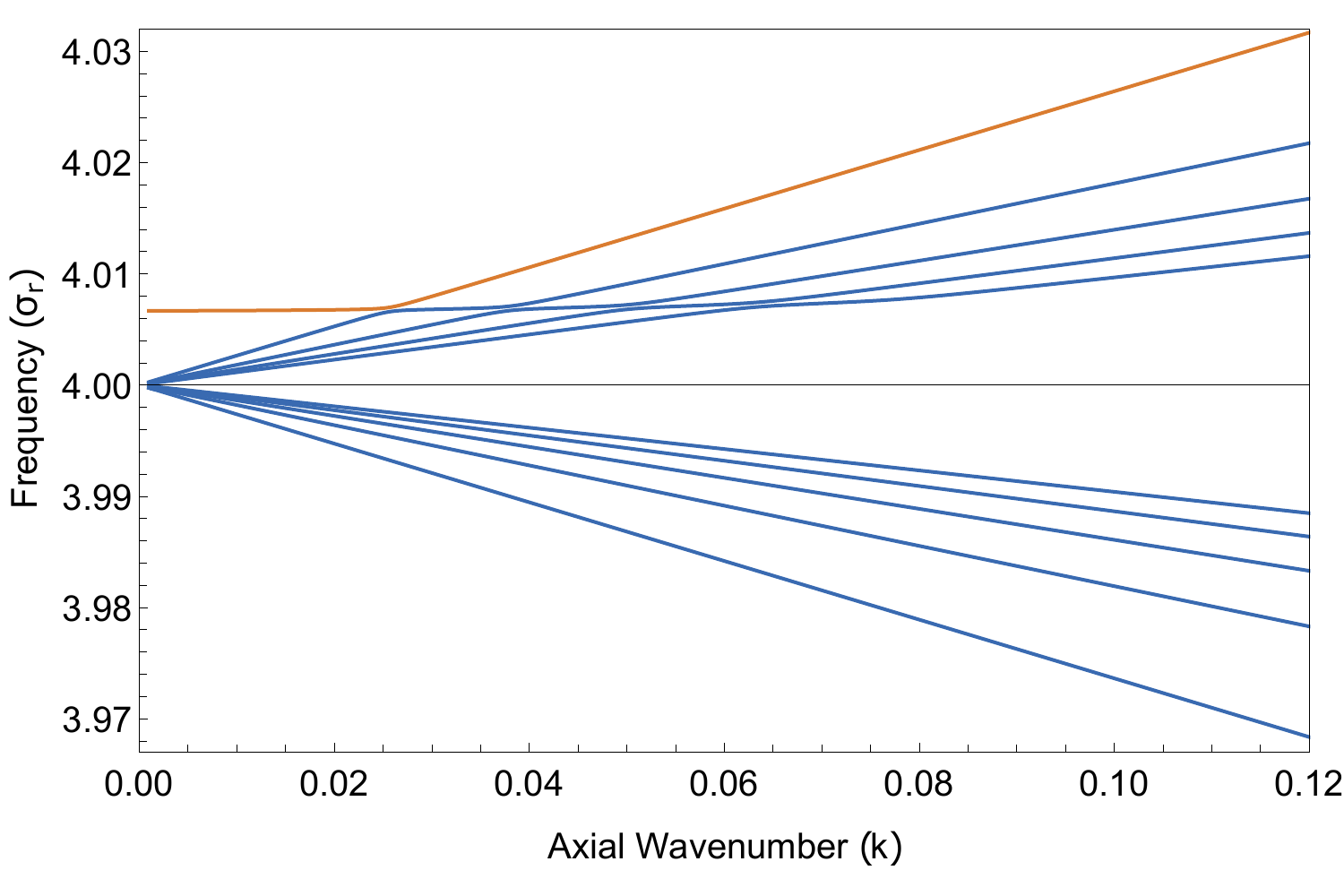}
       \caption{$We = 14.95$}
       \label{fig:mode_splitting-c}
    \end{subfigure}
       \hfill
    \begin{subfigure}{.48\textwidth}
       \centering
       \includegraphics[width=\linewidth]{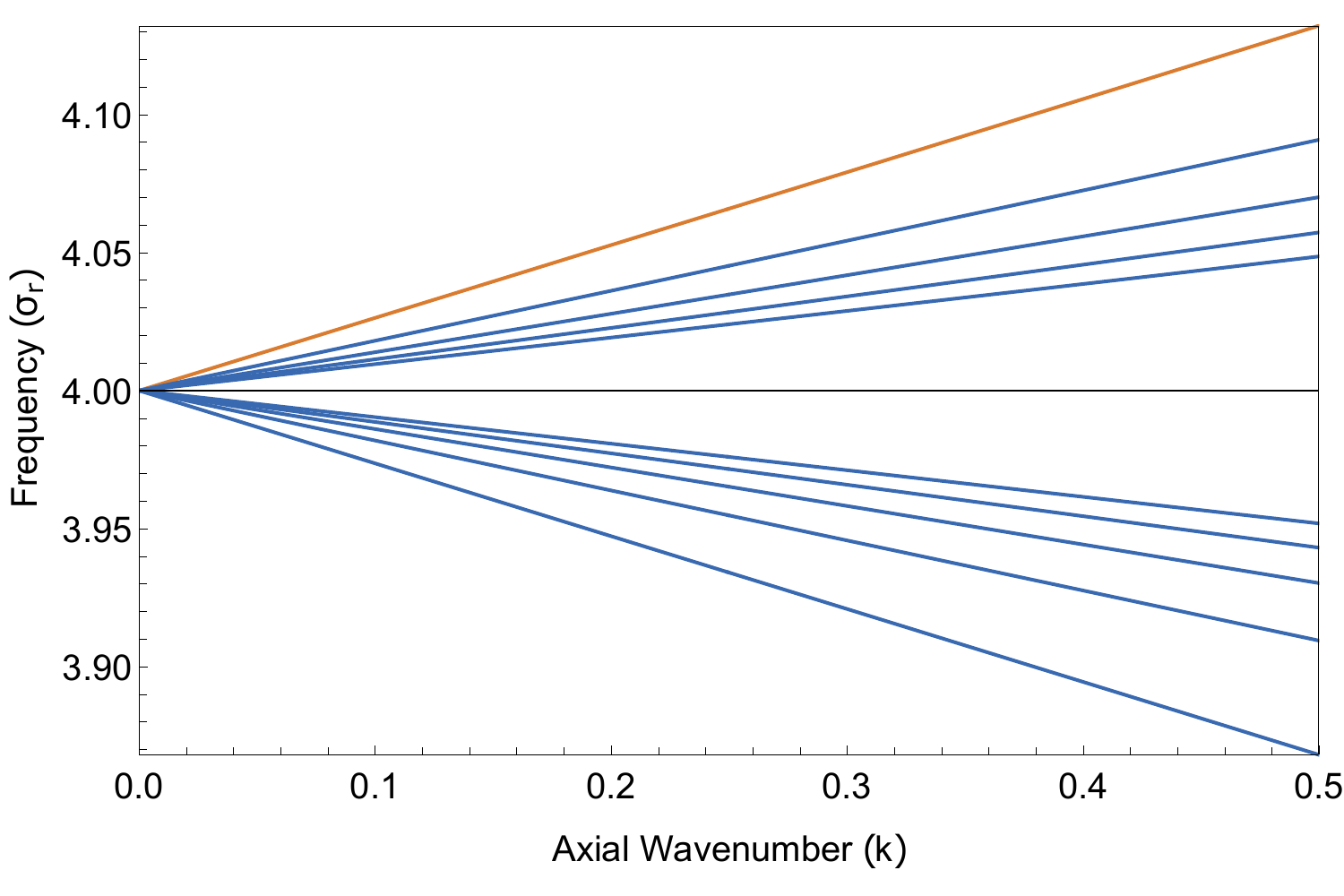}
       \caption{$We = 15$}
        \label{fig:mode_splitting-d}
    \end{subfigure}
       \hfill
    \begin{subfigure}{.48\textwidth}
       \centering
       \includegraphics[width=\linewidth]{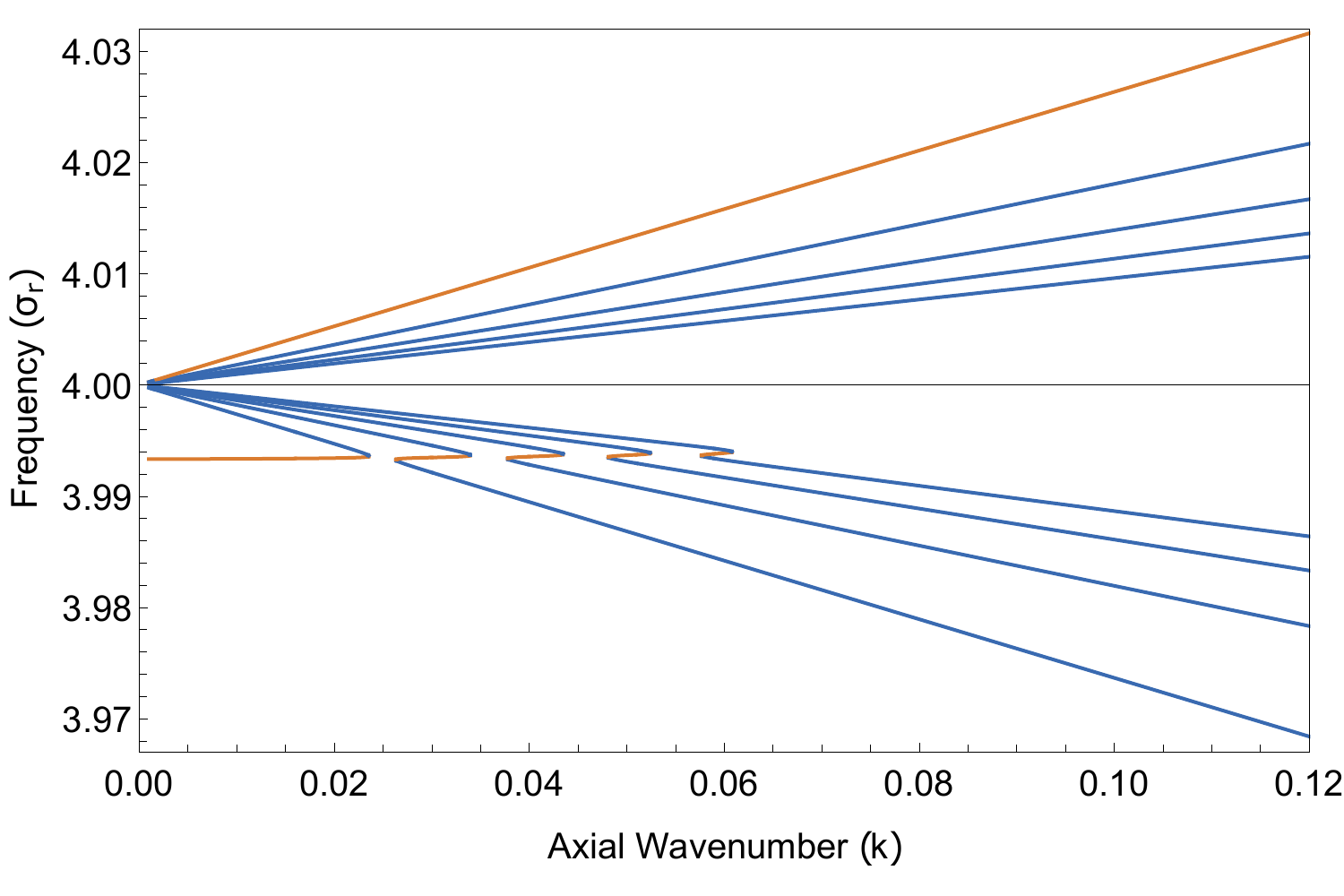}
       \caption{$We = 15.05$}
       \label{fig:mode_splitting-e}
    \end{subfigure}
       \hfill
    \begin{subfigure}{.48\textwidth}
       \centering
       \includegraphics[width=\linewidth]{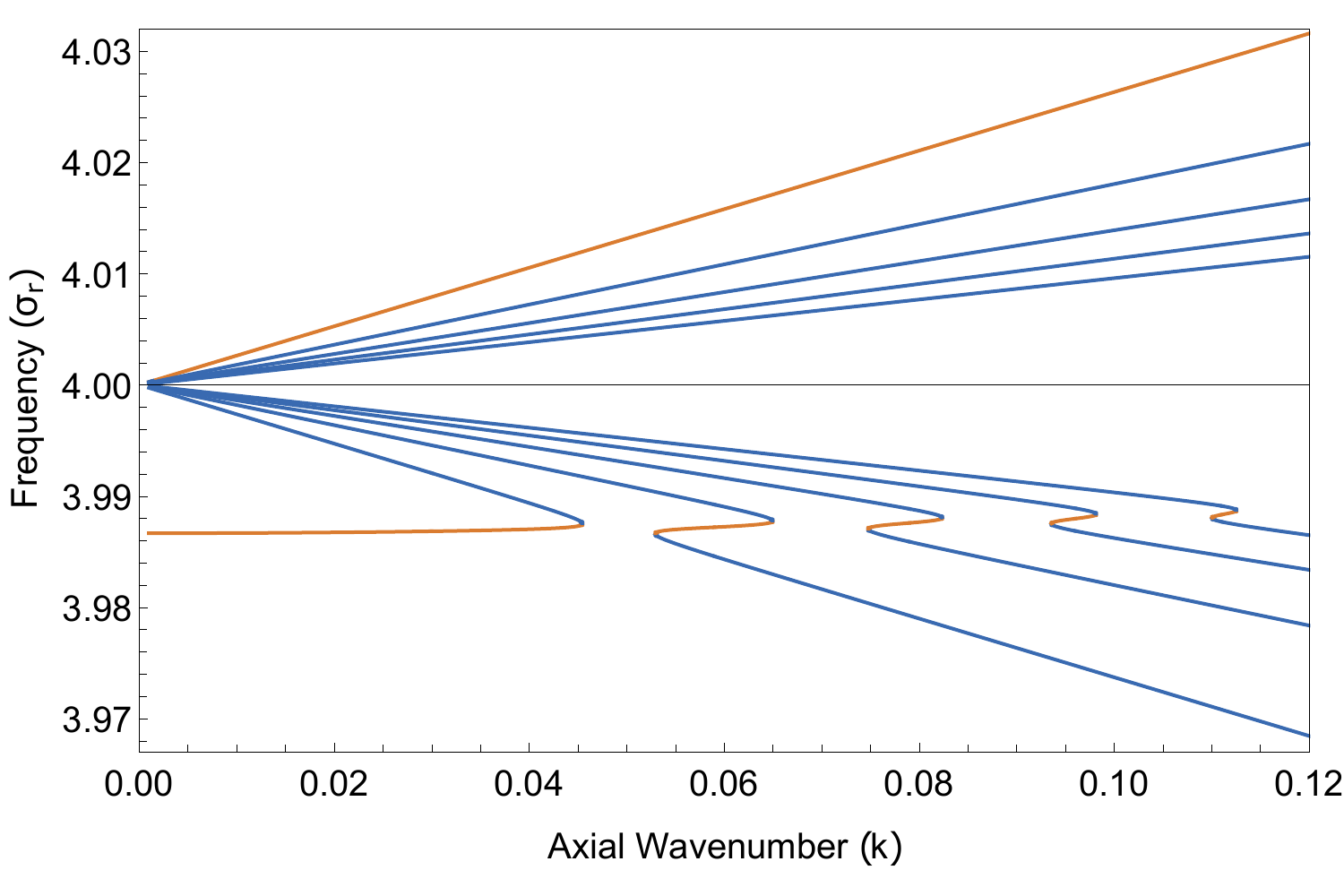}
       \caption{$We = 15.1$}
       \label{fig:mode_splitting-f}
    \end{subfigure}
    \caption{Dispersion curves ($\sigma$ vs $k$) for $n = 4$ and $We$ close to
    $n^2 - 1$. At $We = n^2 - 1$ the capillary mode splits into two and yields
    a new retrograde mode (also shown in orange) which subsequently undergoes an infinite sequence of 
    coalescences with the retrograde Coriolis modes for higher $We$.}
    \label{fig:mode_splitting}
 \end{figure}
 
 The discussion along with the preceding figures establish the following
 behavior. With increasing $We$, the upper capillary branch moves down to lower
 frequencies, appearing to cross the zero-Doppler-frequency line $\sigma = n$ in
 the process, and thereafter, undergoes a pair of coalescences (one with the
 lowest Coriolis mode, and the other with the lower capillary branch). These
 coalescences lead to intermediate unstable ranges of wavenumbers which, with
 varying $We$, trace out a stable island in the $We-k$ plane (see Fig.
 \ref{fig:stability_islands}). There are two subtle aspects with regard to this
 general behavior that need amplification, however. The first is that the upper
 capillary branch does not, in fact, end up crossing $\sigma = n$ (hence, the
 usage 'appears to' in all the instances above). Instead, as shown in Fig.
 \ref{fig:mode_splitting}, at $We = n^2-1$, when the zero-$k$ frequency of this
 branch equals $\sigma = n$ (Fig. \ref{fig:mode_splitting-d}), the capillary
 branch stops moving downward as a whole; instead, a new discrete mode emanates
 from $\sigma = n$, and continues down into the retrograde frequency range with
 further increase in $We$ (Figs. \ref{fig:mode_splitting-e} and
 \ref{fig:mode_splitting-f}). Further, even as the upper capillary branch
 descends towards $\sigma = n$, for $We$ just below $n^2-1$, it never crosses the
 lower cograde Coriolis branches. These `avoided crossings' are illustrated via
 suitably magnified views in Figs.
 \ref{fig:mode_splitting-a}-\ref{fig:mode_splitting-c}, and arise from the Krein
 signature criterion, required for unstable coalescences between different modal
 branches, not being satisfied for cograde modes [\cite{mackay_1987,
 chernyavsky_2018, fukumoto_2003}].

 % Increasing no of hysteretic curves for small k
 
 \begin{figure}
    \centering
    \begin{subfigure}{0.49\textwidth}
       \includegraphics[width=\textwidth]{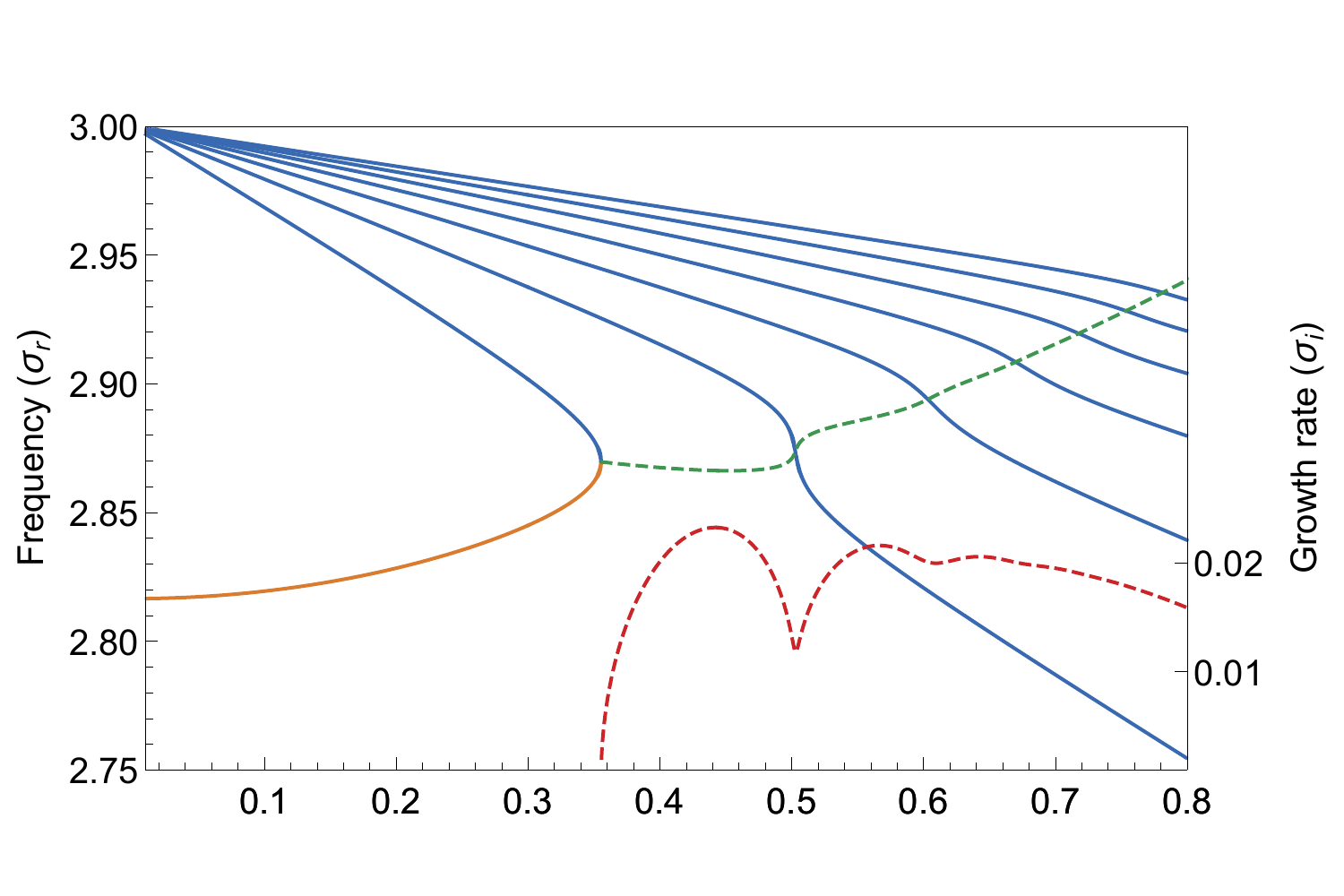}
       \caption{$We = 9$: A single coalescence between a capillary and a Coriolis mode.}
       \label{fig:increasing_no_of_hysteretic_dispCurves-a}
    \end{subfigure}
    \hfill
    \begin{subfigure}{0.49\textwidth}
       \includegraphics[width=\textwidth]{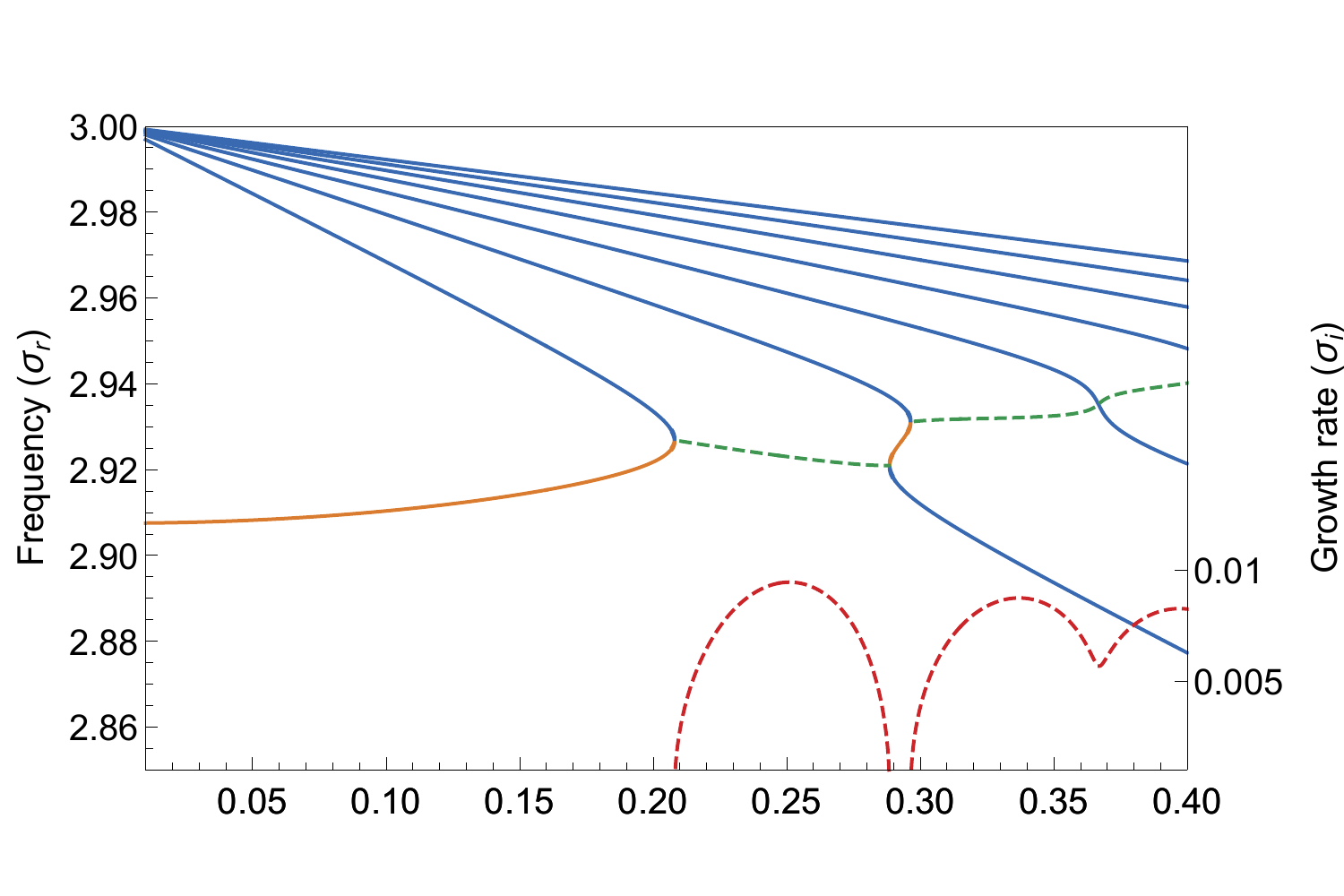}
       \caption{$We = 8.5$: The capillary-Coriolis coalescence plus a single hysteretic dispersion curve.}
       \label{fig:increasing_no_of_hysteretic_dispCurves-b}
    \end{subfigure}
    \begin{subfigure}{0.49\textwidth}
       \includegraphics[width=\textwidth]{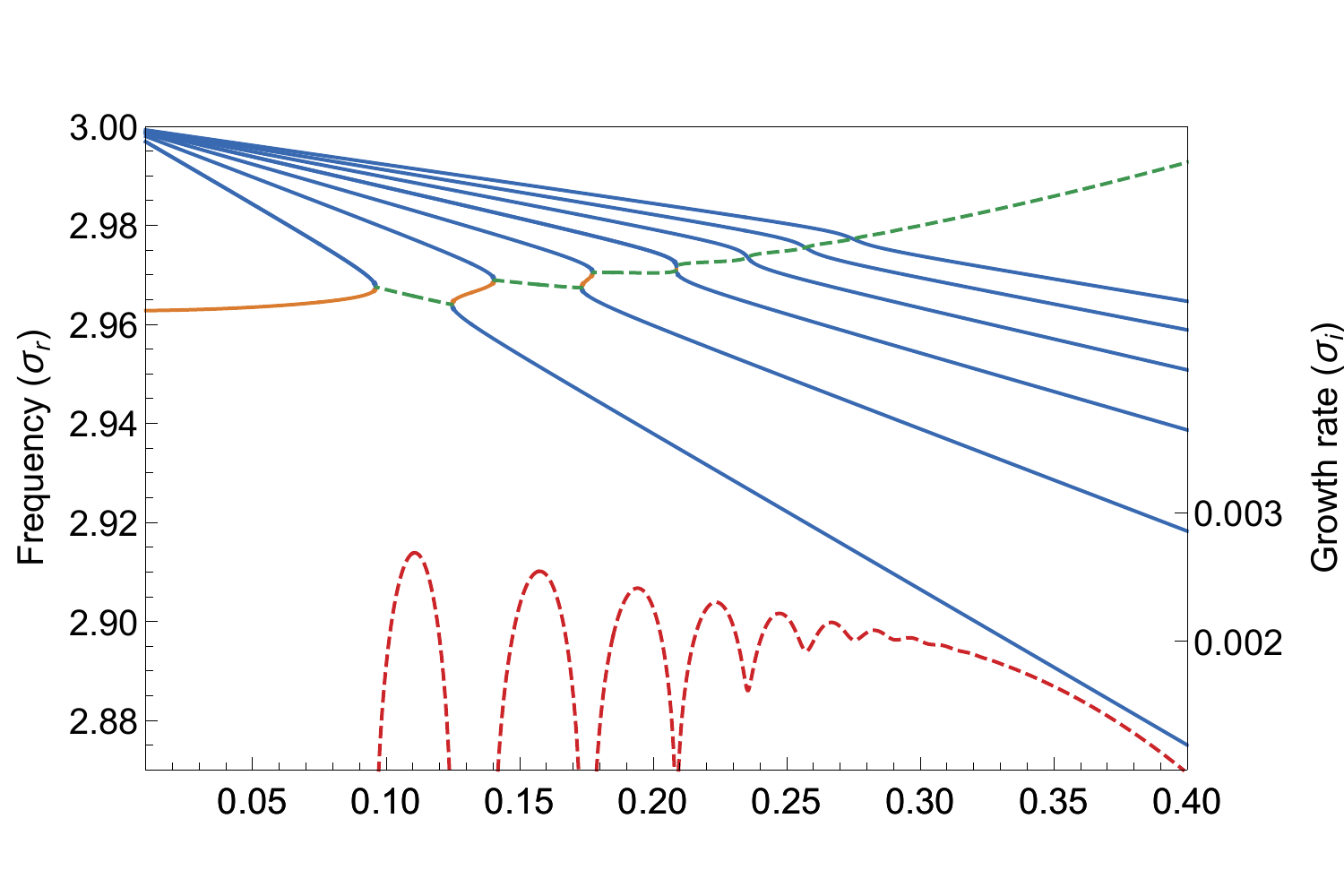}
       \caption{$We = 8.2$: The capillary-Coriolis coalescence plus three hysteretic dispersion curves.}
       \label{fig:increasing_no_of_hysteretic_dispCurves-c}
    \end{subfigure}
    \hfill
    \begin{subfigure}{0.49\textwidth}
       \includegraphics[width=\textwidth]{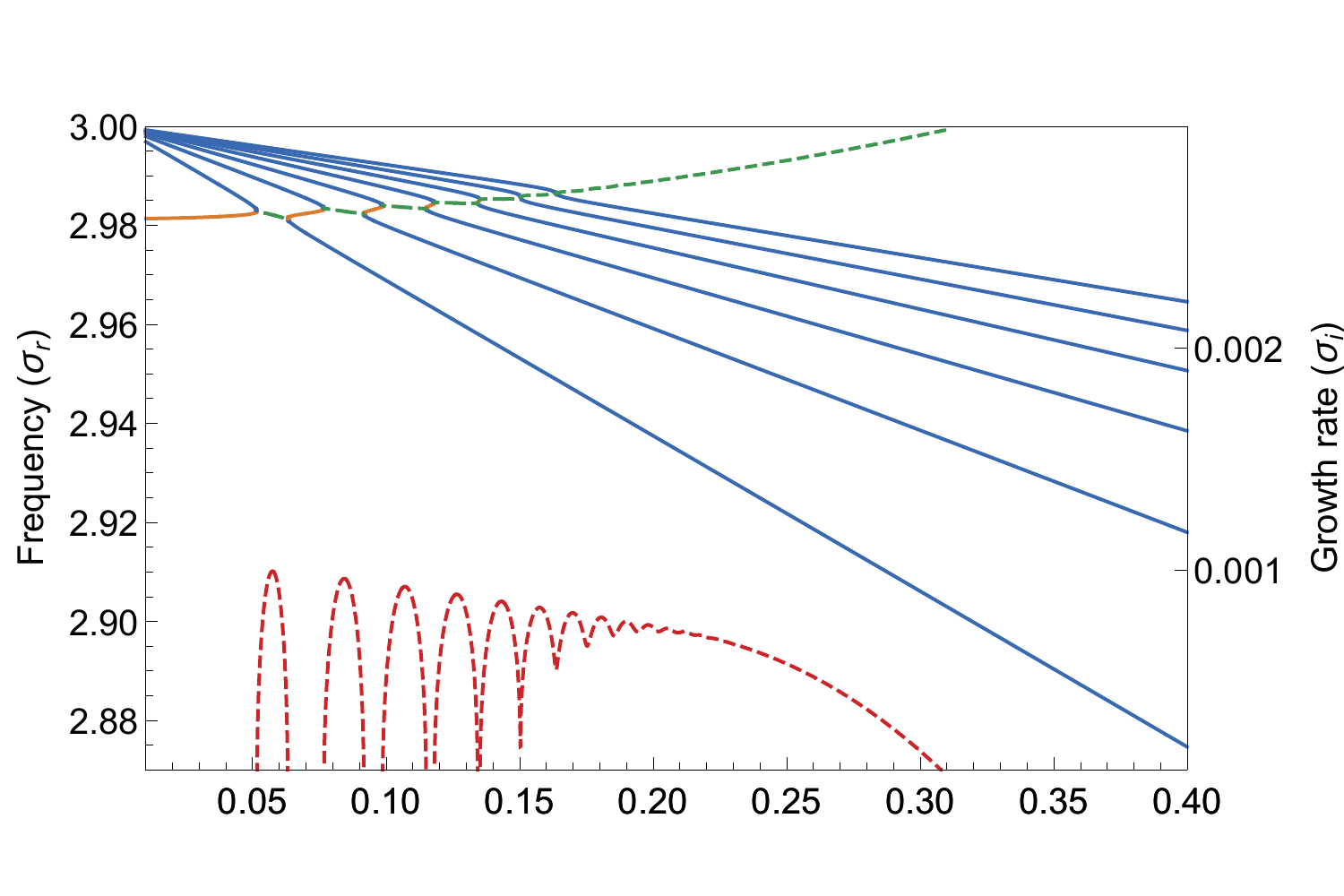}
       \caption{$We = 8.1$. The capillary-Coriolis coalescence plus four hysteretic dispersion curves.}
       \label{fig:increasing_no_of_hysteretic_dispCurves-d}
    \end{subfigure}
    \caption{An increasing number of retrograte dispersion curves exhibit a
    hysteretic character for $We \to (n^2-1)^+$; the figures show this
    behavior for $n = 3$.}
    \label{fig:increasing_no_of_hysteretic_dispCurves}
 \end{figure}

\begin{figure}
     \includegraphics[width=\textwidth]{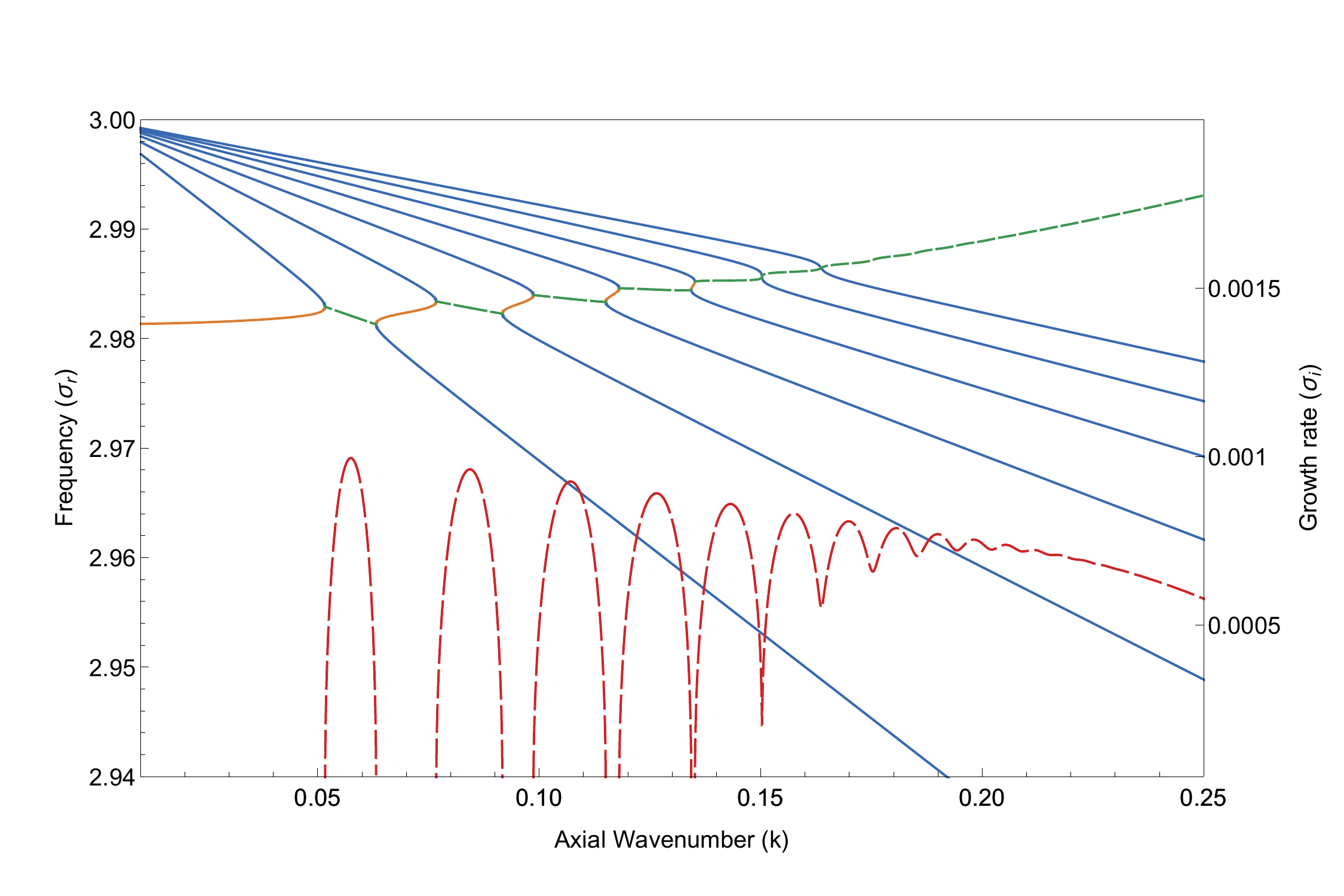}
     \caption{A magnified view of Fig.
     \ref{fig:increasing_no_of_hysteretic_dispCurves-d}; $We = 8.1; n = 3$. Note the multiple
     hysteretic dispersion curves (four, besides the original) and corresponding
     oscillations in the growth rate.}
    \label{fig:multiple_folds_with_growth_rate}
 \end{figure}

\begin{figure}
    \centering
    \begin{subfigure}{.48\textwidth}
       \centering
       \includegraphics[width=\linewidth]{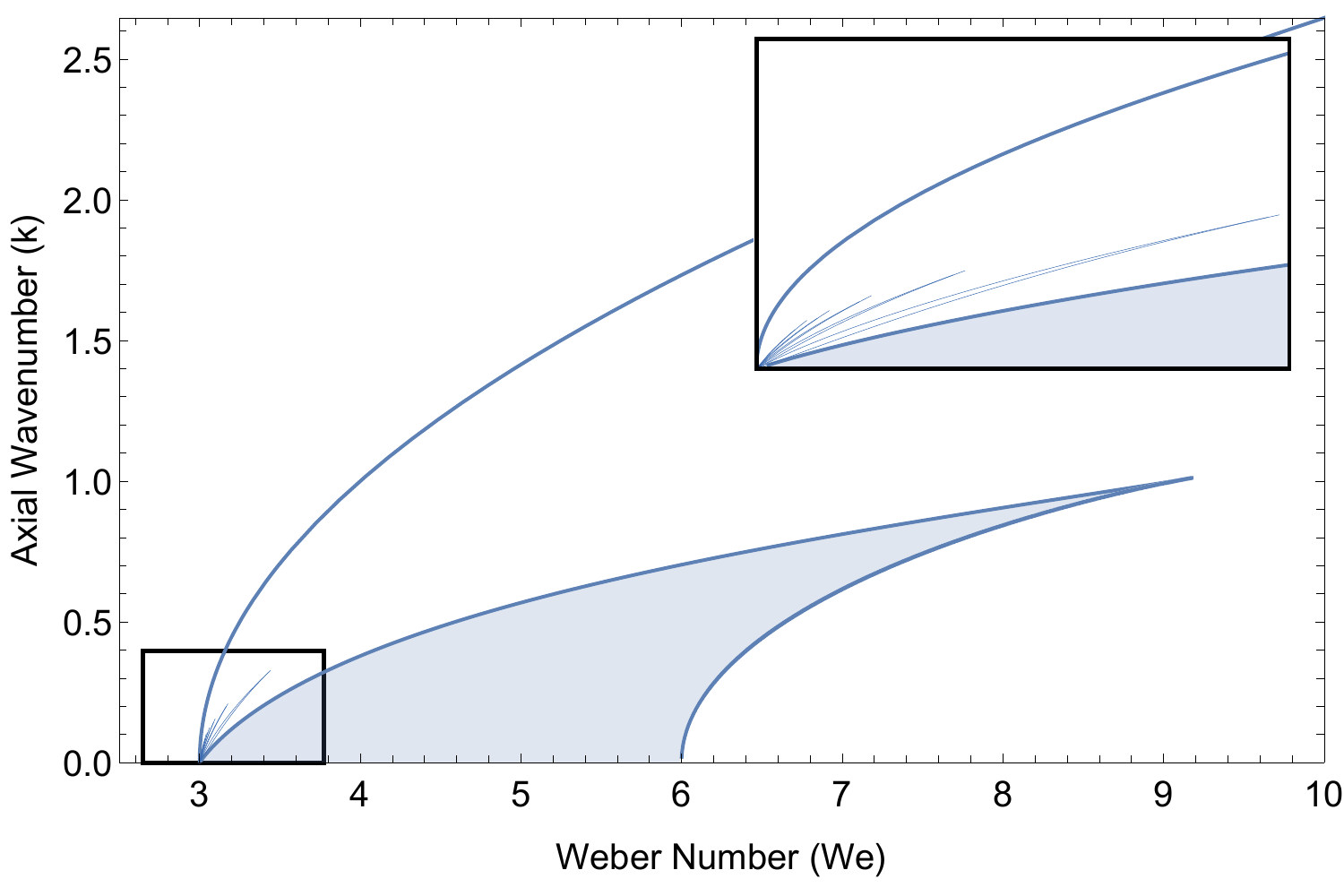}
       \caption{n = 2}
       %  \label{fig:}
    \end{subfigure}
    \hfill
    \begin{subfigure}{.48\textwidth}
       \centering
       \includegraphics[width=\linewidth]{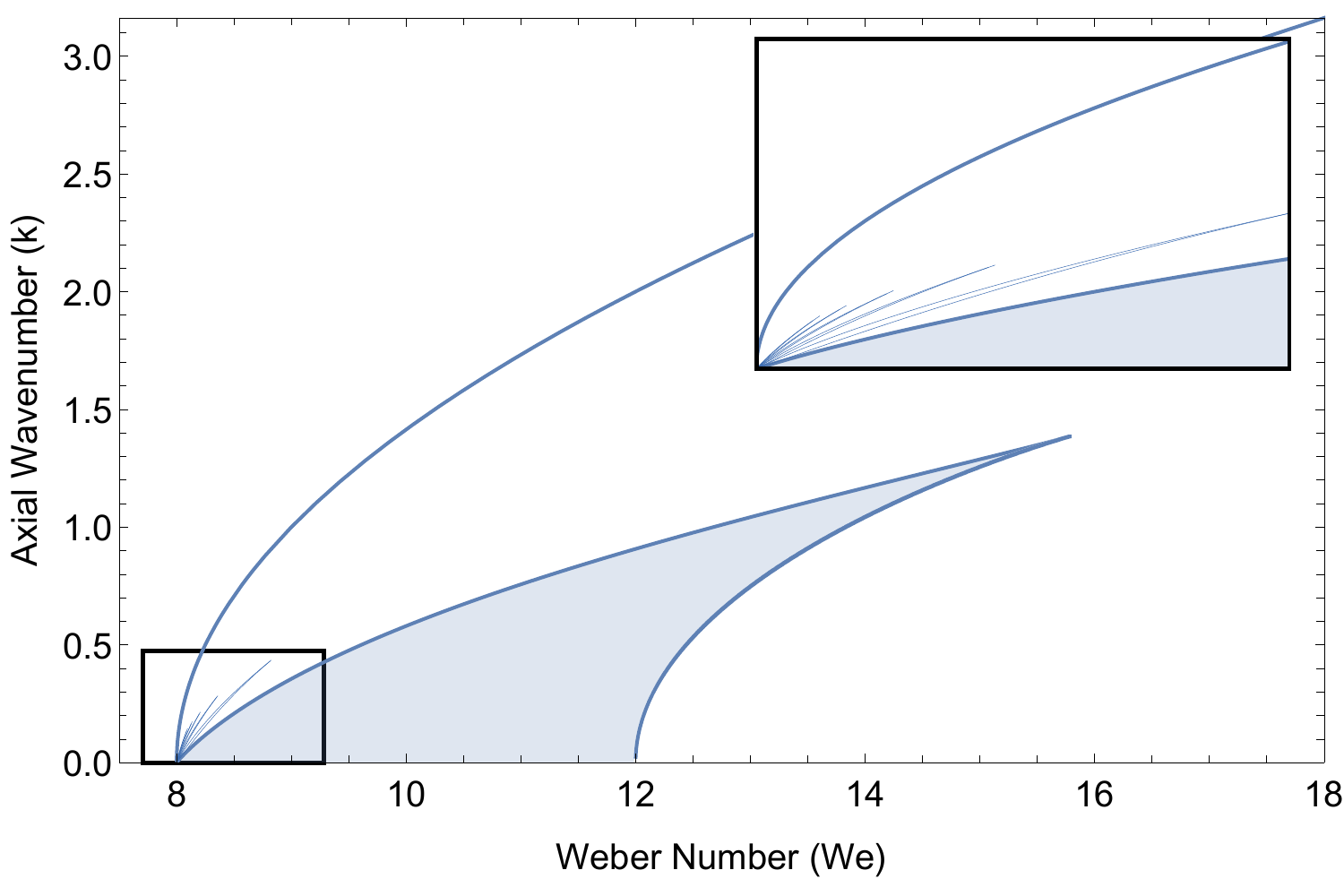}
       \caption{n = 3}
       %  \label{fig:}
    \end{subfigure}
    \hfill
    \begin{subfigure}{.48\textwidth}
       \centering
       \includegraphics[width=\linewidth]{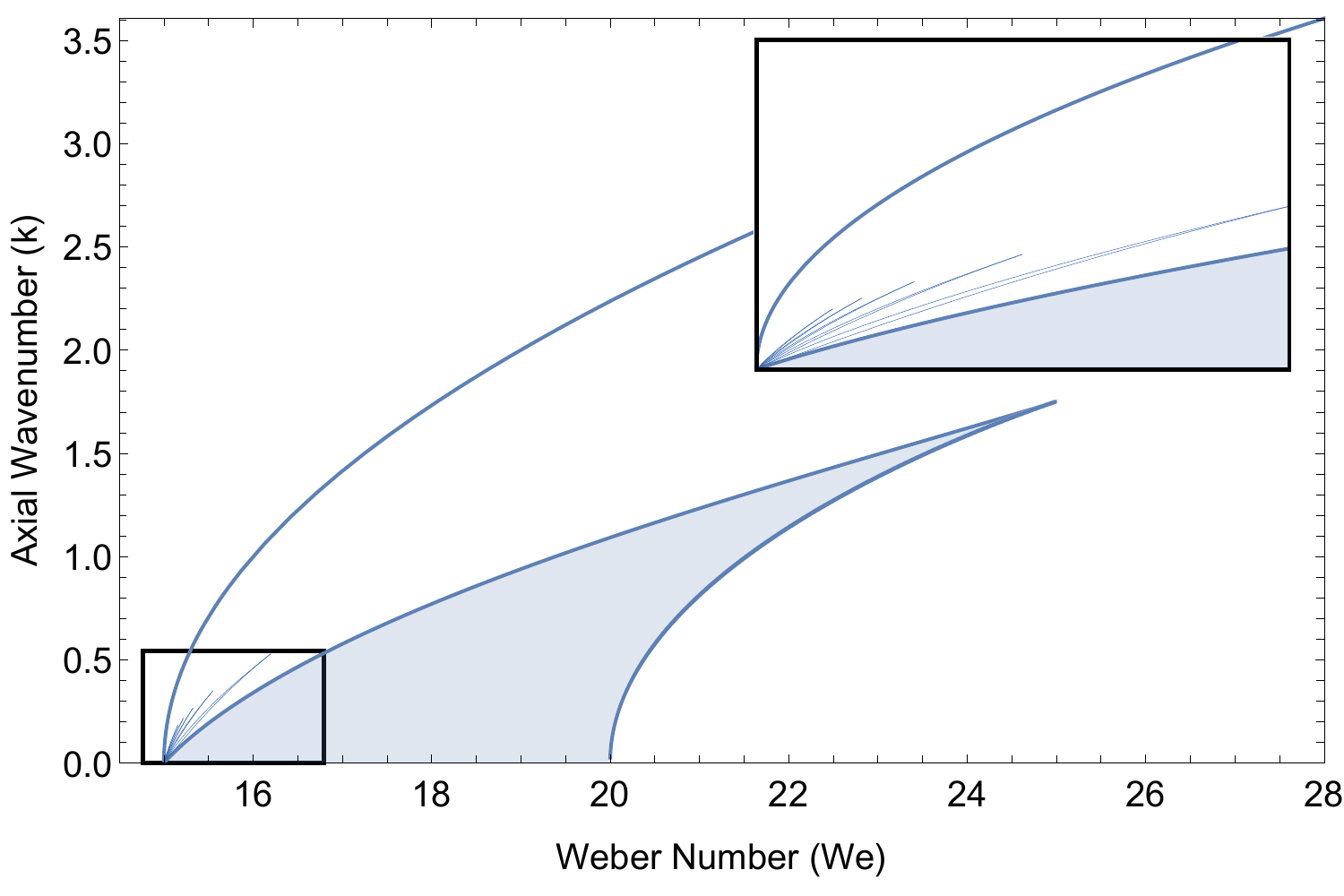}
       \caption{n = 4}
       %  \label{fig:}
    \end{subfigure}
    \hfill
    \begin{subfigure}{.48\textwidth}
       \centering
       \includegraphics[width=\linewidth]{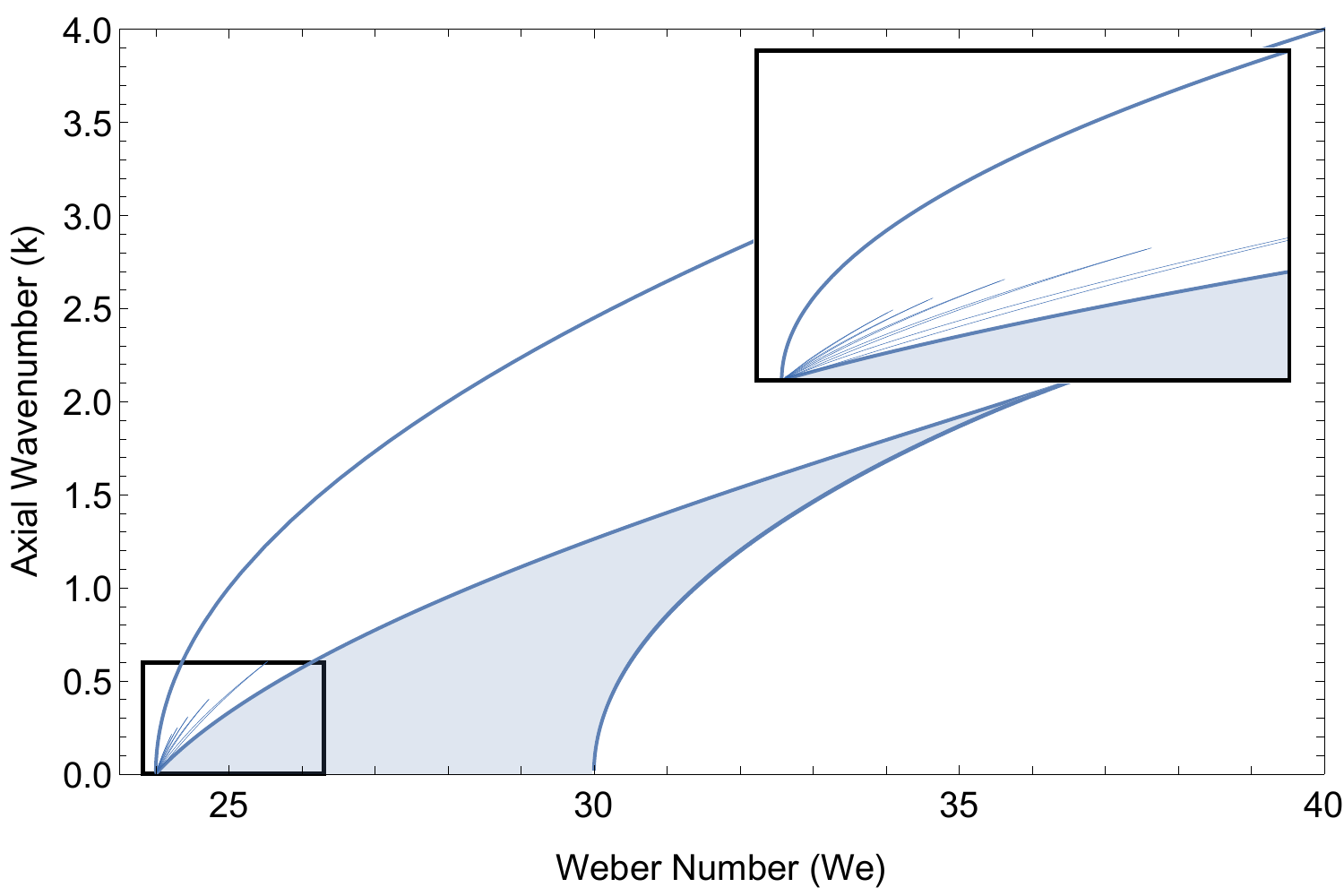}
       \caption{n = 5}
       %  \label{fig:}
    \end{subfigure}
    \caption{The main and satellite islands of inviscid stability, in the $We-k$
    plane, for $n = 2,3,4$ and $5$. Magnified views, of five of the infinite
    hierarchy of satellite islands, appear in the insets.}
    \label{fig:multiple_cusps}
 \end{figure}

 The second subtle aspect is related to the retrograde mode above that bifurcates
 from the cograde capillary mode at $\sigma = n$. In moving further down towards
 $\sigma = n-1$ with increasing $We$ (at which point this mode undergoes a
 coalescence with the lower capillary branch, leading to an unstable wavenumber
 interval $(0, k_2)$, as illustrated in Figs. \ref{fig:3D_regime3} and
 \ref{fig:3D_regimes_n4-c}), the mode must end up crossing an infinite number of
 Coriolis mode branches, in turn implying the possibility of an infinite
 hierarchy of coalescences, instead of just the single one with the lowermost
 (retrograde) Coriolis branch shown in Figs. \ref{fig:3D_regime2} and
 \ref{fig:3D_regimes_n4-b}. Note that the infinite number of crossings must occur
 in the neighborhood of $\sigma = n$ which, from eq. \ref{eq:planar_dispersion},
 corresponds to $We = n^2-1$. Therefore, one may verify the existence of such a
 hierarchy of crossings by checking for the occurrence of coalescences in the
 eigenspectra in the vicinity of $We = n^2 - 1$. It turns out that all of the
 crossings of the aforementioned retrograde mode with the retrograde Coriolis
 modes lead to coalescences, and thence, hysteretic dispersion curves  with
 intermediate unstable $k$-intervals. The number of such hysteretic curves
 increases rapidly as $We$ approaches $n^2-1$ from above. Fig.
 \ref{fig:increasing_no_of_hysteretic_dispCurves} depicts the increasing number
 of hysteretic dispersion curves that result for $n = 3$ for $We \rightarrow
 8^+$; there is one coalescence for $We = 8.5$, three for $We=8.2$ and four for
 $We=8.1$, besides the original coalescence between the new retrograde mode and
 the lowest Coriolis mode, with successive coalescences occurring at
 progressively smaller $k$. Fig. \ref{fig:multiple_folds_with_growth_rate}
 provides a magnified view of the ensemble of dispersion curves for $We = 8.1$,
 emphasizing the rapid oscillations in the growth rate owing to the multiple
 closely-spaced intervals of stability. Similar to Fig.
 \ref{fig:stability_islands}, each of these hysteretic intervals marks out a
 stable island in the $We-k$ plane. Therefore, in the inviscid limit, there
 appears to be an infinite hierarchy of neutrally stable islands (each of these
 associated with a cusp catastrophe, or a fold in the three-dimensional surface
 characterizing the $We-k-\sigma$ relationship as illustrated in Fig.
 \ref{fig:cusp_formation}) enclosed within the viscously unstable region given by
 $We > n^2+k^2-1$. This infinite hierarchy of inviscidly stable islands only
 appears above, and not below, the viscous threshold since, as already pointed out, the cograde modes
 exhibit avoided collisions\,(Figs.\ref{fig:mode_splitting-a}-\ref{fig:mode_splitting-c}), while the retrograde modes coalesce, yielding complex eigenvalues, consistent with their respective Krein signatures. The higher-order
stable islands are much smaller than the leading one, and decrease in size rapidly, eventually asymptoting to the limit point
 $(We,k) = (n^2-1,0)$. The resulting picture in the $We-k$ plane is illustrated
 in Fig. \ref{fig:multiple_cusps} for $n = 2,3,4$ and $5$. Each of the
 sub-figures shows five leading satellite islands besides the main island, of
 what is likely an infinite hierarchy. As implied by Fig.
 \ref{fig:multiple_folds_with_growth_rate}, a consequence of this infinite
 hierarchy is a rapid alternation of regions of stability and instability as one
 increases $k$ for a fixed $We$, and thence, a rapidly fluctuating growth rate
 with changing $k$.
 
 % Variation of growth-rate in the alternating stable-unstable regions
 
 % Satellite islands resulting due to multiple hysteretic curves
 
\section{Conclusion}
\label{sec:conclusion}
	
In this paper, we have analyzed the inviscid stability of a rotating liquid
column as a function of $We$, and the axial and azimuthal wavenumbers ($k$ and
$n$) of the imposed perturbation, with a focus on the entire eigenspectrum that
includes a pair of capillary modes and an infinite hierarchy of Coriolis modes
in the general case. While it is known that a viscous rotating column becomes
unstable if and only if $We > k^2+n^2-1$, consideration of the full
eigenspectrum highlights the intricate nature of the inviscidly unstable region
in the $We-k$ plane. The intricacy arises from the likelihood of an infinite
hierarchy of coalescences between pairs of dispersion curves in the
neighborhood of the planar viscous threshold $(We = n^2-1)$. As illustrated in
Fig. \ref{fig:multiple_cusps}, these coalescences appear to lead to an infinite
hierarchy of inviscidly stable islands within the viscously unstable region $(We
> n^2+k^2-1)$ in the $We-k$ plane; each of these islands corresponds to a fold in the $We-k-\sigma$
surface in three dimensions\,(see Fig. \ref{fig:cusp_formation}). The existence
of these islands explains why $We < n^2 + k^2 -1$ only serves as a sufficient
condition for stability in the inviscid limit\,[\cite{pedley_1967}]; evidently, one can be stable even when $We > n^2+k^2-1$, provided one is inside any of these islands. Thus, the
necessary and sufficient condition for inviscid instability would require $We >
n^2+k^2 -1$, and in addition, that the $(We,k,n)$ triplet chosen lies outside
the inviscid islands identified in Fig. \ref{fig:multiple_cusps}. It is not
possible to provide a precise expression for this criterion since, as already
pointed out, closed form expressions for the island boundaries are not known;
although, the small-$k$ asymptotes appear to serve as useful approximations
especially for the larger $n$'s (see Fig. \ref{fig:stability_islands}). Interestingly, it has been shown in the Appendix of \cite{weidman_1997} that one may again obtain only a sufficient
condition for inviscid stability for the analogous two-fluid system. Thus, for a configuration consisting of a central column of a denser liquid and an annular domain of a lighter one, both being in a state of rigid-body rotation, the sufficient condition for inviscid stability becomes
\begin{equation}
\mathit{We}_{1} - \mathit{We}_{2} < k^2 + n^2 -1
\label{eq:two_fluid_stability_condition}
\end{equation}
where $\mathit{We}_{1}$ is the Weber number defined using the density of the inner fluid and $\mathit{We}_{2}$ is that defined using the density of the outer fluid.
Although the viscous problem, and therefore, the viscous stability criterion for the two-fluid system problem has not yet been examined, the discussion presented here suggests an analogous relationship between the viscous and inviscid criteria.

The discovery of an infinite hierarchy of inviscidly stable islands in the
$We-k$ plane also has implications for viscous stability for large but finite
$Re$. As shown in Fig. \ref{fig:multiple_folds_with_growth_rate}, for the case
$n =3$ and $We = 8.1$, the inviscid growth rate oscillates rapidly between zero
and order-unity values, with the oscillations becoming increasingly dense and
rapid for $We \rightarrow 8^+$ (the value of $n^2-1$ for $n = 3$);
interestingly, although the growth rates decrease for $We \rightarrow
(n^2-1)^+$, as expected, the amplitude of the oscillations, for a fixed $We$,
does not appear to decay with increasing order of the modal coalescence, the
order here referring to the modal index of the (retrograde) Coriolis mode
involved in the particular coalescence. Now, for large but finite $Re$, one
expects only a finite number of oscillations regardless of the proximity of $We$
to $n^2-1$. This is because, as mentioned in section \ref{sec:planar_ptb}, the
viscous decay rates for planar perturbations can attain arbitrarily high values
for large enough modal indices, owing to the vanishingly small radial scale
associated with the eigenfunction. Thus, for finite $Re$ however large, one
expects the (inviscid) growth rate associated with modal coalescences above a
certain threshold order to be overwhelmed by viscous decay. Nevertheless, for an
$Re$ large enough that the spacings between adjacent islands is greater than
$O(Re^{-2})$, one expects rapid oscillations in the growth rate between
order-unity values between the islands, and $O(Re^{-1})$ values within them.
While the $Re$'s required to see substantial growth-rate oscillations requires a
full viscous calculation, it does appear that the $Re$'s involved might be very
large. In contrast, experiments on the rotating liquid column have only accessed a maximum
$Re$ of $O(10)$; see \cite{kubitschek_2008}.

The question of $Re$ being large enough for one to be able to observe the
aforementioned oscillations in the (viscous) growth rate leads us to the
astrophysical analog of the configuration examined thus far - that of a rotating
self-gravitating fluid column, which may be likened to a large-scale filamentary structure; such structures appear to have a ubiquitous presence in the interstellar medium [\cite{andre17}]. Chandrasekhar and Fermi performed one of the earliest studies on the stability of a
self-gravitating column\,[see \cite{chandrafermi53,
chandra64}]. Their calculations were done in the incompressible
limit and revealed an axisymmetric instability. Analogous to the
Rayleigh-Plateau instability of the liquid column, the axisymmetrically deformed
fluid column has a lower gravitational potential energy than the original columnar
configuration for sufficiently long-wavelength perturbations and is therefore
unstable to all axisymmetric perturbations with $k<1.0668$. Subsequently,
Ostriker extended Chandrasekhar and Fermi’s analysis to a compressible base
state, evaluating the base-state density profiles for different values of the
polytropic index [\cite{ostriker64a}]. Compressibility leads to an inhomogeneous base state of
a finite radius, the isothermal case being an exception in leading to an infinite radius. Thus,
\cite{ostriker64b} analyzed the effects of compressible perturbations
on a base state that is a solution of the incompressible equations\,(a homogeneous finite-radius cylinder).
Later, \cite{nagasawa87} carried out a stability calculation for the inhomogeneous density profile corresponding
to the aforementioned isothermal base state, and concluded that the isothermal
problem is only unstable to axisymmetric disturbances, similar to Chandrasekhar
and Fermi’s analysis above. The critical wavelength obtained by
\cite{nagasawa87} was also found earlier by \cite{stodolkiewicz63}, who had carried out a restricted
stability analysis of the isothermal base state only to axisymmetric disturbances. Due to their
prominent role in star formation, studies on the gravitational instability of
filamentary molecular clouds continue to garner attention
[\cite{mckeeostriker07}]. Recently \cite{motiei21} have revisited the problem of
axisymmetric stability of self-gravitating fluid cylinders, including effects of an
external pressure, an axial magnetic field and a wide array of non-isothermal
equations of state that are a better representation of observations.

Rotation is ubiquitous in self-gravitating filaments, and \cite{hansen76} were
one of the first to examine the equilibrium and stability of an isothermal and
uniformly rotating cylinder. The inclusion of rotation introduced an unusual
feature to the equilibrium state - a (spatially)\,damped oscillatory density
profile. \cite{hansen76} analyzed the stability of the rotating base states of a
finite radius to two-dimensional disturbances. They found that rotation renders
cylinders, exceeding a critical size, unstable to non-axisymmetric disturbances.
\cite{jog14} have recently studied the stability of a polytropic rotating
cylinder using a local stability analysis. \cite{sagar16} have
extended the local stability analysis to polytropic rotating cylinders, with the
addition of a magnetic field, and observe that the background rotation acts to
reduce the unstable growth rates. Thus, there exist instances in the literature of rotation having 
conflicting roles - in terms of both stabilizing and destabilizing
self-gravitating masses. To comprehensively understand the role that rotation plays
in the stability of filamentary molecular clouds, there is an imminent need for
a global three-dimensional stability analysis of rotating polytropic cylinders.
In this section, however, we only look at the limiting case of an
incompressible cylinder - the rotating version of Chandrasekhar and Fermi’s
seminal study. This is because our focus here is to primarily draw an analogy
between the cohesive forces of surface tension and gravitation; we show below that a rotating self-gravitating incompressible fluid column exhibits an infinite hierarchy of cusp-catastrophes analogous to the rotating liquid column
above. Although beyond the scope
of the present calculation, we expect our findings to continue to be relevant for compressible
self-gravitating columns.

\begin{figure}
   \centering
   \begin{subfigure}{.48\textwidth}
      \centering
      \includegraphics[width=\linewidth]{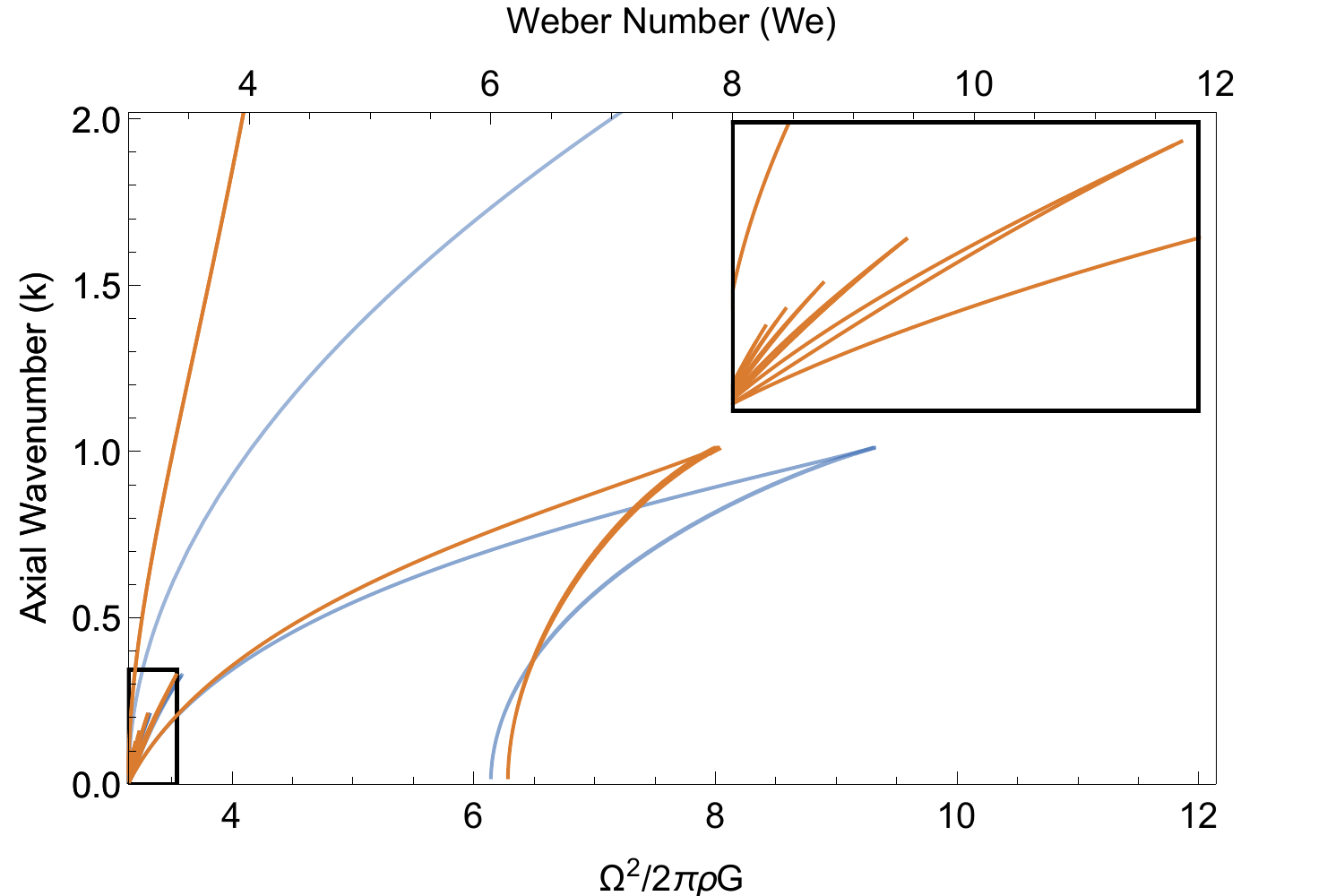}
      \caption{n = 2}
      %  \label{fig:}
   \end{subfigure}
   \hfill
   \begin{subfigure}{.48\textwidth}
      \centering
      \includegraphics[width=\linewidth]{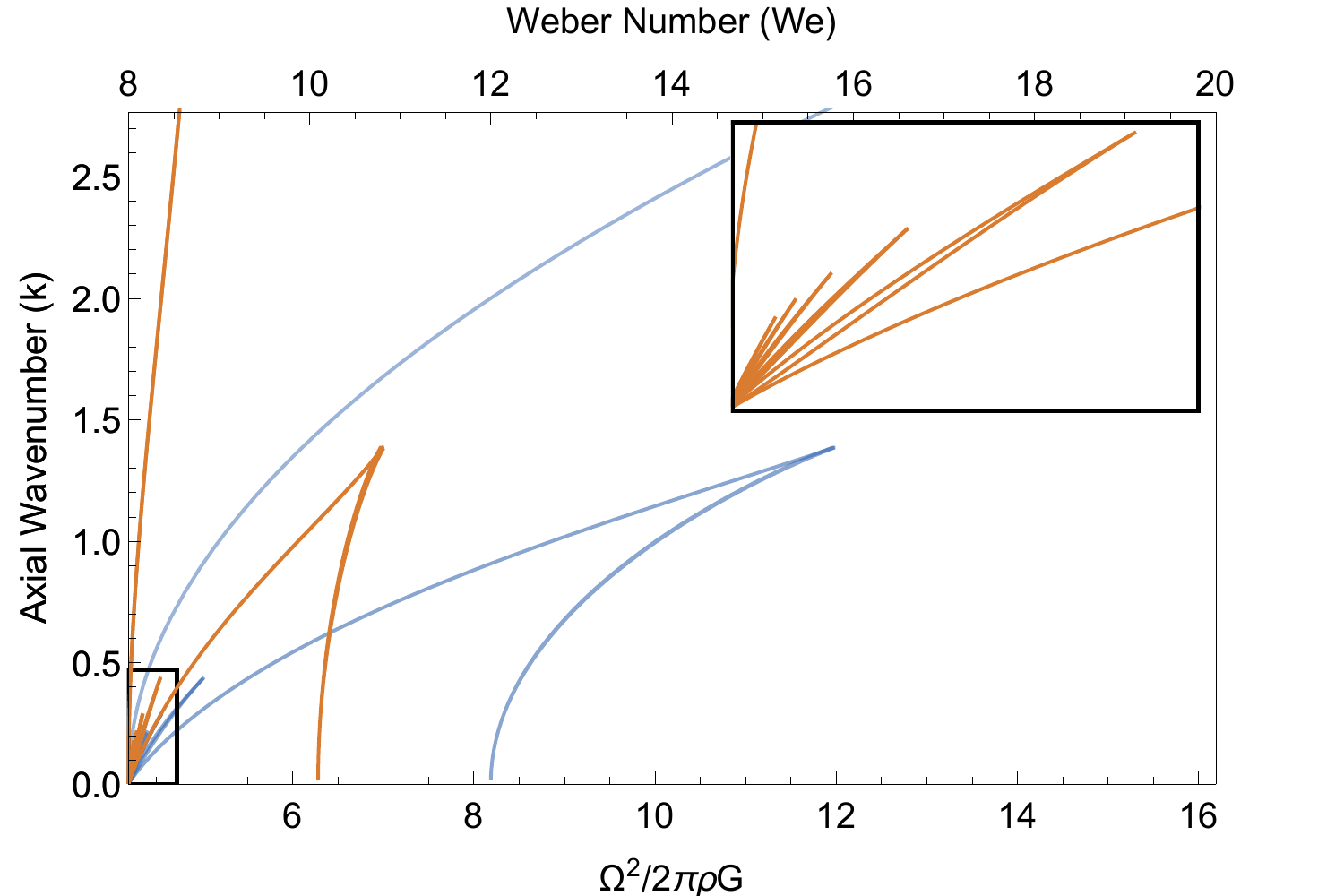}
      \caption{n = 3}
      %  \label{fig:}
   \end{subfigure}
   \hfill
   \begin{subfigure}{.48\textwidth}
      \centering
      \includegraphics[width=\linewidth]{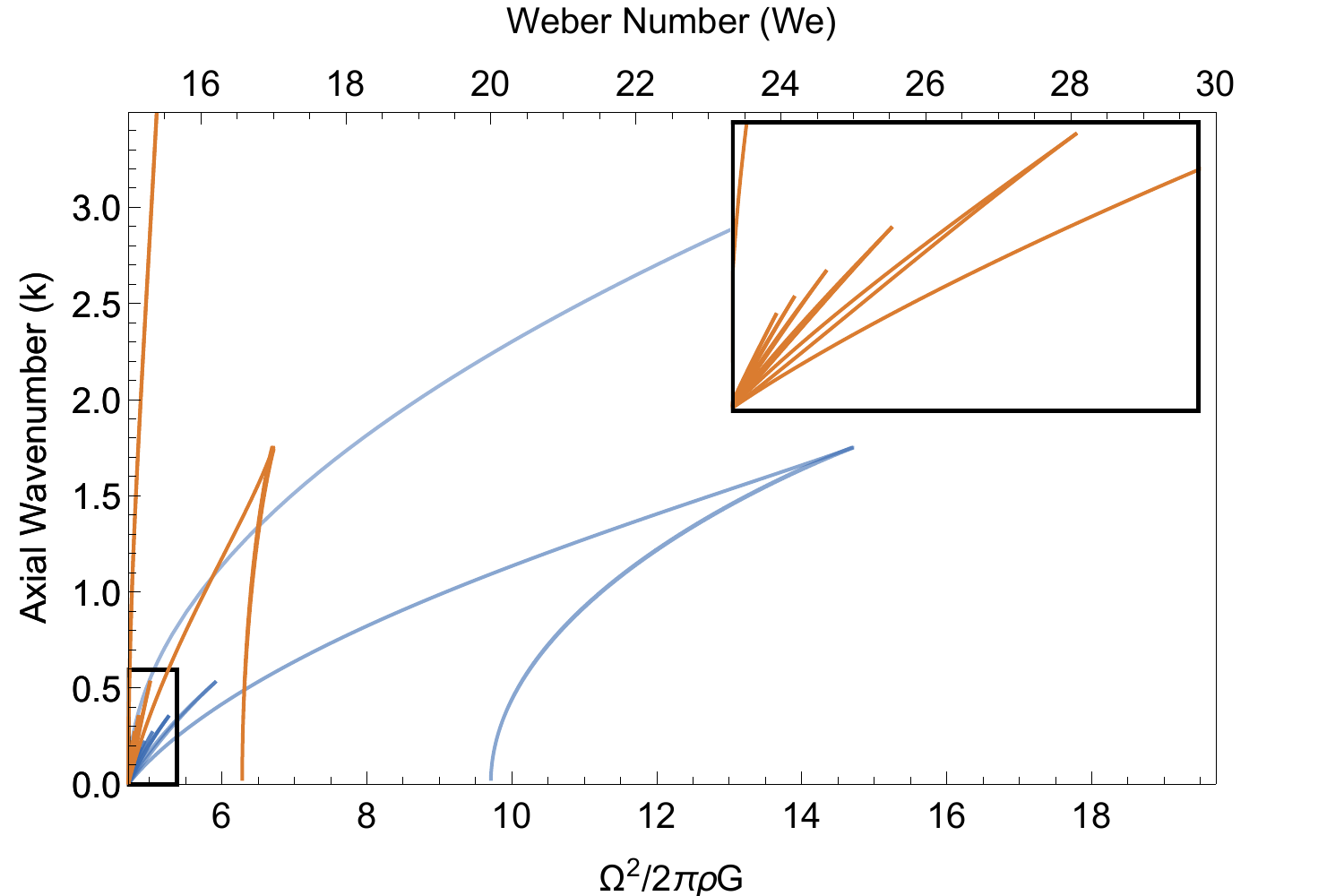}
      \caption{n = 4}
      %  \label{fig:}
   \end{subfigure}
      \hfill
   \begin{subfigure}{.48\textwidth}
      \centering
      \includegraphics[width=\linewidth]{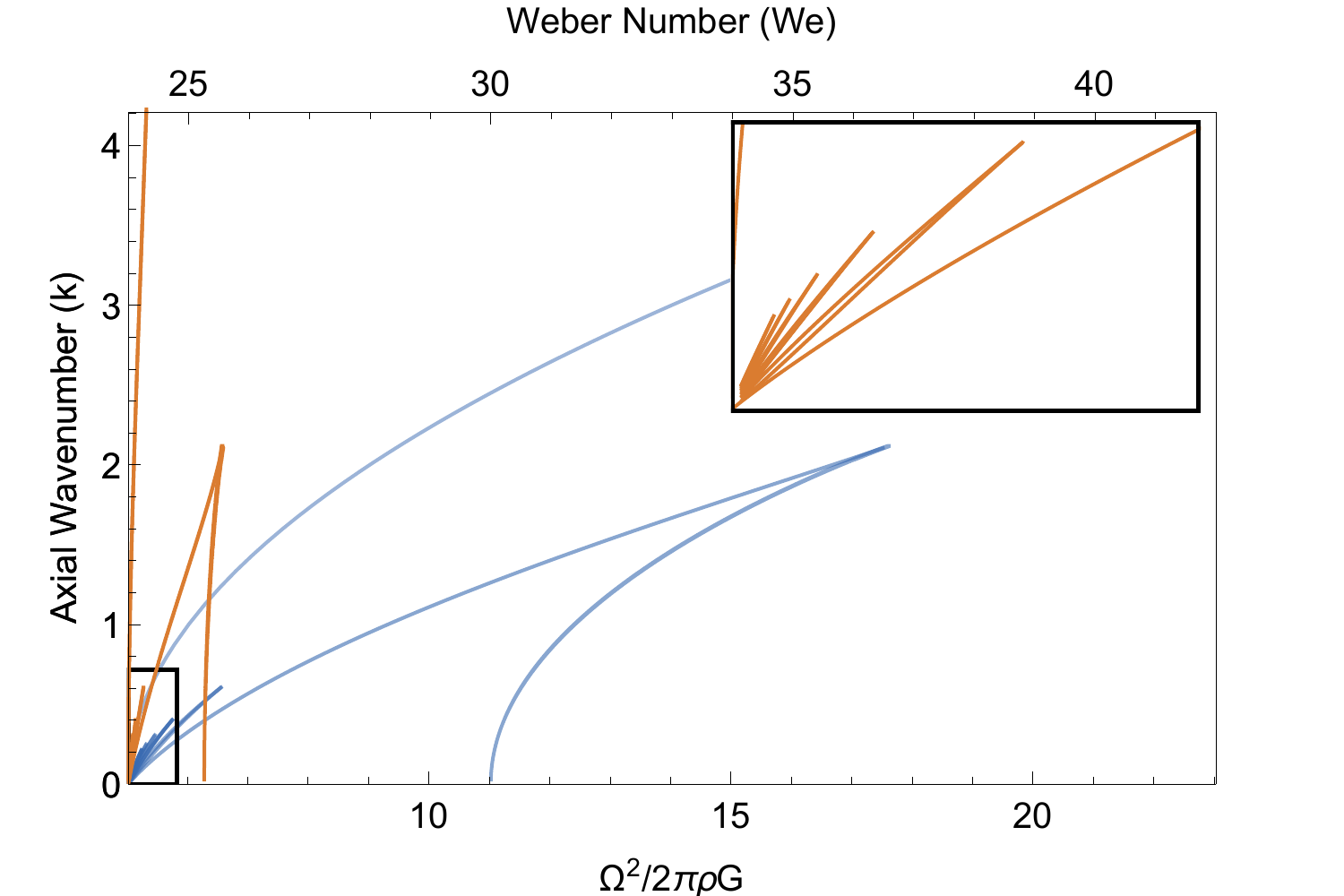}
      \caption{n = 5}
      %  \label{fig:}
   \end{subfigure}
   \caption{Comparison of the inviscidly stable islands for the surface-tension
   (blue) and self-gravitating (orange) cases for $n=2,3,4$ and $5$. The insets
   in each figure show magnified views of the first five satellite islands for
   the self-gravitating column.} 
   \label{fig:sg_comparison}
\end{figure}

A linear stability analysis of a self-gravitating cylinder, in a state of
rigid-body rotation, readily furnishes the following dispersion relation: 

\begin{equation}
   \alpha \frac{J_{n-1}(\alpha)}{J_{n}(\alpha)} - 
   \frac{\frac{\Omega^2}{2\pi\rho G}(4-(\sigma-n)^2)}{\frac{\Omega^2}{2\pi\rho G} - (1 - 2K_n(k)I_n(k))} - 
   n\left( 1+\frac{2}{(\sigma-n)} \right)=0,
   \label{eq:sg_dispersion}
\end{equation}
where $\frac{\Omega^2}{2\pi\rho G}$ is the analog of the Weber
number for the self-gravitating case, in measuring the relative importance of
gravitational and centrifugal forces, $G$ being the gravitational constant. Let
us now define

\begin{equation}
   \mathit{We_G} = f(n,k)\,\frac{\Omega^2}{2\pi\rho G},
   \label{eq:sg_We}
\end{equation}
where $f(n,k) = \frac{k^2+n^2-1}{1 - 2I_n(k)K_n(k)}$, so that we have, in terms
of $\mathit{We_G}$, a dispersion relation identical to that in eq.
\ref{eq:3D_rotating_column_dispersion}. Since $f(n,k)$ is a monotonically
increasing function of $k$ for a given $n$, one has a unique value of
$\frac{\Omega^2}{2\pi\rho G}$ for every $\mathit{We_G}$. The
$\frac{\Omega^2}{2\pi\rho G}-k$ plane will then be topologically equivalent to
the $\mathit{We_G}-k$ plane, and thence, to the $\mathit{We}-k$ plane of the
rotating liquid column seen earlier. The resulting infinite hierarchy of
cusp-catastrophes for the rotating self-gravitating fluid column, and its comparison with
the hierarchy already seen above, for the rotating liquid column (with surface tension), is shown in Fig. \ref{fig:sg_comparison}.

% comparison of self-gravitating and surface-tension driven stable islands

\clearpage

\appendix

\section{Stability criteria for $n = 1$}
\label{sec:appendix}
	For planar perturbations with $n = 1$, one sees from eq. \ref{eq:planar_dispersion} that $\sigma$ vanishes identically
for all non-zero $We$. Therefore, the inviscid threshold, $We = n(n+1)$, is of no relevance in this case, and there can be no analog of the different $We$-regimes that exist for $n \geq 2$\,(see Figs. \ref{fig:3D_regime1}-\ref{fig:3D_regime4}). As already stated in the main text, this is expected since a planar
perturbation for $n=1$ corresponds to a mere translational displacement. Further, for $n \geq 2$, the pair of planar frequencies, corresponding to the two capillary branches, are widely separated, and symmetrically distributed about $\sigma = n-1$, for small $We$. The motion of the upper capillary branch down towards $\sigma = n$, with increasing $We$, and the subsequent birthing of a new retrograde mode at $\sigma = n$, is essential for the infinite hierarchy of coalescences in the
$\sigma-k$ plane, and thence, for the infinite hierarchy of inviscidly stable islands in the $We-k$ plane. In contrast, for $n = 1$, $\sigma = 0$ is a double root of the governing quadratic. Thus, a second mode emanates from $\sigma = 0$, with increasing $k$, in addition to the retrograde capillary mode, and may be likened to the retrograde mode above, for $n \geq 2$, that undergoes coalescences with the Coriolis modes. But, for $n =1$, this mode is already below the infinite hierarchy of retrograde Coriolis modes even at the smallest $We$\,(see Fig. \ref{fig:n=1_dispCurves-1} below) , and one may only expect a merger with the lowest member of the Coriolis hierarchy. As a result, one expects a leading stable island, but no satellite islands, in the $We-k$ plane.

In fact, the nature of the dispersion curves for $n=1$ does not change qualitatively with increasing $We$. One observes a single coalescence, and thence, a single unstable interval bounded away from $k = 0$, for any finite $We$. Figs. \ref{fig:n=1_dispCurves-1}-\ref{fig:n=1_dispCurves-4} confirm this behavior for $We$ ranging from 1 to 10. As $We=10$ is significantly greater than the hypothetical threshold\,($n(n+1) = 2$), we do not expect this picture to alter
for greater $We$'s and the resulting stable island in the $We-k$ plane would thus extend to infinity (Fig. \ref{fig:n1_islands}). This, once again, points to the absence of distinct $We$ regimes observed for $n \geq 2$. The growth rate
and range of unstable wavenumbers shown in \ref{fig:n=1_dispCurves-4} are
consistent with Fig. 4 of \cite{weidman_1997}. Finally, from the growth rate perspective, the $n=1$ mode
remains subdominant for all $Re$ greater than $1.166$ (approx). As shown by \cite{kubitschek_weidman_2007} the azimuthal wavenumber of the dominant perturbation jumps from $n=0$ to $n=2$ beyond $Re \approx 1.166$, and then to higher $n$'s with increasing $We$. Thus, the anomalous behavior of $n=1$ is less important in the limit of large $Re$ as, for instance, in the astrophysical context above.

% Dispersion curves and the stable island for n = 1

\begin{figure}
   \centering
   \begin{subfigure}{0.48\textwidth}
      \centering
      \includegraphics[width=\textwidth]{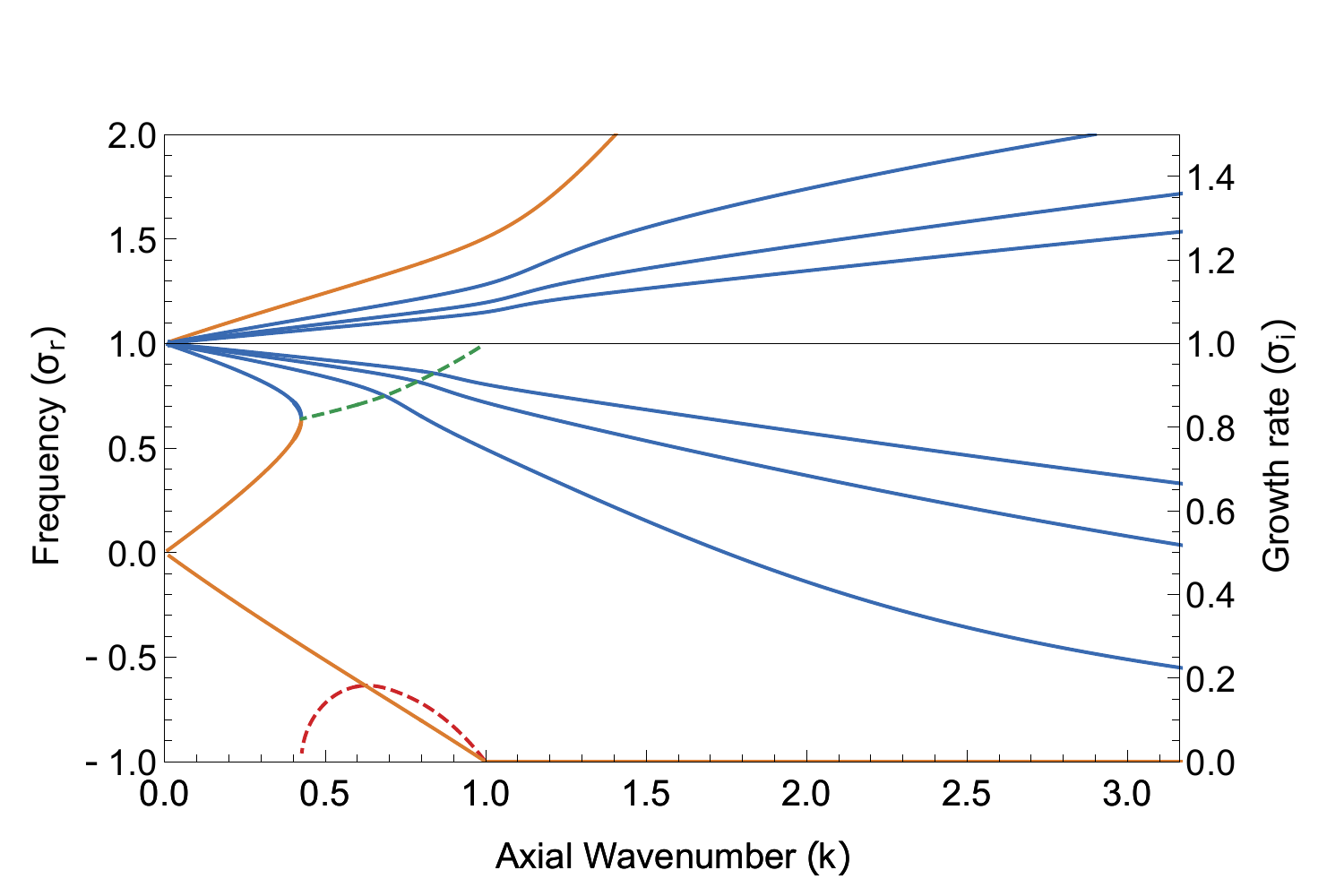}
      \caption{$We = 1$}
      \label{fig:n=1_dispCurves-1}
   \end{subfigure}
   \hfill
   \begin{subfigure}{0.48\textwidth}
      \centering
      \includegraphics[width=\textwidth]{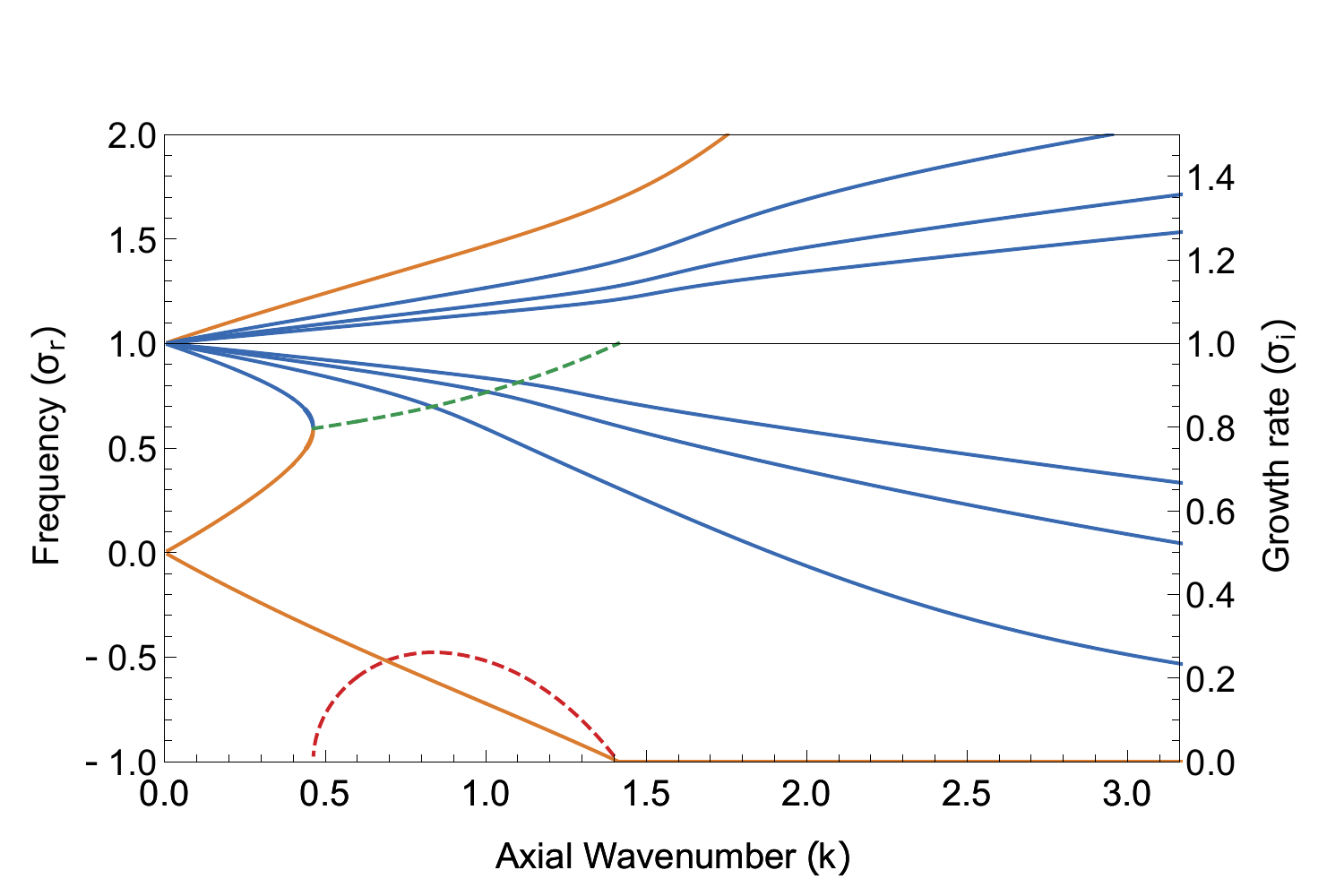}
      \caption{$We = 2$}
      \label{fig:n=1_dispCurves-2}
   \end{subfigure}
   \hfill
   \begin{subfigure}{0.48\textwidth}
      \centering
      \includegraphics[width=\textwidth]{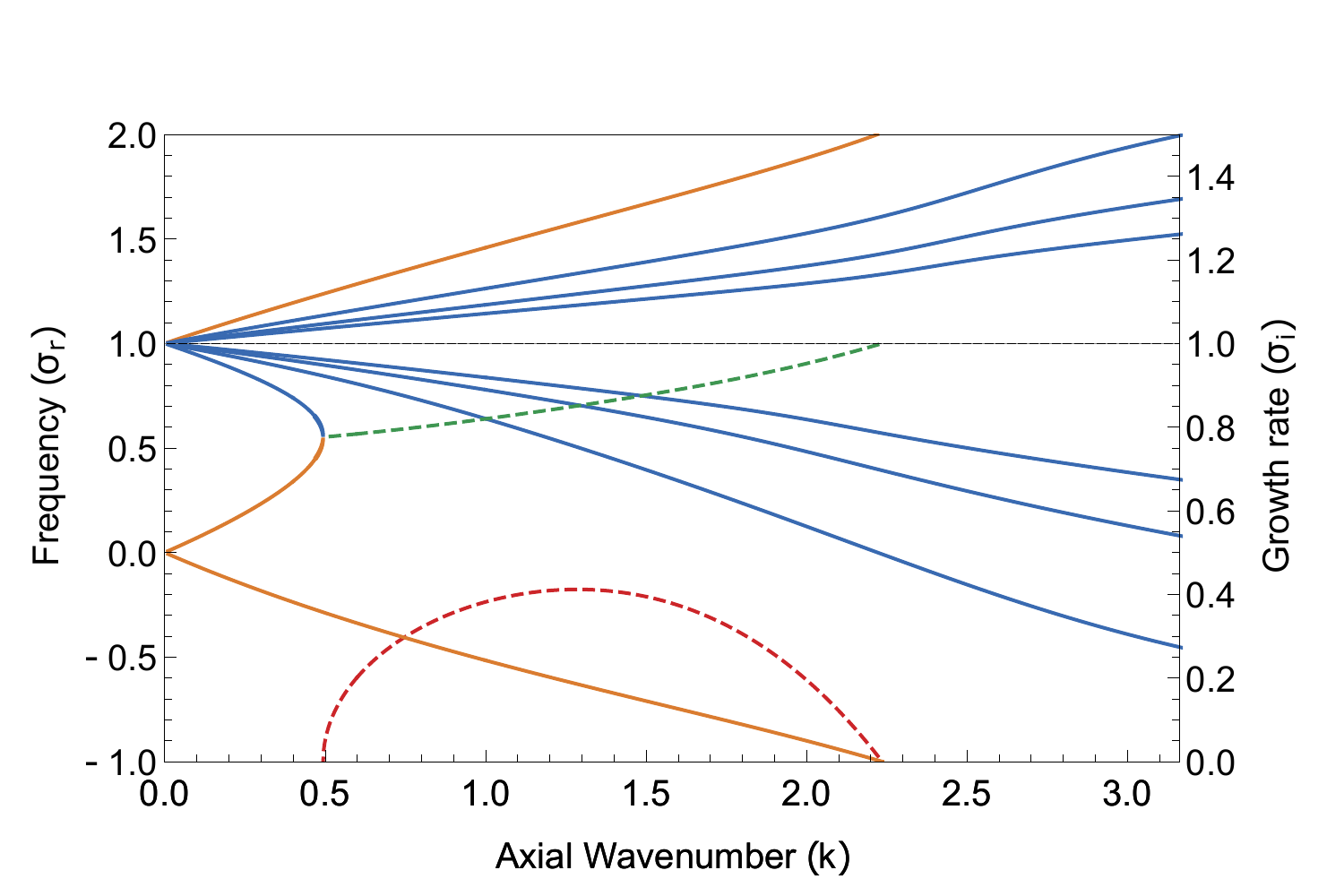}
      \caption{$We = 5$}
      \label{fig:n=1_dispCurves-3}
   \end{subfigure}
   \hfill
   \begin{subfigure}{0.48\textwidth}
      \centering
      \includegraphics[width=\textwidth]{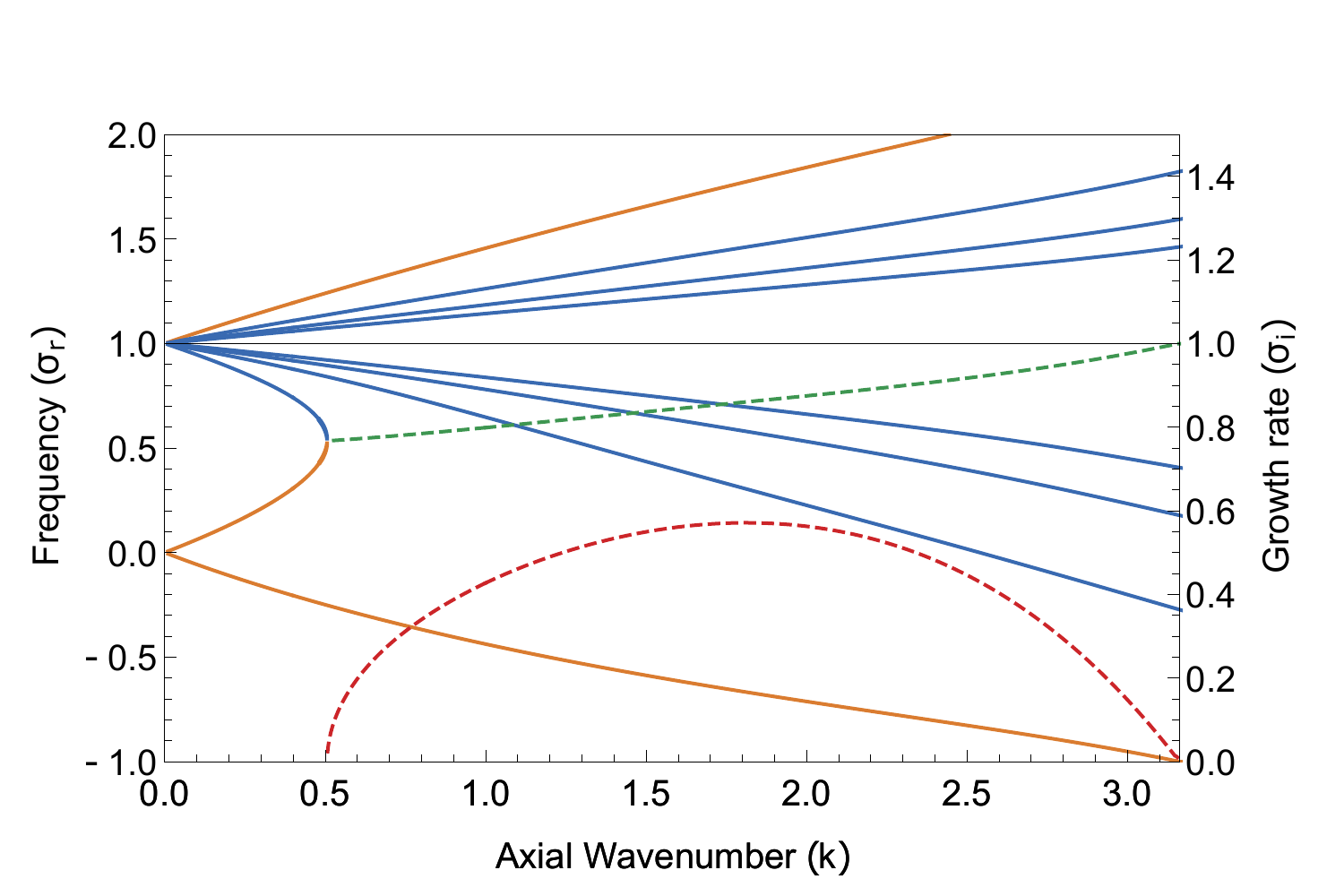}
      \caption{$We = 10$}
      \label{fig:n=1_dispCurves-4}
   \end{subfigure}
   \vfill
   \centering
   \begin{subfigure}{0.96\textwidth}
      \centering
      \includegraphics[width=\textwidth]{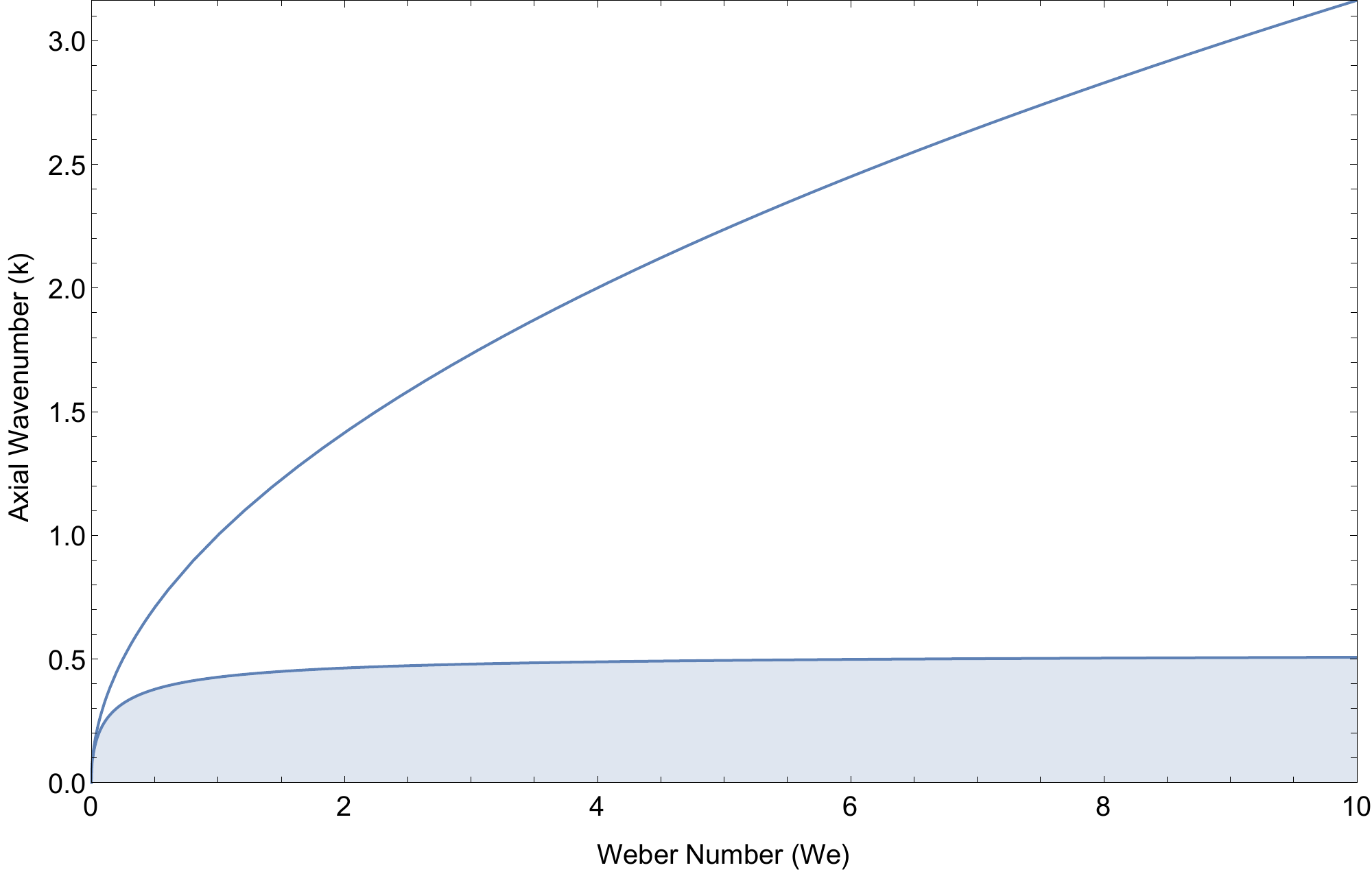}
      \caption{Region of stability for n=1}
      \label{fig:n1_islands}
   \end{subfigure}
   \caption{Dispersion curves for $n = 1$ (Figs. (a-d)); the solid blue curves
   indicate the Coriolis modes, the solid orange curves indicate the capillary
   modes and the dashed curves indicate real (red) and imaginary (green) parts
   of the unstable modes. Fig. (e) depicts the stable island generated as a
   result of coalescence of the lowest Coriolis mode and the retrograde
   capillary mode.}
   \label{fig:n=1_dispCurves}
\end{figure}

\clearpage

\bibliography{references}
\bibliographystyle{jfm}

\end{document}